%% file: main.tex
\input harvmacMv3.tex
\input epsf.tex

\def\frac#1#2{{#1\over #2}}


\def\centretable#1{ \hbox to \hsize {\hfill\vbox{
                    \offinterlineskip \tabskip=0pt \halign{#1} }\hfill} }

\preprint{}

\vskip -.5in

\title Tree-level 11D supergravity amplitudes
from the pure spinor worldline

\author
Max Guillen\email{{\dagger\ddagger*}}{maxgui@chalmers.se}$^{\dagger\ddagger*}$,
Marcelo dos Santos\email{{\#*}}{mafsantos@ucdavis.edu}$^{\#*}$ and
Eggon Viana\email{{\star\circ*}}{eggon.viana@unesp.br}$^{\star\circ*}$

\address
${\dagger}$ Department of Physics and Astronomy, Uppsala University, 75108 Uppsala, Sweden

\vskip .1in
$\ddagger$ Department of Mathematical Sciences, Chalmers University of Technology and the University of Gothenburg, SE-412 96 Gothenburg, Sweden

\vskip .1in
${*}$ ICTP South American Institute for Fundamental Research
Instituto de F\'{i}sica Te\'{o}rica, UNESP-Universidade Estadual Paulista
R. Dr. Bento T. Ferraz 271, Bl. II, S\~{a}o Paulo 01140-070, SP, Brazil

\vskip .1in
${\#}$ Center for Quantum Mathematics and Physics (QMAP)
Department of Physics \& Astronomy, University of California, Davis, CA 95616 USA

\vskip .1in
${\star}$ Instituto Gallego de F\'{i}sica de Altas Energ\'{i}as (IGFAE), Spain

\vskip .1in
${\circ}$ Department of Mathematical Sciences, Durham University, Durham DH1 3LE, UK

\vskip -.1in

\abstract
We develop a pure spinor worldline formalism for computing tree-level scattering amplitudes in 11D supergravity. Focusing first on the 4-point amplitude, we demonstrate that our prescription is consistent with BRST symmetry and gauge invariance, and that the resulting expression is invariant under permutation of the external particles. Remarkably, the amplitude admits a compact representation in pure spinor superspace and agrees precisely with the result obtained via perturbiner methods. We further extend our construction to the N-point case, proposing a general correlator that preserves BRST closure and gauge invariance, thereby offering a systematic framework for higher-point computations in 11D supergravity.


\bigskip
\bigskip
\bigskip
\bigskip
\vskip -.2in
\Date {August 2025}

\newif\iffig
\figfalse



\lref\ICTP{
N.~Berkovits,
``ICTP lectures on covariant quantization of the superstring,''
ICTP Lect. Notes Ser. 13 (2003) 57 [hep-th/0209059].
}

\lref\pselevenparticle{
M.~Guillen,
``Equivalence of the 11D pure spinor and Brink-Schwarz-like superparticle cohomologies,''
Phys. Rev. D {\bf 97}, no.6, 066002 (2018).
[arXiv:1705.06316 [hep-th]].
}

\lref\superpoincarequantization{
N.~Berkovits,
``Super Poincare covariant quantization of the superstring,''
JHEP {04}, 018 (2000). [arXiv:hep-th/0001035 [hep-th]].
}

\lref\pssupermembrane{
N.~Berkovits,
``Towards covariant quantization of the supermembrane,''
JHEP {\bf 09}, 051 (2002).
[arXiv:0201151 [hep-th]].
}

\lref\neight{
M.~Cederwall,
``N=8 superfield formulation of the Bagger-Lambert-Gustavsson model,''
JHEP {\bf 09}, 116 (2008).
[arXiv:0808.3242 [hep-th]].
}

\lref\nsix{
M.~Cederwall,
``Superfield actions for N=8 and N=6 conformal theories in three dimensions,''
JHEP {\bf 10}, 070 (2008).
[arXiv:0809.0318 [hep-th]].
}

\lref\nfour{
M.~Cederwall,
``An off-shell superspace reformulation of D=4, N=4 super-Yang-Mills theory,''
Fortsch. Phys. {\bf 66}, no.1, 1700082 (2018).
[arXiv:1707.00554 [hep-th]].
}

\lref\Mizera{
S.~Mizera and B.~Skrzypek
``Perturbiner Methods for Effective Field Theories and the Double Copy''
JHEP 10 (2018) 018.
}

\lref\pssugra{
M.~Cederwall,
``D=11 supergravity with manifest supersymmetry,''
Mod. Phys. Lett. A {\bf 25}, 3201-3212 (2010).
[arXiv:1001.0112 [hep-th]].
}

\lref\psborninfeld{
M.~Cederwall and A.~Karlsson,
``Pure spinor superfields and Born-Infeld theory,''
JHEP {\bf 11}, 134 (2011).
[arXiv:1109.0809 [hep-th]].
}

\lref\mafraone{
N.~Berkovits and C.~R.~Mafra,
``Some Superstring Amplitude Computations with the Non-Minimal Pure Spinor Formalism,''
JHEP {\bf 11}, 079 (2006).
[arXiv:hep-th/0607187 [hep-th]].
}

\lref\mafratwo{
H.~Gomez and C.~R.~Mafra,
``The Overall Coefficient of the Two-loop Superstring Amplitude Using Pure Spinors,''
JHEP {\bf 05}, 017 (2010).
[arXiv:1003.0678 [hep-th]].
}

\lref\mafrathree{
H.~Gomez and C.~R.~Mafra,
``The closed-string 3-loop amplitude and S-duality,''
JHEP {\bf 10}, 217 (2013).
[arXiv:1308.6567 [hep-th]].
}

\lref\rnspsone{
N.~Berkovits,
``Covariant Map Between Ramond-Neveu-Schwarz and Pure Spinor Formalisms for the Superstring,''
JHEP {\bf 04}, 024 (2014).
[arXiv:1312.0845 [hep-th]].
}

\lref\rnspstwo{
N.~Berkovits,
``Manifest spacetime supersymmetry and the superstring,''
JHEP {\bf 10}, 162 (2021).
[arXiv:2106.04448 [hep-th]].
}

\lref\maxmaor{
M.~Ben-Shahar and M.~Guillen,
``10D super-Yang-Mills scattering amplitudes from its pure spinor action,''
JHEP {\bf 12}, 014 (2021).
[arXiv:2108.11708 [hep-th]].
}

\lref\MafraNpoint{
C.~R.~Mafra, O.~Schlotterer, and S.~Stieberger,
``Complete N-Point Superstring Disk Amplitude I. Pure Spinor Computation''
Nucl. Phys. B {\bf 873} (2013) 419-460.
[arXiv:1106.2645 [hep-th]].
}

\lref\mafraoli{
C.~R.~Mafra and O.~Schlotterer,
``multi-particle SYM equations of motion and pure spinor BRST blocks,''
JHEP {\bf 07}, 153 (2014).
[arXiv:1404.4986 [hep-th]].
}

\lref\dynamical{
N.~Berkovits,
``Dynamical twisting and the b ghost in the pure spinor formalism,''
JHEP {\bf 06}, 091 (2013).
[arXiv:1305.0693 [hep-th]].
}

\lref\xiyin{
C.~M.~Chang, Y.~H.~Lin, Y.~Wang and X.~Yin,
``Deformations with Maximal Supersymmetries Part 2: Off-shell Formulation,''
JHEP {\bf 04}, 171 (2016).
[arXiv:1403.0709 [hep-th]].
}

\lref\chiralmax{
M.~Guillen,
``Green-Schwarz and pure spinor formulations of chiral strings,''
JHEP {\bf 12}, 029 (2021).
[arXiv:2108.11724 [hep-th]].
}

\lref\bcjone{
Z.~Bern, J.~J.~M.~Carrasco, and H.~Johansson,
``New Relations for Gauge-Theory Amplitudes,''
Phys. Rev. D {\bf 78}, 085011 (2008).
[arXiv:0805.3993 [hep-ph]].
}

\lref\bcjtwo{
Z.~Bern, J.~J.~M.~Carrasco, and H.~Johansson,
``Perturbative Quantum Gravity as a Double Copy of Gauge Theory,''
Phys. Rev. Lett. {\bf 105}, 061602 (2010).
[arXiv:1004.0476 [hep-th]].
}

\lref\bcjthree{
Z.~Bern, J.~J.~Carrasco, M.~Chiodaroli, H.~Johansson, and R.~Roiban,
``The Duality Between Color and Kinematics and its Applications,''
[arXiv:1909.01358 [hep-th]].
}

\lref\maximalloopcederwall{
M.~Cederwall and A.~Karlsson,
``Loop amplitudes in maximal supergravity with manifest supersymmetry,''
JHEP {\bf 03}, 114 (2013).
[arXiv:1212.5175 [hep-th]].
}

\lref\maxnotesworldline{
M.~Guillen,
``Notes on the 11D pure spinor wordline vertex operators,''
JHEP {\bf 08}, 122 (2020).
[arXiv:2006.06022 [hep-th]].
}

\lref\OdaTonin{
I.~Oda and M.~Tonin,
``On the Berkovits covariant quantization of GS superstring,''
Phys. Lett. B {\bf 520}, 398-404 (2001).
[arXiv:hep-th/0109051 [hep-th]].
}

\lref\perturbiner{
A.~A.~Rosly and K.~G.~Selivanov,
``On amplitudes in selfdual sector of Yang-Mills theory,''
Phys. Lett. B {\bf 399}, 135-140 (1997).
[arXiv:9611101 [hep-th]].
}
\lref\NMPS{
N.~Berkovits, ``Pure spinor formalism as an N=2 topological string,''
JHEP {\bf 10}, 089 (2005). [arXiv:0509120 [hep-th]].
}

\lref\elevendsimplifiedb{
N.~Berkovits and M.~Guillen,
``Simplified $D = 11$ pure spinor $b$ ghost,''
JHEP {\bf 07}, 115 (2017).
[arXiv:1703.05116 [hep-th]].
}
\lref\brinkschwarz{
L.~Brink and J.~H.~Schwarz,
``Quantum Superspace,''
Phys. Lett. B {\bf 100}, 310-312 (1981).
}

\lref\brinkhowe{
L.~Brink and P.~S.~Howe,
``Eleven-Dimensional Supergravity on the Mass-Shell in Superspace,''
Phys. Lett. B {\bf 91}, 384-386 (1980).
}

\lref\quartet{
T.~Kugo and I.~Ojima,
``Local Covariant Operator Formalism of Nonabelian Gauge Theories and Quark Confinement Problem,''
Prog. Theor. Phys. Suppl. {\bf 66}, 1-130 (1979).
}

\lref\cederwallequations{
N.~Berkovits and M.~Guillen,
``Equations of motion from Cederwall's pure spinor superspace actions,''
JHEP {\bf 08}, 033 (2018).
[arXiv:1804.06979 [hep-th]].
}

\lref\pssreview{
M.~Cederwall,
``Pure spinor superfields -- an overview,''
Springer Proc. Phys. {\bf 153}, 61-93 (2014).
[arXiv:1307.1762 [hep-th]].
}

\lref\tendsupertwistors{
N.~Berkovits,
``Ten-Dimensional Super-Twistors and Super-Yang-Mills,''
JHEP {\bf 04}, 067 (2010).
[arXiv:0910.1684 [hep-th]].
}

\lref\maxdiegoone{
D.~Garc\'\i{}a Sep\'ulveda and M.~Guillen,
``A pure spinor twistor description of the $D = 10$ superparticle,''
JHEP {\bf 08}, 130 (2020).
[arXiv:2006.06023 [hep-th]].
}

\lref\maxdiegotwo{
D.~G.~Sep\'ulveda and M.~Guillen,
``A Pure Spinor Twistor Description of Ambitwistor Strings''. [arXiv:2006.06025 [hep-th]].
}

\lref\nmmax{
N.~Berkovits, M.~Guillen, and L.~Mason,
``Supertwistor description of ambitwistor strings,''
JHEP {\bf 01}, 020 (2020).
[arXiv:1908.06899 [hep-th]].
}

\lref\maxmasoncasaliberkovits{
N.~Berkovits, E.~Casali, M.~Guillen, and L.~Mason,
``Notes on the $D=11$ pure spinor superparticle,''
JHEP {\bf 08}, 178 (2019).
[arXiv:1905.03737 [hep-th]].
}

\lref\maxthesis{
M.~Guillen,
``Pure spinors and $D=11$ supergravity''. [arXiv:2006.06014 [hep-th]].
}

\lref\Berkovitsparticle{
N.~Berkovits, ``Covariant quantization of the superparticle using pure spinors,'' 
JHEP {\bf 09}, 016 (2001). [arXiv:0105050 [hep-th]].
}

\lref\BenShahar{
M.~Ben-Shahar and M.~Guillen, ``Superspace expansion of the 11D linearized superfields in the pure spinor formalism, and the covariant vertex operator,'' 
JHEP {\bf 09}, 018 (2023). [arXiv:2305.19898 [hep-th]].
}

\lref\mafracohomology{
C.~R.~Mafra,
``Towards Field Theory Amplitudes From the Cohomology of Pure Spinor Superspace,''
doi.10.1007/JHEP11(2010)096.
}

\lref\Berkovitstopological{
N.~Berkovits,
``Pure Spinor Formalism as an N = 2 Topological String,'' JHEP
0510(2005) 089, hep-th/0509120
}

\lref\selivanov{
K.~G.~Selivanov,
`` On tree form-factors in (supersymmetric) Yang-Mills theory,'' Commun.Math.Phys. 208 (2000) 671-687. [arXiv:9809046 [hep-th]].
}

\lref\Mafraalgorithm{
E.~Bridges and C.~R.~Mafra,
``Algorithmic construction of SYM multi-particle superfields in the BCJ gauge,'' JHEP 10 (2019) 022
}

\lref\ghostnumberzero{
M.~Guillen, M.~dos~Santos, and E.~Viana,
``The 11D pure spinor ghost number zero vertex operator''. [arXiv:2508.19744 [hep-th]].
}

\lref\mafraphd{
C.~R.~Mafra,
``Superstring Scattering Amplitudes with the Pure Spinor Formalism'', PhD thesis, Sao Paulo IFT
}

\lref\maframultiparticles{
S.~Lee, C.~R.~Mafra, and O.~Schlotterer
``Non-linear gauge transformations in D=10 SYM theory and the BCJ duality,'' JHEP 03 (2016) 090
}

\lref\Diana{
Y.~Du and D.~Vaman.
``Tree-level Graviton Scattering in the Worldline Formalism."
8 2023. 2308.11326 [hep-th]
}

\lref\wittentwistor{
E.~Witten,
``Twistor-Like Transform In Ten-Dimensions''
Nucl.Phys. B 266, 245 (1986).
}

\lref\Siegelsuperfields{
W.~Siegel,
``Superfields in Higher Dimensional Space-time''
Phys. Lett. B 80, 220
(1979).
}

\lref\MafratowardsI{
C.~R.~Mafra and O.~Schlotterer,
``Towards the N-point one-loop superstring amplitude. Part I. Pure spinors and superfield kinematics,''
JHEP {\bf 08}, 090 (2019). [arXiv:1812.10969 [hep-th]].
}

\lref\MafratowardsII{
C.~R.~Mafra and O.~Schlotterer,
``Towards the N-point one-loop superstring amplitude.
Part II. Worldsheet functions and their duality to kinematics,''
JHEP {\bf 08}, 091 (2019). [arXiv:1812.10970 [hep-th]].
}

\lref\Mafratwoloops{
E.~D’Hoker, C.~R.~Mafra, B.~Pioline, and O.~Schlotterer,
``Two-loop superstring fivepoint amplitudes. Part I. Construction via chiral splitting and pure spinors,''
JHEP {\bf 08}, 135 (2020). [arXiv:2006.05270 [hep-th]].
}

\lref\MafraGomes{
H.~Gomez and C.~R.~Mafra,
``The closed-string 3-loop amplitude and S-duality,''
JHEP {\bf 10}, 217 (2013). [arXiv:1308.6567 [hep-th]].
}

\lref\Siegelworldline{
P.~Dai, Y.~Huang, and W.~Siegel,
``Worldgraph approach to Yang-Mills amplitudes from N= 2 spinning particle''
Journal of High Energy Physics  (2008) 010
}

\lref\MafraNMoneloop{
C.~R.~Mafra and C.~Stahn,
``The One-loop Open Superstring Massless Five-point Amplitude
with the Non-Minimal Pure Spinor Formalism,''
JHEP {\bf 03}, 126 (2009). [arXiv:0902.1539 [hep-th]].
}

\lref\BerkovitsPS{
N.~Berkovits,
``Super Poincare covariant quantization of the superstring,''
JHEP {\bf 04}, 018 (2000).
[arXiv:0001035 [hep-th]].
}

\lref\tamingbghos{
M.~Guillen,
``Taming the 11D pure spinor b-ghost,'' 
JHEP {\bf 03}, 135 (2023). [arXiv:2212. 13653 [hep-th]].
}

\lref\BG{
F.~A.~Berends and W.~T.~Giele,
``Recursive Calculations for Processes with n Gluons'',
Nucl. Phys. B306 (1988) 759.
}

\lref\measureeleven{
M.~Cederwall,
``Towards a manifestly supersymmetric action for 11-dimensional supergravity,''
JHEP {\bf 01} (2010) 117. [arXiv:0912.1814 [hep-th]].
}

\lref\Nahm{
W.~Nahm,
``Supersymmetries And Their Representations,''
Nucl. Phys. B {\bf 135} (1978) 149.
}

\lref\Scherk{
E.~Cremmer, B.~Julia, and J.~Scherk,
``Supergravity Theory In 11 Dimensions,''
Phys. Lett. 76B (1978) 409.
}

\lref\TownsendM{
C.~Hull and P.~Townsend,
``Unity of superstring dualities,''
Nucl. Phys. B 438 (1995) 109–137. [arXiv:9410167 [hep-th]].
}

\lref\WittenM{
E.~Witten,
``String theory dynamics in various dimensions,''
Nucl. Phys. B 443 (1995) 85–126. [arXiv:9503124 [hep-th]].
}

\lref\tendimensions{
M.~Guillen, M.~dos~Santos, and E.~Viana.
``The pure spinor superparticle and 10D super-Yang-Mills amplitudes''. [arXiv:2508.19601 [hep-th]].
}

\lref\Duff{
M.~Duff, P.~S.~Howe, T.~Inami, and K.~Stelle,
``Superstrings in D=10 from Supermembranes in D=11,''
Phys. Lett. B 191 (1987) 70.
}

\lref\Bergshoeff{
E.~Bergshoeff, E.~Sezgin, and P.~Townsend,
``Supermembranes and Eleven-Dimensional Supergravity,''
Phys. Lett. B 189 (1987) 75–78.
}

\lref\Grassi{
L.~Anguelova, P.~A.~Grassi, and P.~Vanhove,
``Covariant One-loop Amplitudes in D=11,''
Nucl. Phys. B {\bf 702} (2004) 269–306. [arXiv:0408171 [hep-th]].
}

\lref\Bjornsson{
J.~Bjornsson, 
``Multi-loop Amplitudes in Maximally Supersymmetric Pure
Spinor Field Theory,"
JHEP {\bf 01}, 002 (2011). [arXiv:1009.5906 [hep-th]].
}

\lref\perturbinereleven{
M.~Guillen, M.~dos~Santos, and E.~Viana.
``Tree-level 11D supergravity amplitudes from the pure spinor worldline,'' to appear.
}

\lref\Strassler{
M.~J.~Strassler.
``Field Theory Without Feynman diagrams: One-Loop Effective Actions."
Nuclear Physics B, 385(1–2):145–184, October 1992.
}

\lref\RennanI{
Hu.~Gomez and R.~L.~Jusinskas.
``Multi-particle Solutions to Einstein’s Equations."
Phys.Rev.Lett. 127 (2021) 18, 181603.
}

\lref\RennanII{
H.~Gomez, R.~L.~Jusinskas, and A.~Q.~Velez.
``One-Loop N-Point Correlators in Pure Gravity."
Phys.Rev.Lett. 134 (2025) 11, 111602.
}

\lref\berkovitsnekrasov{
N.~Berkovits and N.~Nekrasov,
``Multiloop superstring amplitudes from non-minimal pure spinor formalism,''
JHEP {\bf 12}, 029 (2006). [arXiv:0609012 [hep-th]].
}

\lref\guillenchiral{
M.~Guillen,
``Green-Schwarz and pure spinor formulations of chiral strings,''
JHEP {\bf 12}, 029 (2021).
[arXiv:2108.11724 [hep-th]].
}

\font\mbb=msbm10 
\newfam\bbb
\textfont\bbb=\mbb

\def\startcenter{%
  \par
  \begingroup
  \leftskip=0pt plus 1fil
  \rightskip=\leftskip
  \parindent=0pt
  \parfillskip=0pt
}
\def\stopcenter{%
  \par
  \endgroup
}

\listtoc
\writetoc
\filbreak

\newsec Introduction

\seclab\secone

Eleven-dimensional supergravity was first proposed in \Nahm\ and explicitly constructed in \Scherk. It was later conjectured to arise as the low-energy limit of M-theory \refs{\WittenM,\TownsendM}, an eleven-dimensional quantum framework unifying all of the ten-dimensional superstring theories. Within this context, the supermembrane was identified as a candidate for the fundamental object of the theory \refs{\Duff,\Bergshoeff}. Despite its importance, eleven dimensional supergravity remains notoriously difficulty to analyze, and progress on M-theory crucially depends on advancing our understanding of this unique maximal supergravity.

\medskip
A supersymmetric formulation of the supermembrane based on the pure spinor formalism was introduced in \pssupermembrane. When the worldvolume fields are taken to be independent of the spatial coordinates, this reduces to the eleven-dimensional pure spinor superparticle, which provides a Batalin-Vilkovisky description of eleven-dimensional supergravity. In this setting, physical states correspond to ghost number three in the pure spinor BRST cohomology \pssupermembrane. An action principle with manifest supersymmetry reproducing eleven-dimensional supergravity has also been constructed using non-minimal pure spinor variables \refs{\pssreview,\pssugra, \cederwallequations}.




\medskip
\noindent Despite the extensive efforts over the past years to further develop the 11D pure spinor program, many aspects of this formalism remain unclear or incomplete. In particular, a functional and concrete prescription for computing 11D scattering amplitudes is still unknown. It is worth mentioning that the authors of \Grassi\ have proposed a candidate for the N-point correlation function within the minimal pure spinor setting. However, the correct vertex operators needed for the explicit evaluation of such a correlator (including the ghost number zero vertex operator) were never completely constructed. In fact, it was later shown that the original proposal, constructed with the minimal pure spinor variables, is incompatible with 11D supergravity \maxmasoncasaliberkovits.

\medskip
In contrast, the use of pure spinors in the context of the superstring \BerkovitsPS\ has proven to be quite successful for studying different properties of superstring theory. In particular, it provides a far more efficient and elegant framework for computing string scattering amplitudes than the traditional Ramond-Neveu-Schwarz and Green-Schwarz formalisms \refs{\MafraNpoint,\MafratowardsI, \MafratowardsII,\Mafratwoloops, \MafraGomes}. This remarkable feature is a direct consequence of the manifest supersymmetry realized by the pure spinor formulation. The construction was also adapted to the ten-dimensional superparticle in \Berkovitsparticle, where it has likewise proven invaluable for deriving scattering amplitudes in ten-dimensional super-Yang-Mills theory \refs{\Bjornsson,\maxmaor,\tendimensions}.

\medskip
In recent years, several techniques for computing scattering amplitudes have been developed or revived as alternatives to the traditional Feynman diagrams approach. Notable among these are the perturbiner formalism and worldline framework. The latter utilizes the path integral of a single particle, with operator insertions representing the effects of background fields. N-particle interactions are then obtained from N-point correlation functions of single- and multi-particle vertex operators, defined through consistency conditions and expanded order by order in the number of fields.

\medskip
In this paper, we start a program that aims to find a consistent prescription for computing 11D supergravity scattering amplitudes from 11D worldline pure spinor correlators. We make use of the ghost number one and three single- and multi-particle vertex operators extensively studied by one of the authors in \refs{\maxnotesworldline, \maxmasoncasaliberkovits, \maxthesis, \BenShahar, \tamingbghos}, as well as of the newly discovered ghost number zero vertex operator \ghostnumberzero. These building blocks together with a proper generalization of the ideas presented in \tendimensions\ to the 10D setting, are then used to introduce a pure spinor prescription for the 11D 4-point function at tree level. We show that our pure spinor correlator successfully passes a number of consistency checks, namely invariance under BRST and gauge transformations. Furthermore, we discuss and display in detail the full evaluation of this correlator in pure spinor superspace, and verify that it satisfies BRST closure and exhibits full permutation symmetry. Remarkably, our formula is found to reproduce the 4-point scattering amplitude obtained via perturbiner methods. Based on our findings, we also propose a tree-level N-point correlator within the 11D pure spinor worldline formalism, and argue for the validity and feasibility of our ansatz.
 

\medskip
The paper is organized as follows. In Section 2, we review the 11D pure spinor superparticle and discuss its BRST cohomology. We then introduce the non-minimal pure spinor variables, and present the b-ghost as well as the pure spinor vertex operators of ghost numbers one and zero. Next, we introduce the multi-particle vertex operators, with special focus on the ghost number one two-particle vertex operator, which proves to be essential for later computations. In Section 3, we provide a tree-level prescription for the 4-point pure spinor correlator, and evaluate it explicitly. We provide an alternative representation of our 11D amplitude which manifestly exhibits BRST invariance and full permutation symmetry. We also verify that our expression is in complete agreement with results obtained from perturbiners. In Section 4, we use our previous results as well as insights from the 10D setup \tendimensions, and propose a general N-point correlation function in 11D pure spinor superspace, consistent with BRST symmetry. We close with discussions and future work. This paper also contains three Appendices: Appendix A discusses integration-by-parts (IBP) techniques used in pure spinor correlators; Appendix B reviews 11D supergravity at linear and second orders, and finally Appendix C displays detailed calculations of the algebraic manipulations carried out in the correlation functions studied in Section 3.

\newsec Pure spinor superparticle in 11D

\seclab\sectwo

The 11D pure spinor superparticle action in a general background is given by \refs{\pselevenparticle,\pssupermembrane}
\eqnn \elevendpsaction
$$ \eqalignno{
S &= \int d\tau \left[ P_{M}\partial_{\tau}Z^{M} + w_{\alpha}\left(\partial_{\tau}\lambda^{\alpha} + \partial_{\tau}Z^{M}\Omega_{M,\beta}{}^{\alpha}\lambda^{\beta}\right) - \half P^2\right]. & \elevendpsaction
}
$$

\noindent We are using capital letters at the beginning/middle of the Latin alphabet to denote tangent/curved space indices. Likewise, Latin/Greek letters at the (middle) beginning of the alphabet indicate (curved) flat vector/spinor indices. The variables $Z^{M} = (X^{m},\theta^{\mu})$ represent the bosonic and fermionic coordinates of the 11D ordinary superspace, and their corresponding conjugate momenta are denoted by $P_{M} = (P_m, p_{\mu})$. The 11D vielbeins in superspace will be indicated by $E_{A}{}^{M}$ and $E_{M}{}^{A}$, and the spin-connection by $\Omega_{M,A}{}^{B}$. The 11D gamma matrices are represented by $(\gamma^{a})_{\alpha\beta}$, $(\gamma^{a})^{\alpha\beta}$, and they satisfy the Clifford algebra $(\gamma^{(a})_{\alpha\beta}(\gamma^{b)})^{\beta\delta} = \eta^{ab}\delta_{\alpha}^{\delta}$. The field $\lambda^{\alpha}$ is an 11D bosonic spinor satisfying the pure spinor constraint $\lambda\gamma^a \lambda = 0$. Due to this constraint, its conjugate momentum $w_{\alpha}$ is only defined up to the gauge transformation $\delta w_{\alpha} = (\gamma^{a}\lambda)_{\alpha}\rho_a$, for any vector $\rho_{a}$. The fields $\lambda^\alpha$ and $w_\alpha$ are ghost fields, and will be assigned to carry ghost numbers $1$ and $-1$, respectively. In a flat background, the superparticle action \elevendpsaction\ simplifies to
\eqnn \elevendpsactionflat
$$ \eqalignno{
S &= \int d\tau \left[ P_{a}\partial_{\tau}X^{a} + p_{\alpha}\partial_{\tau}\theta^{\alpha} + w_{\alpha}\partial_{\tau}\lambda^{\alpha}  - \half P^2\right]. & \elevendpsactionflat
}
$$

In this work, we will employ the path integral formalism. In this setup, the fields satisfy the following free field contractions
\eqnn\wick
$$\eqalignno{
P_a(\tau_1) X^b(\tau_2) & \sim - \delta_a^{b} \sigma_{12}, & \cr
p_{\alpha}(\tau_1) \theta^{\beta}(\tau_2) & \sim \delta_{\alpha}^{\beta} \sigma_{12} , & \cr
\omega_{\alpha}(\tau_1) \lambda^{\beta}(\tau_2) & \sim - \delta_{\alpha}^{\beta}\sigma_{12}, &\wick
}
$$
where $\sigma_{ij} = \half$sign$(\tau_i -\tau_j)$. These contractions are equivalent to the canonical commutation relations. This can be checked by using the map between graded commutators and operator insertions on the path integral, given by
\eqnn\mapcommutatorspathintegral
$$\eqalignno{
[A(\tau), B(\tau)\} & \sim A(\tau+\epsilon)B(\tau) \mp B(\tau)A(\tau-\epsilon), &\mapcommutatorspathintegral
}
$$
where $\epsilon \to 0$ and the sign is $-$ for commutators, and $+$ for anticommutators.
\medskip
The action \elevendpsactionflat\ is supplemented by the BRST charge $Q_{0} = \lambda^{\alpha} d_{\alpha}$, where $d_{\alpha}=p_\alpha - \half(\gamma^a\theta)_\alpha P_a$
is the Brink-Schwarz fermionic constraint \brinkschwarz. It has been shown that the quantization of this model reproduces all the fields and antifields of the Batalin-Vilkovisky description of linearized 11D supergravity \pssupermembrane. In particular, the antifield of the ghost-for-ghost-for-ghost associated to the gauge symmetry in 11D supergravity, is located at the ghost number seven sector, and it represents the top cohomology of the BRST charge. Due to this, it has been suggested as a candidate for defining manifestly supersymmetric correlation functions in 11D, as we will discuss in next section.

\subsec Non-minimal variables and the physical operators

\subseclab\sectwoone

The action \elevendpsactionflat\ admits a topological extension preserving the space of physical states. This modification of the pure spinor worldline is obtained after adding the pair of conjugate bosonic spinors $(\bar{\lambda}_{\alpha}, \bar{w}^{\alpha})$, and the pair of conjugate fermionic spinors $(r_{\alpha}, s^{\alpha})$, subject to the constraints $\bar\lambda \gamma^{a}\bar\lambda = \bar{\lambda}\gamma^{a}r = 0$. The non-minimal version of the pure spinor superparticle \elevendpsactionflat\ then reads \refs{\elevendsimplifiedb,\pssupermembrane}
\eqnn \nonminimalpssuperparticle
$$
\eqalignno{
S &= \int d\tau \left[P_{a}\partial_{\tau}X^{a} + p_{\alpha}\partial_{\tau}\theta^{\alpha} + w_{\alpha}\partial_{\tau}\lambda^{\alpha} + \bar{w}^{\alpha}\partial_{\tau}\bar{\lambda}_{\alpha} + s^{\alpha}\partial_{\tau}r_{\alpha} - \half P^2\right]. & \nonminimalpssuperparticle
}
$$
The non-minimal BRST charge takes the new form
\eqnn \nonminimalbrstcharge
$$
\eqalignno{
Q &= Q_{0} + r_{\alpha}\bar{w}^{\alpha}, & \nonminimalbrstcharge
}
$$
so that the cohomologies of $Q$ and $Q_{0}$ are equivalent to each other. The action \nonminimalpssuperparticle\ is invariant under the global symmetry $J = w_{\alpha}\lambda^{\alpha} -\bar{w}^{\alpha}\bar{\lambda}_{\alpha}$, referred to as the non-minimal ghost number charge. Hence, $\bar{\lambda}_{\alpha}$ and $r_{\alpha}$ carry ghost numbers $-1$ and $0$, respectively.

\medskip
One of the main advantages of the non-minimal framework is that it allows for the construction of Lorentz covariant operators of negative ghost number. In particular, one can construct the so-called b-ghost, which satisfies the standard relation
\eqnn\Qb
$$\eqalignno{
&\{Q, b\} = \half P^{2}.&\Qb
}$$
The explicit form of the b-ghost was originally found in \maximalloopcederwall. The solution contains intricate algebraic expressions involving pure spinors and gamma matrices, which make it untractable for practical applications. This quite complicated formula was dramatically simplified in \refs{\elevendsimplifiedb,\tamingbghos} after introducing the so-called physical operators \psborninfeld. Since these operators will be relevant for our computations later, we shortly review them next.

\medskip
The 11D pure spinor physical operators will be denoted by ${\bf C}^a, {\bf C}_\alpha, {\bf \Phi}_c, {\bf C}^{ab}, {\bf \Omega}_{ab}, {\bf T}_{a}{}^{\alpha}$. These are defined as linear operators which carry negative ghost number and satisfy the following set of relations
\eqnn \drchatalpha
\eqnn \drchata
\eqnn \drphihata
\eqnn \drphihatalpha
\eqnn \dromegaab
$$ \eqalignno{
&[Q, {\bf C}_{\alpha}] = -{1\over3}d_{\alpha} - (\gamma^a \lambda)_{\alpha}{\bf C}_{a},  & \drchatalpha \cr
&\{Q, {\bf C}_{a}\} = {1\over3}P_{a} -(\lambda\gamma^{ab}\lambda){\bf \Phi}_{b}, 
 & \drchata \cr
&[Q, {\bf \Phi}^{a}] = (\lambda\gamma^a {\bf \Phi}), & \drphihata \cr
&\{Q, {\bf \Phi}^{\alpha}\} = {1\over 4}(\lambda\gamma^{ab})^{\alpha}{\bf \Omega}_{ab}, & \drphihatalpha\cr
&[Q, {\bf \Omega}_{ab}] = - (\lambda\gamma_{b})_{\delta}{\bf T}_{a}{}^{\delta}. & \dromegaab
}
$$
The physical operators are defined so that their action on the ghost number three vertex operator $U^{(3)}$ reproduces the field content of linearized supergravity, up to shift symmetry and BRST-exact terms. For instance, ${\bf C}_{\alpha} U^{(3)} = C_{\alpha}$ up to shift-symmetry. They can be used to construct superfield descriptions of supergravity in the pure spinor formalism. The reader might review linearized gravity in Appendix B.

The system of equations \drchatalpha-\dromegaab\ can be solved with the use of the non-minimal variables. Indeed, one finds
\eqnn \chatalphaexp
\eqnn \chataexp
\eqnn \phihataexp
\eqnn \phihatalphaexp
\eqnn \omegaabexp
\eqnn \taalphaexp
$$
\eqalignno{
{\bf C}_{\alpha} =& {w_{\alpha} \over 3} + {1\over 3\eta}(\lambda\gamma^{abc}w)(\bar{\lambda}\gamma_{bc}\bar{\lambda})(\lambda\gamma_{a})_{\alpha}, & \chatalphaexp \cr
{\bf C}_{a} =& {1\over 3\eta}(\bar{\lambda}\gamma^{bc}\bar{\lambda})(\lambda\gamma_{abc}d) - {2\over 3\eta}(\bar{\lambda}\gamma^{bc}r)(\lambda\gamma_{abc}w) + {2\over 3\eta^2}\phi (\bar{\lambda}\gamma^{bc}\bar{\lambda})(\lambda\gamma_{abc}w) & \cr
& +{4\over 3 \eta^{2}}(\lambda\gamma_{ac}\lambda)(\bar{\lambda}\gamma^{bc}\bar{\lambda})(\bar{\lambda}\gamma^{de}r)(\lambda\gamma_{bde}w),  & \chataexp \cr
{\bf \Phi}^{a} =& -{2 \over 3}\bigg[{1 \over \eta}(\bar{\lambda}\gamma^{ab}\bar{\lambda})P_{b} - {2 \over \eta^{2}}(\bar{\lambda}\gamma^{ab}\bar{\lambda})(\bar{\lambda}\gamma^{cd}r)(\lambda\gamma_{bcd}d) - \{s, {2\over \eta^2}(\bar{\lambda}\gamma^{ab}\bar{\lambda})(\bar{\lambda}\gamma^{cd}r)\}(\lambda\gamma_{bcd}w) & \cr
&\ \ \ \ \ - {8 \over \eta^3}(\lambda\gamma^{a}\xi_{b})(\bar{\lambda}\gamma^{cb}r)(\bar{\lambda}\gamma^{de}r)(\lambda\gamma_{cde}w)\bigg], & \phihataexp \cr
{\bf \Phi}^{\alpha} =& -{8 \over 3}\xi_{a}^{\alpha}\bigg[{1\over \eta^{2}}(\bar{\lambda}\gamma^{ab}r)P_{b} - {2 \over \eta^{4}}(\bar{\lambda}\gamma^{ab}r)(\bar{\lambda}\gamma^{cd}r)(\lambda\gamma_{bcd}d) & \cr
&\ \ \ \ \ - \bigg({8 \over \eta^{4}} (\bar{\lambda}\gamma^{ab}r)(\lambda\gamma_{cb}\lambda)(\bar{\lambda}\gamma^{cd}r)(\bar{\lambda}\gamma^{ef}r) + {8 \over \eta^{5}}\phi(\bar{\lambda}\gamma^{ab}r) (\bar{\lambda}\gamma^{cd}r)\bigg)(\lambda\gamma_{bcd}w)\bigg], &  \phihatalphaexp\cr
{\bf\Omega}_{ab} =& -{32\over \eta}\bigg[({\bf \Phi}\gamma_{[ak}\lambda)(\bar{\lambda}\gamma_{b]}{}^{k}r) + {1\over 8}({\bf \Phi}\gamma_{abcd}\lambda)(\bar{\lambda}\gamma^{cd}r)\bigg], &\omegaabexp\cr
{\bf T}_{a}{}^{\alpha} = &\ \ {2\over \eta}(\gamma^{bcd}\lambda)^{\alpha}(\bar{\lambda}\gamma_{cd}r){\bf \Omega}_{ab}, & \taalphaexp
}
$$
where $\phi = (\lambda\gamma^{ab}\lambda)(\bar{\lambda}\gamma_{ab}r)$, $\xi_{a}^{\beta} = \half (\gamma_{abc}\lambda)^{\beta}(\bar{\lambda}\gamma^{bc}\bar{\lambda})$ and $\eta = (\lambda\gamma^{ab}\lambda)(\bar\lambda\gamma_{ab}\bar\lambda)$. Although these might seem very complicated, most of the manipulations can be performed with ease by using the defining properties of the physical operators.


As shown in \tamingbghos, one can use the physical operators to write the 11D b-ghost in the simple and compact way
\eqnn\bghost
$$\eqalignno{
& b = {3\over2}P^a{\bf C}_a + {3\over2}(\lambda\gamma^ad){\bf \Phi}_a - {3\over2}(\lambda\gamma^aw)(\lambda\gamma_a{\bf \Phi}). &\bghost
}$$
In this form, it is straightforward to check that $\{Q, b\} = \half P^2$, and $b$ is nilpotent up to BRST-exact terms. Throughout this paper, we will use this version of the b-ghost. 
\subsec Single-particle vertex operators

\subseclab\sectwotwo

The pure spinor superfield in the BRST cohomology which describes the physical fields of linearized 11D supergravity carries ghost number three. It can be written as
\eqnn\operatorthree
$$\eqalignno{
&U^{(3)} = \lambda^{\alpha}\lambda^{\beta}\lambda^{\delta}C_{\alpha\beta\delta},&\operatorthree
}$$
where $C_{\alpha \beta \delta}$ is the lowest-dimensional component of the 3-form superfield of 11D supergravity. This fact can easily been seen from the physical state conditions
\eqnn \closed
\eqnn \exact
$$ \eqalignno{
\{Q,U^{(3)}\} = 0  &\Rightarrow D_{(\alpha}C_{\beta\delta\epsilon)} = (\gamma^{a})_{(\alpha\beta}C_{a\delta\epsilon)}, & \closed \cr
\delta U^{(3)} = [Q,\Sigma] &\Rightarrow \delta C_{\alpha\beta\delta} = D_{(\alpha}\Sigma_{\beta\delta)}, & \exact
}
$$
where $\Sigma = \lambda^{\alpha}\lambda^{\beta}\Sigma_{\alpha\beta}$, and $\Sigma_{\alpha\beta}$ is any superfield. Eqns. \closed\ and \exact\ above are nothing but the physical superspace constraints of linearized 11D supergravity. For more details, see Appendix B.1. 

\medskip
For later reference, $U^{(3)}$ can also be shown to satisfy 
\eqnn\partialUthree
\eqnn\DUthree
$$\eqalignno{
&\partial_a U^{(3)} - 3[Q,C_a] = 3(\lambda\gamma_{ab}\lambda)\phi^b,&\partialUthree\cr
&D_\alpha U^{(3)} + 3\{Q,C_\alpha\} = -3(\lambda\gamma^a)_\alpha C_a,&\DUthree
}$$
where $C_a = \lambda^\alpha\lambda^\beta C_{a\alpha\beta}$, $C_\alpha = \lambda^\beta\lambda^\gamma C_{\alpha\beta\gamma}$, with $(C_{a\alpha\beta}, C_{\alpha\beta\gamma})$ being components of the linearized 11D super 3-form, and $\phi^a = \lambda^\alpha h_\alpha{}^a$, with $h_\alpha{}^a$ being the first-order correction of the vielbein.


\medskip
The ghost number one vertex operator was introduced in \maxmasoncasaliberkovits\ as the first-order perturbation of the superparticle BRST charge coupled to a curved background. It takes the explicit form
\eqnn \ghostnumberonevo
$$\eqalignno{
U^{(1)} &= P_a\phi^a + d_\alpha \phi^\alpha - \half \phi_{ab}N^{ab}, &\ghostnumberonevo
}
$$
where $N^{ab} = \lambda\gamma^{ab}w$ is the pure spinor Lorentz generator. We also define 
$\phi^\alpha = \lambda^\beta h_{\beta}{}^{\alpha}$, and $\phi_{ab} = \lambda^\alpha(\gamma_{ab})^\beta{}_\delta\Omega_{\alpha\,\beta}{}^{\delta}$, where $h_{\alpha}{}^{\beta}$ and $ \Omega_{\alpha\,\beta}{}^{\delta}$ are the linearized versions of the respective components of the vielbeins and spin-connection. By construction, this operator is BRST-closed
\eqnn\descendone
$$\eqalignno{
& \{Q,U^{(1)}\}=0. &\descendone
}$$
As a double check, one can use the equations of motion listed in Appendix B to prove the validity of eqn. \descendone.

\medskip
More recently, the ghost number zero vertex operator was constructed by using the physical operators studied above. It takes the strikingly compact form \ghostnumberzero
\eqnn \vertexoperatorgzero
$$
\eqalignno{
U^{(0)} &= 3\Bigg[P^{a}{\bf C}^{b}\Omega_{ab} + (\lambda\gamma^{bc}\lambda)P^{a}h_{ab}{\bf \Phi}_{c} - (\lambda\gamma_{a})_{\delta}\lambda^{\alpha}T_{b\alpha}{}^{\delta}{\bf C}^{ab} + P^{a}{\bf C}_{b}\partial_{a}\phi^{b} - P^{a}{\bf C}_{\alpha}\partial_{a}\phi^{\alpha}\Bigg]. &\cr 
& \ \ \ \ \ \ \ \ \ \ &\vertexoperatorgzero
}
$$
The numerical factor in front of the square brackets was chosen for convenience. Fundamentally, this operator satisfies the standard descent equation with $U^{(1)}$, namely
\eqnn\descendzero
$$\eqalignno{
& [Q,U_i^{(0)}] = \partial_{\tau} U_i^{(1)}.&\descendzero
}$$
One can easily check this via eqns. \drchatalpha-\dromegaab. This relation will be crucial when showing that our amplitude prescription satisfies a number of consistency checks.

\subsec Multi-particle vertex operators

\subseclab\sectwothree

Following the ideas developed in the 10D case \tendimensions, our worldline prescription for computing 11D correlators will require the introduction of the so-called multi-particle vertex operators. These objects are essential when defining BRST invariant worldline correlation functions. The multi-particle vertices with ghost numbers zero and one will be denoted by $\bU^{(0)}$ and $\bU^{(1)}$, respectively. The superfield $\bU^{(0)}$ can be determined from the deformation of the free Hamiltonian, induced by the supergravity background at all orders in the coupling constant. Likewise, the superfield $\bU^{(1)}$ can be defined from the curved background deformation of the BRST operator $Q$, at all orders in the coupling constant.


\medskip
These multi-particle operators satisfy a number of constraints which can be deduced from the properties of nilpotency and time conservation of the curved BRST charge. Explicitly,
\eqnn\hamiltonianbrst
$$\eqalignno{
\left\{\tilde{Q}, \tilde{Q}\right\} &= 0, \ \ \ \ \ \  \quad \quad \left[\tilde{H}, \tilde{Q}\right] = 0, &\hamiltonianbrst
}
$$
where $\tilde{Q} = Q + \bU^{(1)}$ is the curved BRST charge, and $\tilde{H}= H + \bU^{(0)}$ is the curved Hamiltonian, with $H=\half P^{2}$.
Eqns. \hamiltonianbrst\ demand the following consistency relations  
\eqnn\consistenceBRSTone
\eqnn\consistenceBRSTzero
$$\eqalignno{
&\left\{Q, \bU^{(1)}\right\} = - \half \left\{\bU^{(1)}, \bU^{(1)}\right\},&\consistenceBRSTone\cr
&\left[Q, \bU^{(0)}\right] = \left[H,\bU^{(1)}\right] + \left[\bU^{(0)}, \bU^{(1)}\right].&\consistenceBRSTzero
}$$

\noindent These equations can be solved order by order in the coupling constant by employing the perturbiner method of Selivanov \perturbiner. In this approach, we expand the multi-particle vertex operators in terms of multi-particle asymptotic states labeled by words $P = p_1 \dots p_m$ carrying momentum $k_P = k_{p_1} + \dots + k_{p_m}$. Each letter $p_i$ in the word $P$, represents one particle in the corresponding state. These states can be written as plane wave factors multiplying linear operators depending on external polarizations. Concretely,
\eqnn\expansionsuperfieldseleven
$$\eqalignno{
&\bU^{(a)} = \sum_P U^{(a)}_Pe^{ik_P\cdot X(\tau)} = U^{(a)}_{i}e^{ik_i\cdot X(\tau)} + U^{(a)}_{ij}e^{ik_{ij}\cdot X(\tau)} + \cdots , & \expansionsuperfieldseleven
}$$
where $a=0,1$ is the ghost number. We refer to the number of particles in a letter $P$ as the rank of the associated operator. After plugging the expansion \expansionsuperfieldseleven\ into eqns. \consistenceBRSTone\ and \consistenceBRSTzero, and properly matching the powers of the respective exponentials, one can find a set of relations obeyed by these operators for any specific value of the rank. As usual, the action of the Hamiltonian on the exponential factors introduces $\tau$-derivatives in these relations.

\medskip
The application of this procedure to the rank-1 superfields $U_i^{(1)}$ and $U^{(0)}_i$, yields the familiar relations \descendone\ and \descendzero, satisfied by the ghost number one and zero single-particle vertex operators.

\medskip
In the case of rank-2 superfields, one finds for the ghost number one two-particle operator $U_{ij}^{(1)}$
\eqnn\Qranktwo
$$\eqalignno{
& \big\{Q,U^{(1)}_{ij}\big\} = -\big\{U^{(1)}_i,U^{(1)}_j\big\}. &\Qranktwo
}$$
One possible solution to eqn. \Qranktwo\ is given by the collision of two single-particle operators, one at ghost number one and the other at ghost number zero, as follows
\eqnn\ranktwo
$$\eqalignno{
& U_{ij}^{(1)} = {2\over s_{ij}}\left[U^{(1)}_{i},U^{(0)}_{j}\right], &\ranktwo
}$$
where $s_{ij} = (k_{i} + k_{j})^{2} = 2 k_i\cdot k_j$. Acting with the BRST charge and using the descent eqn. \descendone, it is not hard to see that \Qranktwo\ indeed holds. This solution can be shown to be unique up to BRST transformations. In addition, one can explicitly verify that the antisymmetric part of the operator \ranktwo\ is BRST-trivial
\eqnn\Uijsym
$$\eqalignno{
& \left[Q,{4\over s^{2}_{ij}}\left[U^{(0)}_i,U^{(0)}_j\right]\right] = \left(U^{(1)}_{ji} - U^{(1)}_{ij}\right).&\Uijsym
}$$
In contrast, the symmetrization of the rank-2 operator \ranktwo\ does not admit this representation. It will be denoted by $V^{(1)}_{ij}$
\eqnn\secondorderop
$$\eqalignno{
&V_{ij}^{(1)} \equiv -U_{(ij)}^{(1)} = -\half\left(U^{(1)}_{ij} + U^{(1)}_{ji}\right). &\secondorderop
}$$
and referred to as {\it pinching operator}. Under BRST transformations, the pinching operator transforms according to
\eqnn\QVij
$$\eqalignno{
&\{Q,V_{ij}^{(1)}\} = \{U^{(1)}_i,U^{(1)}_j\}.&\QVij
}$$

\medskip
After plugging eqns. \ghostnumberonevo, \vertexoperatorgzero\ into the definition of the rank-2 operator \ranktwo, one finds that the pinching operator \secondorderop\ can be expressed in the following compact way
\eqnn\pinchinexpression
$$\eqalignno{
&V^{(1)}_{23} = {2\over s_{23}}\Big(P_a\tilde\phi^a_{23} + d_\alpha\tilde\phi_{23}^\alpha  - \half\tilde\phi_{23,ab}N^{ab}\Big),&\pinchinexpression
}
$$
with $\tilde\phi_{23}^a = \phi_{23}^a - \frac{s_{23}}{2}\psi_{23}^a$, $\tilde\phi_{23}^\alpha = \phi_{23}^\alpha - \frac{s_{23}}{2}\psi_{23}^\alpha$ and $\tilde\phi_{23,ab} = \phi_{23,ab}-\frac{s_{23}}{2}\psi_{23,ab}$. These superfields correspond to two-particle configurations in 11D supergravity, and are explicitly constructed in Appendix B.2,
\eqnn\phiA
\eqnn\psia
\eqnn\psialpha
\eqnn\psiab
$$\eqalignno{
& \phi_{23}^a = \lambda^\alpha h_{23,\alpha}{}^a,\ \ \phi_{23}^\alpha = \lambda^\beta h_{23,\beta}{}^\alpha,\ \ (\lambda\gamma^{ab})^\alpha\phi_{23,ab} = \Omega_{23}^{\alpha}, &\phiA\cr
& \psi_{23}^{a} = \phi_{2}^{c}h_{3,c}{}^{a} + \phi_{3}^{c}h_{2,c}{}^{a} + \phi_{2}^{\delta}h_{3,\delta}{}^{a} + \phi_{3}^{\delta}h_{2,\delta}{}^{a},&\psia\cr
&\psi_{23}^{\alpha} = \phi_{2}^{c}h_{3,c}{}^{\alpha} + \phi_{3}^{c}h_{2,c}{}^{\alpha} + \phi_{2}^{\delta}h_{3,\delta}{}^{\alpha} + \phi_{3}^{\delta}h_{2,\delta}{}^{\alpha},&\psialpha\cr
& (\lambda\gamma^{ab})^\alpha\psi_{23,ab} = - 2\phi^{a}_{(2}\Omega_{3),a}{}^{\alpha} + 2\phi^{c}_{(2}\lambda^{\beta}T_{3),\beta c}{}^{\alpha},&\psiab
}$$
where the superfields $h_{23,\alpha}{}^{a}, h_{23,\beta}{}^{\alpha}, \Omega_{23}{}^{\alpha}$ are second order deformations of the 11D vielbeins and spin-connection, and they satisfy non-linear equations of motion consistent with 11D supergravity. The superfields $\tilde\phi_{23}$ in \pinchinexpression, obey the following BRST transformation rules, 
\eqnn\Qphia
\eqnn\Qphialpha
$$\eqalignno{
&\{Q,\tilde\phi^a_{23}\} = \lambda\gamma^a\tilde\phi_{23} - s_{23}\big(\phi_{(2}^c\partial_c\phi_{3)}^a + \phi_{(2}^\delta D_\delta\phi_{3)}^a - \Omega_{(2}^\alpha h^a_{3),\alpha}\big),&\Qphia\cr
&[Q,\tilde\phi^\alpha_{23}] = (\lambda\gamma^{ab})^\alpha\tilde\phi_{23,ab}- s_{23}\big(\phi_{(2}^c\partial_c\phi_{3)}^\alpha + \phi_{(2}^\delta D_\delta\phi_{3)}^\alpha - \Omega_{(2}^\beta h^\alpha_{3),\beta}\big).&\Qphialpha
}$$

\medskip
It will also be useful to introduce a rank-2 operator carrying ghost number three, represented by $U^{(3)}_{ij}$ and defined as
\eqnn\contrationsUUone
$$\eqalignno{
&U_{ij}^{(3)} = \left[U^{(3)}_{i},U_{j}^{(0)}\right] - \frac{3s_{ij}}{2}\bigg(C_{i,a}\phi_{j}^{a} + C_{i,\alpha}\phi_{j}^{\alpha}\bigg).&\contrationsUUone
}
$$ 
In terms of the linearized 11D supergravity superfields, $U^{(3)}_{ij}$ takes the simple form
\eqnn\Uthreeij
$$
\eqalignno{
&\ \ U_{ij}^{(3)} = 3\Big[k_{i,a}\Omega_i^{ab}C_{j,b} - (\lambda\gamma_{cb}\lambda)h_i^{ab}k_{j,a}\phi_j^c + (\lambda\gamma_a)_\alpha T_{i,b}{}^\alpha C_j^{ab}\Big].&\Uthreeij
}$$
Under BRST symmetry, $U_{ij}^{(3)}$ transforms as 
\eqnn\QUij
$$\eqalignno{
&\{Q,U_{ij}^{(3)}\} = \frac{3s_{ij}}{2}(\lambda\gamma^{ab}\lambda)\phi^a_{i}\phi^b_{j}.&\QUij
}
$$
Likewise, one can show that this ghost number three rank-2 operator obeys the following equations of motion,
\eqnn\eqmUijone
\eqnn\eqmUijtwo
$$\eqalignno{
&D_{\alpha}U^{(3)}_{12} + 3\{Q,C_{\alpha,12}\} =-3(\gamma^{a}\lambda)_{\alpha}C_{a,12}& \cr &+\frac{3s_{12}}{2}\Big(2(\lambda\gamma_{ab})_{\alpha}\phi^{a}_{1}\phi_{2}^{b} + (\lambda\gamma_{ab}\lambda)h_{1,\alpha}^{a}\phi_{2}^{b} + (\lambda\gamma_{ab}\lambda)h_{2,\alpha}^{a}\phi_{1}^{b}\Big), & \eqmUijone\cr
&\partial_{a}U^{(3)}_{12} - 3[Q,C_{a,12}] =3(\lambda\gamma_{ba}\lambda)\phi^{b}_{12}&\cr &+\frac{3s_{12}}{2}\Big((\lambda\gamma_{ab}\phi_{1})\phi_{2}^{b} + (\lambda\gamma_{ab}\phi_{2})\phi_{1}^{b} + (\lambda\gamma_{bc}\lambda)\phi_1^ch^b_{2,a} + (\lambda\gamma_{bc}\lambda)\phi_2^ch^b_{1,a}\Big), &\eqmUijtwo
}$$
where the superfields $C_{ij}$ are two-particle superfields obtained from the 11D supergravity super 3-form at second order in perturbation theory.

\medskip
The vertex operators studied in this section will form the basis for the evaluation of the 11D 4-point correlation function in the next section.

\newsec Tree-level 4-point scattering amplitude in 11D

\seclab\secthree

In this section, we introduce a worldline prescription for computing tree-level 4-point correlation functions in 11D pure spinor superspace, and test its consistency under BRST invariance and gauge transformations. We then evaluate the proposed correlator, and demonstrate that it reproduces the 4-particle scattering amplitude at tree level derived via perturbiner methods. 

\subsec Worldline prescription

\subseclab\secthreeone

Following the approach of \tendimensions, we define the 11D pure spinor worldline 4-point function as
\eqnn\fourpointeleven
$$\eqalignno{
&{\cal A}_4 = \int_{-\infty}^{\infty}d\tau\left\langle U_1^{(3)}(\infty) U_2^{(0)}(\tau)U_3^{(1)}(0)U_4^{(3)}(-\infty) \right\rangle  + \left\langle U_1^{(3)}(\infty)V_{23}^{(1)}(0)U_4^{(3)}(-\infty)\right\rangle.\ \ \ \ \ \ \ \ &\fourpointeleven
}$$
where the angle brackets denote the path integral over the spacetime and diffeomorphism ghost fields, as well as the pure spinor projector $\langle \lambda^{7}\theta^{9}\rangle = 1$. In this formula \fourpointeleven, the worldline is connected by two particles (1 and 4) represented by the ghost number three vertex operators, and located at $\pm \infty$. The residual one-dimensional translational symmetry has been used to fix the position of the ghost number one vertex operators to $\tau = 0$. The remaining particle is represented by a ghost number zero vertex operator, which is integrated over the entire worldline. Eqn. \fourpointeleven\ has a simple diagrammatic interpretation, as drawn in fig. 1.

\ifig\diagrameleven{The three diagrams contributing to the 4-particle scattering process at tree level. The first diagram represents the first correlator in \fourpointeleven\ when $\tau$ varies from $0$ to $\infty$, while the second diagram corresponds to this very same correlator when $\tau$ is in the region from $-\infty$ to $0$. The third diagram presents an insertion corresponding to the pinching operator $V_{23}^{(1)}$, defined in eqn. \secondorderop.}
{\epsfxsize=.8\hsize\epsfbox{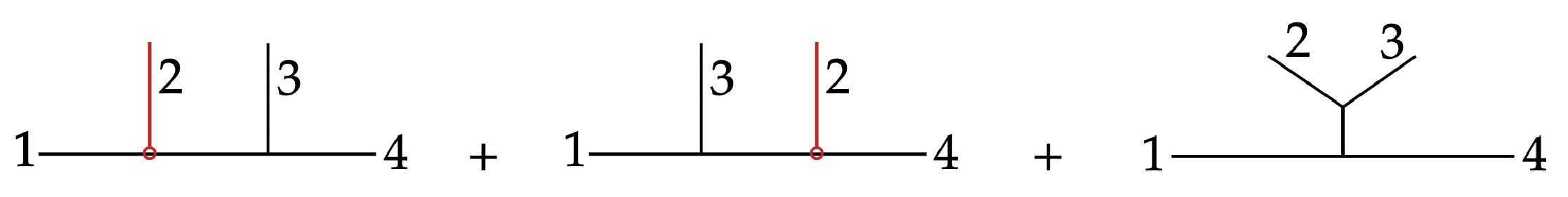}}
\medskip 
Next, we show that the 4-point function \fourpointeleven\ is BRST-closed. To this end, let us first compute the BRST variation of the first term in eqn. \fourpointeleven. This variation gives rise to a contact term, which emerges from the collision of two ghost number one vertex operators, as follows
\eqnn\brstcloseone
$$\eqalignno{
& \left\{Q, \left\langle U_1^{(3)}\int_{-\infty}^{\infty} U_2^{(0)}(\tau)U_3^{(1)}(0)U_4^{(3)} \right\rangle\right\} = -\left\langle U_1^{(3)}\int_{-\infty}^{\infty} \partial_\tau U_2^{(1)}(\tau)U_3^{(1)}(0)U_4^{(3)} \right\rangle&\cr
=\ &\Big\langle U_1^{(3)}U_4^{(3)}\left(U_2^{(1)}(\epsilon)U_3^{(1)}(0) - U_2^{(1)}(-\epsilon)U_3^{(1)}(0)\right)\Big\rangle, \ \ \ \ &\brstcloseone\cr
}$$
where $\epsilon$ was introduced as an arbitrarily small parameter, and we drop the terms $U_2^{(1)}(\pm \infty)$ in the second line. The subtraction inside the parentheses in \brstcloseone\ can be written as an operator insertion on the worldline after appropriate normal ordering,
\eqnn\normalordering
$$\eqalignno{
&U_2^{(1)}(\epsilon)U_3^{(1)}(0) - U_2^{(1)}(-\epsilon)U_3^{(1)}(0) = \{U_2^{(1)},U_3^{(1)}\}(0), &\normalordering
}$$
as per eqn. \mapcommutatorspathintegral. The BRST variation of the first term in \fourpointeleven\ is hence given by
\eqnn\brstcloseonevv
$$\eqalignno{
& \left\{Q, \left\langle U_1^{(3)}\int_{-\infty}^{\infty} U_2^{(0)}(\tau)U_3^{(1)}(0)U_4^{(3)} \right\rangle\right\} = \Big\langle U_1^{(3)}U_4^{(3)}\{U_2^{(1)},U_3^{(1)}\}(0)\Big\rangle.\ \ \ \ &\brstcloseonevv\cr
}$$

\medskip
On the other hand, the use of eqn. \QVij\ implies that the second term in eqn. \fourpointeleven\ transforms as
\eqnn\brstclosetwo
$$\eqalignno{
& \left\{Q,\left\langle U_1^{(3)}(\tau_1)V_{23}^{(1)}(0)U_4^{(3)}(\tau_2)\right\rangle\right\} = -\Big\langle U_1^{(3)}\{U_2^{(1)},U_3^{(1)}\}(0)U_4^{(3)}\Big\rangle,&\brstclosetwo
}$$
which thus cancels the first contribution displayed in eqn. \brstcloseonevv. This concludes the proof that the 4-point function \fourpointeleven\ is indeed BRST-closed.

\medskip
The correlator \fourpointeleven\ can also be shown to be invariant under gauge transformations. More precisely, 
the following modifications of the vertex operators
\eqnn \gaugetransformations
$$
\eqalignno{
&\delta U^{(3)} = \{Q,\Sigma\} \ , \ \delta U_3^{(1)}=[Q,\Lambda_3] \ , \ \delta U_2^{(0)}=\partial_{\tau}\Lambda_2+ [Q,\Omega_2] \ , \cr
& \delta V_{23}^{(1)} = [U_{2}^{{(1)}},\Lambda_{3}] + [\Lambda_2, U_3^{(1)}] + [Q,\Omega_{23}],  &\gaugetransformations
}
$$
for arbitrary functions $\Sigma, \Lambda_3, \Lambda_2, \Omega_2, \Omega_{23}$, leave the amplitude \fourpointeleven\ unchanged. These variations are induced by the action of gauge transformations on the physical fields. The proof that \fourpointeleven\ is invariant under the first transformation in \gaugetransformations\ is identical to that elaborated above for BRST closure. Under the second transformation in \gaugetransformations, the correlator \fourpointeleven\ changes as
\eqnn \gtlambdathree
$$
\eqalignno{
 \delta{\cal A}_{4} & = -\left\langle U_{1}^{(3)}\int_{-\infty}^{\infty}\partial_{\tau}U_{2}^{(1)}\Lambda_{3}(0)U_{4}^{(3)} \right\rangle - \left\langle U_{1}^{(3)}U_{2}^{(1)}(0)\Lambda_{3}(0)U_{4}\right\rangle & \cr 
&= \left\langle U_{1}^{(3)}[U_{2}^{(1)},\Lambda_{3}]U_{4}^{(3)}\right\rangle - \left\langle U_{1}^{(3)}[U_{2}^{(1)},\Lambda_{3}]U_{4}^{(3)}\right\rangle &\cr
& = 0. &\gtlambdathree
}
$$
A similar computation reveals that \fourpointeleven\ is also invariant under the transformations labeled by $\Lambda_{2}$. Finally, the gauge transformations with parameters $\Omega_{2}$ and $\Omega_{23}$ can easily be shown to leave \fourpointeleven\ invariant in a trivial manner.

\subsec Simplification

\subseclab \secthreetwo

In order to compute the correlation function \fourpointeleven\ in an efficient and systematic way, we will first simplify it by making use of the following alternative expressions for the ghost number one and zero vertex operators
\eqnn\transformU
$$\eqalignno{
& U^{(0)} = \left\{b,U^{(1)}\right\} + Q(\cdots) ,\ \ \ \  U^{(1)} =  \left[\hat\Sigma,U^{(3)}\right] + Q(\cdots). &\transformU
}$$
These relations were originally presented in \maxnotesworldline, with $b$ being the $b$-ghost of eqn. \bghost, and $\hat\Sigma$ being a ghost number $-2$ non-minimal operator simply defined in terms of the physical operators \chatalphaexp-\taalphaexp
\eqnn\Sigmahat
$$\eqalignno{
& \hat\Sigma = P_a{\bf \Phi}^a + d_\alpha{\bf\Phi}^\alpha - \lambda^\alpha w_\beta (\gamma^{ab})_\alpha{}^\beta{\bf\Omega}_{ab}. &\Sigmahat
}$$
Notice that \transformU\ corresponds to gauge transformations of the form \gaugetransformations, where $\Lambda_3$ and $\Omega_2$ are determined by the BRST-exact terms enclosed in parentheses, while $\Lambda_2 = 0$. In this way, \transformU\ generates a variation in the associated pinching operator $V_{ij}^{(1)}$ consistent with the structure prescribed by eqn. \gaugetransformations. The transformed operator then takes the explicit form 
\eqnn \vtildeoperator
$$
\eqalignno{
& \tilde{V}^{(1)}_{ij} = V_{ij} + [U^{(1)}_{i},\Lambda_{j}]. &\vtildeoperator
}
$$
Since the correlator \fourpointeleven\ has already been shown to be invariant under such transformations, one concludes that
\eqnn\fourpointtilde
$$\eqalignno{
&{\cal A}_4 = \int_{-\infty}^{\infty}d\tau\left\langle U_1^{(3)}(\infty)\tilde U_2^{(0)}(\tau)\tilde U_3^{(1)}(0)U_4^{(3)}(-\infty) \right\rangle  + \left\langle U_1^{(3)}(\infty)\tilde V_{23}^{(1)}(0)U_4^{(3)}(-\infty)\right\rangle,\ \ \ \ \ \ \ \ &\fourpointtilde
}$$
where $\tilde{U}^{(0)}_{2} = \left\{b,U^{(1)}_{2}\right\}$, $\tilde{U}^{(1)}_{3} = \left[\hat\Sigma,U^{(3)}_{3}\right]$, and $\tilde{V}_{23}^{(1)}$ is the transformed pinching operator.

\medskip
We now proceed to analyze eqn. \fourpointtilde. The path integral over the canonical momenta is equivalent to replacing these momenta by their corresponding differential operator representations, up to $\sigma_{ij}$ factors. For instance,
\eqnn\replacingP
$$\eqalignno{
& \langle p_m(\tau_1){\cal O}(x(\tau_2)) \rangle = \sigma_{12}\langle \partial_m{\cal O}(x(\tau_2)) \rangle.&\replacingP
}$$
The application of this substitution rule in eqn. \fourpointtilde, defines the action of the vertex operators as commutators or anticommutators in field space, namely $[\tilde U_i,\ - \}$ and $[\tilde V_{ij},\ - \}$. Using the properties of $b$ and $\hat\Sigma$, as well as of the pure spinor measure, it is possible to use (IBP) integration by parts techniques on these operators, which will prove to be significantly useful for simplifying eqn. \fourpointtilde, see Appendix A for details. Let us illustrate this by considering the first integral in eqn. \fourpointtilde, and restrict the interval of integration to $\infty > \tau >  0$. The piece of the correlator under consideration then reads
\eqnn\intpartsone
$$\eqalignno{
& \int_{0}^{\infty}d\tau\left\langle U_1^{(3)}(\infty)\tilde U_2^{(0)}(\tau)\tilde U_3^{(1)}(0)U_4^{(3)}(-\infty) \right\rangle & \cr
=\ & \half  \int_{0}^{\infty}d\tau\left\langle \Big[U_1^{(3)},\tilde U_2^{(0)}\Big]\tilde U_3^{(1)}U_4^{(3)} - U_1^{(3)}\Big[\tilde U_2^{(0)},\tilde U_3^{(1)}U_4^{(3)}\Big] \right\rangle\cr
=\ & \int_{0}^{\infty}d\tau\left\langle \Big[U_1^{(3)},\tilde U_2^{(0)}\Big]\tilde U_3^{(1)}U_4^{(3)}\right\rangle, & \intpartsone\cr
}$$
where we used IBP for $\tilde U_2^{(0)}$ in the last line. Using the same argument for the action of $\tilde{U}_3^{(1)}$, one obtains
\eqnn\intpartss
$$\eqalignno{
& \int_{0}^{\infty}d\tau\left\langle U_1^{(3)}(\infty)\tilde U_2^{(0)}(\tau)\tilde U_3^{(1)}(0)U_4^{(3)}(-\infty) \right\rangle \cr
& = \int_{0}^{\infty}d\tau\left\langle \left\{\left[U_1^{(3)},\tilde U_2^{(0)}\right],\tilde U_3^{(1)}\right\}U_4^{(3)}\right\rangle.&\intpartss
}$$
\ifig\diagramelevenins{The first diagram that contributes to the 4-point function. Due to BRST invariance and IBP manipulations, one can compute this correlator by considering only the contractions indicated by the arrows, see eqn. \intpartss.}
{\epsfxsize=.3\hsize\epsfbox{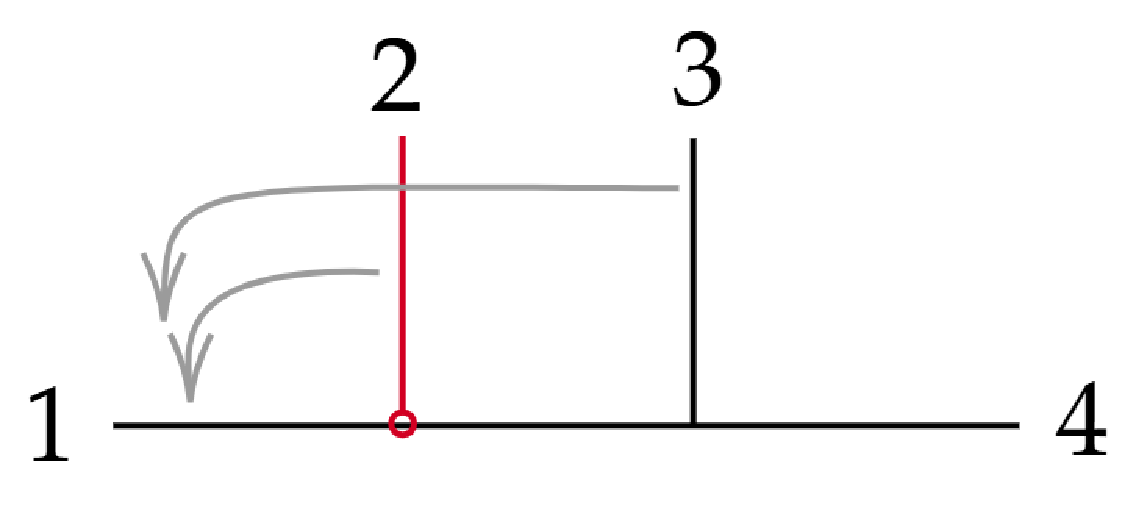}}
\noindent This computation demonstrates that the first term in eqn. \fourpointeleven\ can be derived exclusively from contractions between the worldline-vector dependent vertices $\tilde{U}_2^{(0)}$ and $\tilde{U}_3^{(1)}$, and the unintegrated vertex $U_1^{(3)}$. One can conveniently interpret eqn. \intpartss\ as indicating that the integrated vertex $\tilde{U}_2^{(0)}$ resides on the leg associated to the asymptotic state with which it first collides, to then contract the resulting multi-particle operator with $\tilde{U}_{3}^{(1)}$. See \diagramelevenins.


\medskip
Analogously, one can simplify the correlators representing the other diagrams. The 4-point amplitude \fourpointtilde\ then reduces to
\eqnn\prefourpoint
$$\eqalignno{
{\cal A}_4 =\ & \int_{0}^{\infty}d\tau\left\langle \left\{\left[U_1^{(3)},\tilde U_2^{(0)}\right],\tilde U_3^{(1)}\right\}U_4^{(3)}\right\rangle +  \int_{-\infty}^{0}d\tau\left\langle U_1^{(3)}\left\{\tilde U_3^{(1)},\left[\tilde U_2^{(0)},U_4^{(3)}\right]\right\}\right\rangle\cr
&\ \ \ \ \ \ \ \ \ \ \ \ \ \ \ \ \ \ \ \ \ \ \ \ \ \ \ \ \ +\left\langle \left\{U_1^{(3)},\tilde V_{23}^{(1)}\right\}U_4^{(3)}\right\rangle.&\prefourpoint
}$$
Since the amplitude \prefourpoint\ is invariant under BRST transformations and those consistent with eqn. \transformU, one can remove the tildes in eqn. \prefourpoint\ and write 
\eqnn\postfourpoint
$$\eqalignno{
{\cal A}_4 =\ & \int_{0}^{\infty}d\tau\left\langle \left\{\left[U_1^{(3)}, U_2^{(0)}\right], U_3^{(1)}\right\}U_4^{(3)}\right\rangle +  \int_{-\infty}^{0}d\tau\left\langle U_1^{(3)}\left\{ U_3^{(1)},\left[ U_2^{(0)},U_4^{(3)}\right]\right\}\right\rangle\cr
&\ \ \ \ \ \ \ \ \ \ \ \ \ \ \ \ \ \ \ \ \ \ \ \ \ \ \ \ \ +\left\langle \left\{U_1^{(3)}, V_{23}^{(1)}\right\}U_4^{(3)}\right\rangle.&\postfourpoint
}$$



\subsec Evaluation

\subseclab \secthreethree

We will evaluate now eqn. \postfourpoint\ in pure spinor superspace. The first term is represented by the same diagram in \diagramelevenins, which will be called ${\cal A}_4^{s}$. The superscript $s$ denotes that this configuration corresponds to the $s$-channel, as it reproduces the pole $s=s_{12}$ upon contracting the exponential factors $e^{ik x}$ and performing the integration over $\tau$.
Using the relation \contrationsUUone\ and the equations of motion \eqmUijone\ and \eqmUijtwo, one then obtains
\eqnn\schannel
$$\eqalignno{
{\cal A}_4^s = & \int_{0}^{\infty}d\tau\left\langle \left\{\left[U_1^{(3)}, U_2^{(0)}\right], U_3^{(1)}\right\}U_4^{(3)}\right\rangle& \cr
= &{6 \over s_{12}} \left\langle (\lambda\gamma_{ba}\lambda)\phi^b_{12}\phi_3^a U_4^{(3)}\right\rangle + 3\left\langle \{C_{12}, U_3^{(1)}\} U_4^{(3)}\right\rangle \cr
+ &3\left\langle (\lambda\gamma_{ba}\phi_1)\phi^b_{2}\phi_3^a U_4^{(3)}\right\rangle + 3\left\langle (\lambda\gamma_{ba}\phi_2)\phi^b_{1}\phi_3^a U_4^{(3)}\right\rangle + 6\left\langle (\lambda\gamma_{ba}\phi_3)\phi^b_{1}\phi_2^a U_4^{(3)}\right\rangle \cr
+ &3\left\langle (\lambda\gamma_{cb}\lambda)\phi_1^ch^b_{2,a}\phi_3^a U_4^{(3)}\right\rangle + 3\left\langle (\lambda\gamma_{cb}\lambda)\phi_2^ch^b_{1,a}\phi_3^a U_4^{(3)}\right\rangle\cr
+ & 3\left\langle (\lambda\gamma_{ab}\lambda)\phi_2^a h_{1,\alpha}^{b}\phi_{3}^{\alpha}  U_4^{(3)}\right\rangle + 3\left\langle (\lambda\gamma_{ab}\lambda)\phi_1^ah_{2,\alpha}^b\phi_3^\alpha U_4^{(3)}\right\rangle, & \schannel
}$$
where $C_{ij} = C_{i,a}\phi_{j}^{a} - C_{i,\alpha}\phi_{j}^{\alpha}$. A full derivation of \schannel\ can be found in Appendix C. 

\medskip
\ifig\diagramelevent{The second diagram relevant to the computation of the 4-point function. This configuration gives rise to the pole $t=s_{13}$, which justifies its notation ${\cal A}_{4}^{t}$.}
{\epsfxsize=.3\hsize\epsfbox{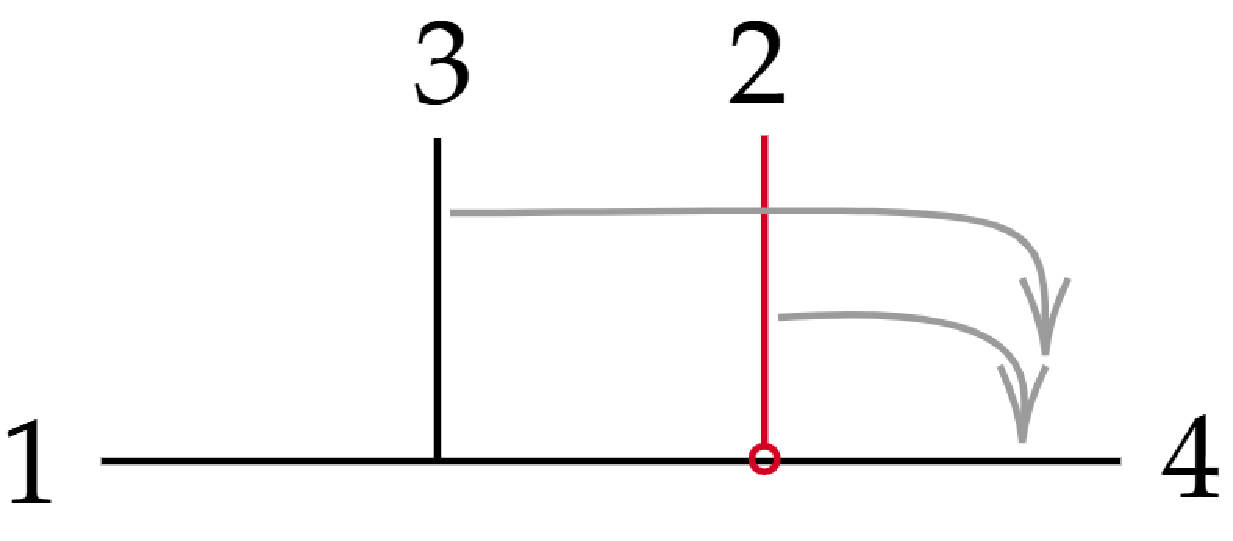}}

 Similarly, the second term in eqn. \postfourpoint\ is diagrammatically displayed in \diagramelevent. In this case, the contraction of the exponential factors $e^{ikx}$ and integration over $\tau$ yield the pole $t=s_{13}$, which indicates that this diagram ${\cal A}_{4}^{t}$ describes the $t$-channel of the 4-particle interaction.

A straightforward strategy for computing this diagram involves observing that it is BRST-equivalent to the diagram studied above up to a particle relabeling. Indeed,
\eqnn\relation
$$\eqalignno{
& \left\langle\left\{Q, \left(U_1^{(3)} \int_{-\infty}^0U_3^{(0)}\int_0^{\infty}U_2^{(0)} U_4^{(3)}\right)\right\}\right\rangle\cr
= &-\left\langle U_1^{(3)} \int_0^{\infty}U_2^{(0)}U_3^{(1)}U_4^{(3)} \right\rangle + \left\langle U_1^{(3)} \int_{-\infty}^0U_3^{(0)}U_2^{(1)}U_4^{(3)} \right\rangle.&\relation
}$$

\ifig\diagramelevenBRST{BRST invariance provides two equivalent formulations for computing the correlator 
 ${\cal A}_{4}^{t}$.}
{\epsfxsize=.6\hsize\epsfbox{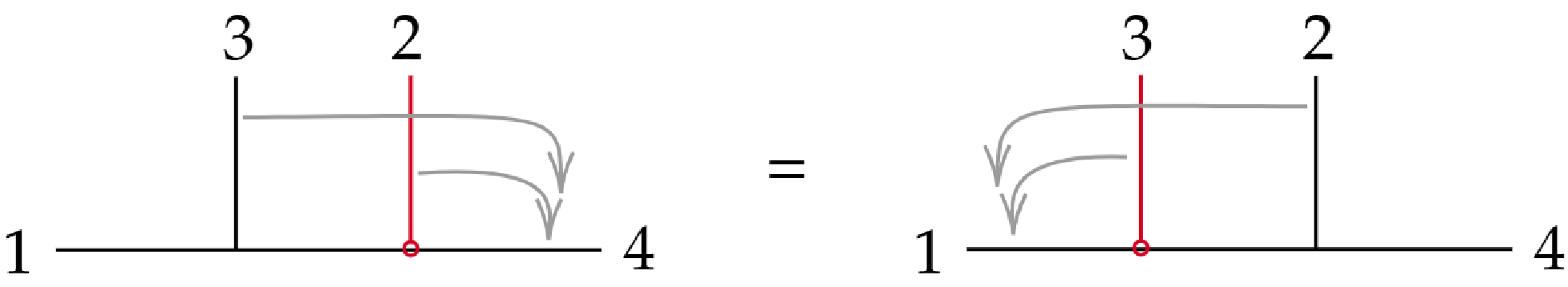}}
\noindent Diagrammatically, this relation corresponds to \diagramelevenBRST. This implies that the $t$-channel contribution can alternatively be computed by first contracting the integrated vertex $U_3^{(0)}$ with the ghost number three vertex $U_1^{(3)}$, and then contracting the resulting expression with the ghost number one operator $U_2^{(1)}$. In this manner, one finds that ${\cal A}^t$ takes the form
\eqnn\tchannel
$$\eqalignno{
{\cal A}_4^t = & \int_{-\infty}^0d\tau\left\langle U_1^{(3)}(\infty)\left\{U_3^{(1)}(0),\left[U_2^{(0)}(\tau),U_4^{(3)}(-\infty)\right]\right\}\right\rangle\cr
= & \int_{0}^{\infty}d\tau\left\langle \Big\{\Big[U_1^{(3)}(\infty),U_3^{(0)}(\tau)\Big],U_2^{(1)}(0)\Big\} U_4^{(3)}(-\infty)\right\rangle\cr
= & {6\over s_{13}} \left\langle (\lambda\gamma_{ba}\lambda)\phi^b_{13}\phi_2^a U_4^{(3)}\right\rangle +  3\left\langle \{C_{13}, U_2^{(1)}\} U_4^{(3)}\right\rangle \cr
+ & 3\left\langle (\lambda\gamma_{ba}\phi_1)\phi^b_{3}\phi_2^a U_4^{(3)}\right\rangle +  3\left\langle (\lambda\gamma_{ba}\phi_3)\phi^b_{1}\phi_2^a U_4^{(3)}\right\rangle
+ 6\left\langle (\lambda\gamma_{ba}\phi_2)\phi^b_{1}\phi_3^a U_4^{(3)}\right\rangle\cr
+ & 3\left\langle (\lambda\gamma_{cb}\lambda)\phi_1^ch^b_{3,a}\phi_2^a U_4^{(3)}\right\rangle + 3\left\langle (\lambda\gamma_{cb}\lambda)\phi_3^ch^b_{1,a}\phi_2^a U_4^{(3)}\right\rangle\cr
+ &  3\left\langle (\lambda\gamma_{ab}\lambda)\phi_3^a h_{1,\alpha}^{b}\phi_{2}^{\alpha}  U_4^{(3)}\right\rangle +  3\left\langle (\lambda\gamma_{ab}\lambda)\phi_1^ah_{3,\alpha}^b\phi_2^\alpha U_4^{(3)}\right\rangle.&\tchannel
}$$

\noindent After summing the two channels \schannel\ and \tchannel, and some algebra (see appendix C for details), one gets
\eqnn\stchannel
$$\eqalignno{
{\cal A}_4^s + {\cal A}_4^t & = {6\over s_{12}} \left\langle (\lambda\gamma_{ba}\lambda)\phi^b_{12}\phi_3^a U_4^{(3)}\right\rangle + {6\over s_{13}} \left\langle (\lambda\gamma_{ba}\lambda)\phi^b_{13}\phi_2^a U_4^{(3)}\right\rangle\cr
+ & 12\left\langle (\lambda\gamma_{ab}\phi_1)\phi^a_{2}\phi_3^b U_4^{(3)}\right\rangle + 12\left\langle (\lambda\gamma_{ab}\phi_{2})\phi_{3}^a\phi_1^b U_4^{(3)}\right\rangle + 12\left\langle (\lambda\gamma_{ab}\phi_{3})\phi_{1}^a\phi_2^b U_4^{(3)}\right\rangle\cr
+ &  6\left\langle (\lambda\gamma_{ab}\lambda)\phi_1^ah^b_{(2,c}\phi_{3)}^c U_4^{(3)}\right\rangle + 6\left\langle (\lambda\gamma_{ab}\lambda)\phi_1^ah^b_{(2,\alpha}\phi_{3)}^\alpha U_4^{(3)}\right\rangle\cr
+ &  6\left\langle C_{1a}\Big(\phi_{(2}^c\partial_c\phi^a_{3)}+\phi_{(2}^\delta D_\delta\phi^a_{3)}-\Omega_{(2}^\alpha h^a_{3),\alpha}\Big)U_4^{(3)}\right\rangle&\cr
- &  6\left\langle C_{1\alpha}\Big(\phi_{(2}^c\partial_c\phi^\alpha_{3)}+\phi_{(2}^\delta D_\delta\phi^\alpha_{3)}-\Omega_{(2}^\beta h^\alpha_{3),\beta}\Big)U_4^{(3)}\right\rangle.&\stchannel
}$$

\medskip

Finally, the $u$-channel can be obtained from the diagram containing the pinching operator in Fig. 5.

\ifig\diagramelevenu{The third diagram entering the 4-point function. Unlike the previous correlators ${\cal A}_{4}^{s}$ and ${\cal A}_{4}^{t}$, this diagram contains the pinching operator $V_{23}^{(1)}$. Its particular configuration describes the $u$-channel of the 4-particle scattering process, which motivates the notation ${\cal A}_{4}^{u}$.}
{\epsfxsize=.3\hsize\epsfbox{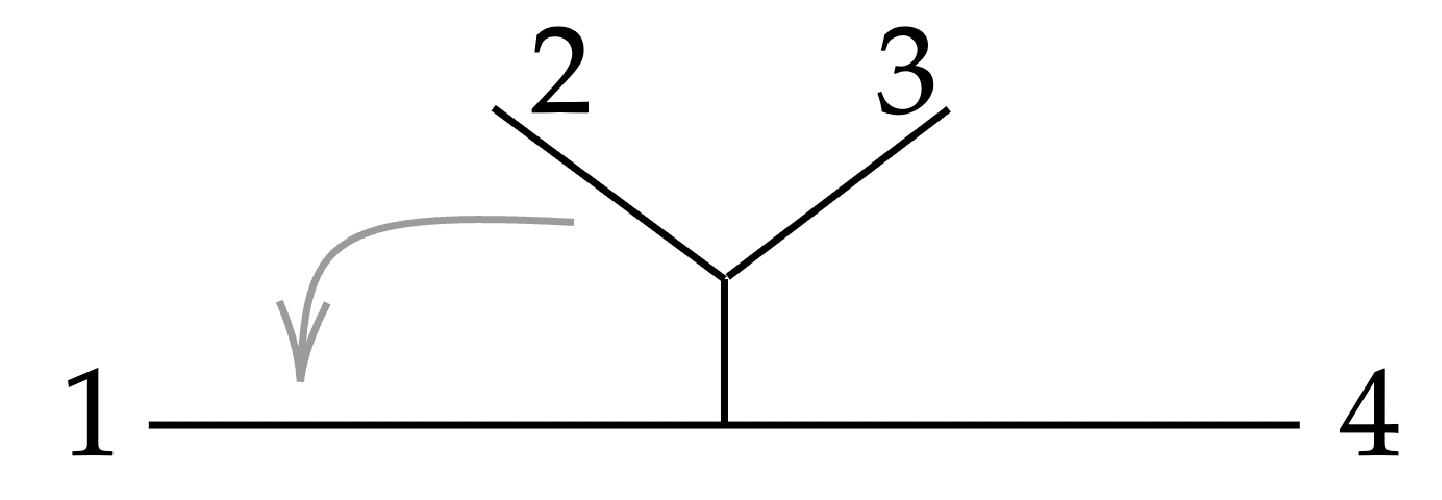}}

We denote this diagram by ${\cal A}_{4}^{u}$ since it reproduces the $u$-channel, as can easily be seen from the definition of the pinching operator $V_{23}^{(1)}$ in eqn. \secondorderop. A detailed computation of this diagram can be found in Appendix C. The result reads 
\eqnn\uchannel
$$\eqalignno{
{\cal A}^u_{4} = & \left\langle \left\{U_1^{(3)},V_{23}^{(1)}\right\}U_4^{(3)}\right\rangle\cr
= &\ {6\over s_{23}}\Big(\left\langle (\lambda\gamma_{ab}\lambda)\phi^a_{1}\tilde\phi_{23}^b U_4^{(3)}\right\rangle + 3\left\langle C_{1,a}(\{Q,\tilde\phi_{23}^a\} - (\lambda\gamma^a\tilde\phi_{23}) )U_4^{(3)}\right\rangle\cr
& - 3\left\langle C_{1,\alpha}([Q,\tilde\phi_{23}^\alpha] + \tilde\Omega^\alpha_{23} )U_4^{(3)}\right\rangle\Big)  \cr
= &\ {6\over s_{23}}\left\langle (\lambda\gamma_{ab}\lambda)\phi_{23}^a\phi^b_{1} U_4^{(3)}\right\rangle - 6\left\langle (\lambda\gamma_{ab}\lambda)\phi^a_{1}h^{b}_{(2,\alpha}\phi^\alpha_{3)} U_4^{(3)}\right\rangle - 6\left\langle (\lambda\gamma_{ab}\lambda)\phi^a_{1}h^{b}_{(2,c}\phi^c_{3)} U_4^{(3)}\right\rangle\cr
& -6\left\langle C_{1,a}\big(\phi_{(2}^c\partial_c\phi_{3)}^a + \phi_{(2}^\delta D_\delta\phi_{3)}^a - \Omega_{(2}^\alpha h^a_{3),\alpha}\big)U_4^{(3)}\right\rangle&\cr
& +6\left\langle C_{1,\alpha}\big( \phi_{(2}^c\partial_c\phi_{3)}^\alpha + \phi_{(2}^\delta D_\delta\phi_{3)}^\alpha - \Omega_{(2}^\beta h^\alpha_{3),\beta} \big)U_4^{(3)}\right\rangle. &\uchannel
}$$

\medskip
After plugging eqns. \stchannel\ and \uchannel\ into eqn. \postfourpoint, one finds the compact formula for the 4-point function in 11D 
\eqnn\allchannels
$$\eqalignno{
{\cal A}_{4} =\ & {\cal A}_4^s + {\cal A}_4^t + {\cal A}_4^u& \cr
= &\ 6\Bigg[\ {1\over s_{12}} \left\langle (\lambda\gamma_{ab}\lambda)\phi^a_{12}\phi_3^b U_4^{(3)}\right\rangle + {1\over s_{13}} \left\langle (\lambda\gamma_{ab}\lambda)\phi^a_{13}\phi_2^b U_4^{(3)}\right\rangle + {1\over s_{23}} \left\langle (\lambda\gamma_{ab}\lambda)\phi^a_{23}\phi_1^b U_4^{(3)}\right\rangle \cr
&\ \ \ \ \ \ \ + 2\left\langle (\lambda\gamma_{ab}\phi_1)\phi^a_{2}\phi_3^b U_4^{(3)}\right\rangle +  2\left\langle (\lambda\gamma_{ab}\phi_3)\phi^a_{1}\phi_2^b U_4^{(3)}\right\rangle +   2\left\langle (\lambda\gamma_{ab}\phi_2)\phi^a_{3}\phi_1^b U_4^{(3)}\right\rangle\Bigg]. & \cr
& &\allchannels
}$$

Next, we demonstrate that our result \allchannels\ is BRST-closed and invariant under the full permutation group of external particles. We also verify that it reproduces the scattering amplitude derived from perturbiner methods applied to 11D supergravity in pure spinor superspace.

\subsec BRST invariance and full permutation symmetry

\subseclab\secthreefour

As shown in Appendix C.1, the amplitude \allchannels\ admits the following compact representation
\eqnn\ciclicity
$$\eqalignno{
&{\cal A}_{4} = 4\left\langle V^{(3)}_{12}\{Q,V^{(3)}_{34}\}\right\rangle + 4\left\langle V^{(3)}_{13}\{Q,V^{(3)}_{24}\}\right\rangle + 4\left\langle V^{(3)}_{23}\{Q,V^{(3)}_{14}\}\right\rangle,&\ciclicity
}
$$
where $V^{(3)}_{ij} \equiv U^{(3)}_{ij}/s_{ij}$, with $U^{(3)}_{ij}$ being the ghost number three vertex operator defined in eqn. \Uthreeij. This form of the amplitude makes both its BRST invariance and full permutation symmetry manifest.

\medskip
Indeed, the use of eqn. \QUij\ results in the following BRST transformation for the amplitude \ciclicity
\eqnn\ciclicityy
$$\eqalignno{
\{Q, A_{4}\} =\ &12\left\langle (\lambda\gamma^{[ab}\lambda)(\lambda\gamma^{c]d}\lambda)\phi_{1,a}\phi_{2,b}\phi_{3,c}\phi_{4,d}\right\rangle . &\ciclicityy
}
$$
BRST invariance then immediately follows from the identity $(\lambda\gamma^{[ab}\lambda)(\lambda\gamma^{c]d}\lambda) = 0$.

\medskip
Moreover, IBP manipulations involving the BRST operator, together with the symmetry of $\{Q,V_{ij}^{(3)}\}$ under the exchange $i \leftrightarrow j$, imply that the amplitude \ciclicity\ is invariant under the full permutation group of external particles. Let us see how this works explicitly. The invariance under the exchange $1 \leftrightarrow 2$ follows from IBP the BRST operator in the first term of \ciclicity
\eqnn\onetotwo
$$\eqalignno{
{\cal A}(1,2,3,4) =\ & 4\left\langle V^{(3)}_{12}\{Q,V^{(3)}_{34}\}\right\rangle + 4\left\langle V^{(3)}_{13}\{Q,V^{(3)}_{24}\}\right\rangle + 4\left\langle V^{(3)}_{23}\{Q,V^{(3)}_{14}\}\right\rangle\cr
=\ & 4\left\langle \{Q,V^{(3)}_{21}\}V^{(3)}_{34}\right\rangle + 4\left\langle V^{(3)}_{23}\{Q,V^{(3)}_{14}\}\right\rangle + 4\left\langle V^{(3)}_{13}\{Q,V^{(3)}_{24}\}\right\rangle  = {\cal A}(2,1,3,4),&\cr
& &\onetotwo
}$$
where in the second line we used the symmetry of the expression $\{Q,V_{12}^{(3)}\}$ under the exchange $1 \leftrightarrow 2$. The same reasoning applies to demonstrate invariance under the exchanges $1 \leftrightarrow 3$, $2 \leftrightarrow 3$. The invariance under the exchanges of indices involving the particle labels $(1,4)$, is established after IBP the BRST operator in the first two terms of \ciclicity 
\eqnn\onetofour
$$\eqalignno{
{\cal A}(1,2,3,4) =\ & 4\left\langle V^{(3)}_{12}\{Q,V^{(3)}_{34}\}\right\rangle + 4\left\langle V^{(3)}_{13}\{Q,V^{(3)}_{24}\}\right\rangle + 4\left\langle V^{(3)}_{23}\{Q,V^{(3)}_{14}\}\right\rangle\cr
=\ & 4\left\langle \{Q,V^{(3)}_{21}\}V^{(3)}_{43}\right\rangle + 4\left\langle \{Q,V^{(3)}_{31}\}V^{(3)}_{42}\right\rangle + 4\left\langle V^{(3)}_{23}\{Q,V^{(3)}_{41}\}\right\rangle= {\cal A}(4,2,3,1).&\cr
&&\onetofour
}$$

Invariance under the exchanges $4 \leftrightarrow 3$, $4 \leftrightarrow 2$ follows directly by analogy with the preceding case. This confirms that the amplitude \ciclicity\ is invariant under the full symmetric group $S_{4}$ acting on the external particles.

\subsec Perturbiner methods

\subseclab\secthreefive

The perturbiner expansion was originally introduced in \selivanov\ as an alternative method for computing gluon scattering amplitudes. In recent years, it has been significantly developed and extended to various settings, including gravity, loop-level interactions, and formulations involving pure spinors \maxmaor. Building on the latter, one can define the following BRST-closed 4-particle amplitude in 11D pure spinor superspace
\eqnn \perturbinerfourpoint
$$
\eqalignno{
{\cal M}_{4} &= \frac{1}{2}\lim_{k_{4}^{2}\rightarrow 0}\left\langle QU_{123}^{(3)}U_{4}^{(3)}\right\rangle, &\perturbinerfourpoint
}
$$
where the normalization constant was chosen for convenience, the angle brackets denote the standard 11D pure spinor measure $\langle \lambda^{7}\theta^{9}\rangle = 1$, and $U_{123}^{(3)}$ is a three-particle operator of ghost number three, defined from the 11D supergravity equations of motion.

\medskip
To evaluate \perturbinerfourpoint, one can expand the supergravity constraint $G_{\alpha\beta\delta\epsilon} = 0$, where $G_{\alpha\beta\delta\epsilon}$ is the lowest mass dimension component of the super 4-form field strength $G_{ABCD}$, at third order in the number of fields. One finds
\eqnn \eomhalphabetadeltaepsilon
$$
\eqalignno{
G_{\alpha\beta\delta\epsilon} &= E_{(\alpha}{}^{M}E_{\beta}{}^{N}E_{\delta}{}^{P}E_{\epsilon)}{}^{Q}G_{QPNM} & \cr
&= H_{\alpha\beta\delta\epsilon} + 4\tilde{E}_{(\alpha}{}^{M}\hat{E}_{\beta}{}^{N}\hat{E}_{\delta}{}^{P}\hat{E}_{\epsilon)}{}^{Q}G_{QPNM} + 6\tilde{E}_{(\alpha}{}^{M}\tilde{E}_{\beta}{}^{N}\hat{E}_{\delta}{}^{P}\hat{E}_{\epsilon)}{}^{Q}G_{QPNM} & \cr 
& + 4\tilde{E}_{(\alpha}{}^{M}\tilde{E}_{\beta}{}^{N}\tilde{E}_{\delta}{}^{P}\hat{E}_{\epsilon)}{}^{Q}G_{QPNM} + \tilde{E}_{(\alpha}{}^{M}\tilde{E}_{\beta}{}^{N}\tilde{E}_{\delta}{}^{P}\tilde{E}_{\epsilon)}{}^{Q}G_{QPNM}=0,& \eomhalphabetadeltaepsilon
}
$$
where $H_{\alpha\beta\delta\epsilon} = \hat{E}_{(\alpha}{}^{M}\hat{E}_{\beta}{}^{N}\hat{E}_{\delta}{}^{P}\hat{E}_{\epsilon)}{}^{Q}G_{QNMP}$, 
($\hat{E}_{A}{}^{M}$, $\hat{E}_{M}{}^{A}$) represent the flat-space values of the corresponding vielbeins, and ($\tilde{E}_{A}{}^{M}, \tilde{E}_{M}{}^{A}$) denote the full vielbeins after removing their background values. As done in \BenShahar, we make the identification $U^{(3)} = \lambda^{\alpha}\lambda^{\beta}\lambda^{\delta}C_{\alpha\beta\delta}$, where $C_{\alpha\beta\delta} = \hat{E}_{(\alpha}{}^{M}\hat{E}_{\beta}{}^{N}\hat{E}_{\delta)}{}^{P}C_{PNM}$, and $C_{PNM}$ is the super 3-form gauge field of 11D supergravity.

\medskip
The perturbiner approach then dictates the following multi-particle expansions
\eqnn \perturbinervielbein
\eqnn \perturbinervielbeininverse
\eqnn \perturbinerfieldstrength
$$
\eqalignno{
\tilde{E}_{A}{}^{M} &= \sum_{{\cal P}}E_{{\cal P}, A}{}^{M}e^{k_{\cal{P}} \cdot X}, & \perturbinervielbein\cr
\tilde{E}_{M}{}^{A} &= \sum_{{\cal P}}E_{{\cal P}, M}{}^{A}e^{k_{\cal{P}} \cdot X}, & \perturbinervielbeininverse\cr
H_{ABCD} &= \hat{H}_{ABCD} + \sum_{{\cal P}}H_{{\cal{P}}, ABCD} e^{k_{\cal{P}} \cdot X}, & \perturbinerfieldstrength
}
$$
where $\hat{H}_{ABCD}$ is the flat-space value of $H_{ABCD}$, and ${\cal P} = p_{1} p_{2}\ldots p_{N}$, represents a ordered sequence of letters, with $p_{i}$ being a single-particle label, and $k_{\cal P} = k_{p_{1}} + k_{p_{2}} + \ldots + k_{p_{N}}$. 

\medskip
After substituting the relations \perturbinervielbein-\perturbinerfieldstrength\ in \eomhalphabetadeltaepsilon, and performing some algebraic manipulations, one finds
\eqnn \eomperturbiners
$$
\eqalignno{
H_{\cal{X},\alpha\beta\delta\epsilon} &= 4\hat{E}_{\alpha}{}^{M}\hat{E}_{\beta}{}^{N}\hat{E}_{\delta}{}^{P}\hat{E}_{\epsilon}{}^{Q}\hat{E}_{Q}{}^{D}\hat{E}_{P}{}^{C}\hat{E}_{N}{}^{B}E_{{\cal X}, M}{}^{A}H_{ABCD} \cr
& + 6\sum_{{\cal P}\cup{\cal Q}={\cal X}}\hat{E}_{\alpha}{}^{M}\hat{E}_{\beta}{}^{N}\hat{E}_{\delta}{}^{P}\hat{E}_{\epsilon}{}^{Q}\hat{E}_{Q}{}^{D}\hat{E}_{P}{}^{C}E_{{\cal P}, N}{}^{B}E_{{\cal Q}, M}{}^{A}H_{ABCD} \cr
& + 4\sum_{{\cal P}\cup{\cal Q}\cup{\cal R}={\cal X}}\hat{E}_{\alpha}{}^{M}\hat{E}_{\beta}{}^{N}\hat{E}_{\delta}{}^{P}\hat{E}_{\epsilon}{}^{Q}\hat{E}_{Q}{}^{D}E_{{\cal P},P}{}^{C}E_{{\cal Q}, N}{}^{B}E_{{\cal R}, M}{}^{A}H_{ABCD} \cr
& + \sum_{{\cal P}\cup{\cal Q}\cup{\cal R}\cup{\cal S}={\cal X}}\hat{E}_{\alpha}{}^{M}\hat{E}_{\beta}{}^{N}\hat{E}_{\delta}{}^{P}\hat{E}_{\epsilon}{}^{Q}E_{{\cal P}, Q}{}^{D}E_{{\cal Q},P}{}^{C}E_{{\cal R}, N}{}^{B}E_{{\cal S}, M}{}^{A}H_{ABCD}. & \eomperturbiners
}
$$

Using the definition ${\it h}_{{\cal P}, A}{}^{B} = \hat{E}_{A}{}^{M}\tilde{E}_{{\cal P},M}{}^{B}$, one finds the following equation of motion at cubic order in the number of fields
\eqnn \eomwithh
$$
\eqalignno{
H_{\cal{X},\alpha\beta\delta\epsilon} &= -6(\gamma_{ab})_{(\alpha\beta}\sum_{{\cal P}\cup {\cal Q}={\cal X}}h_{{\cal P},\delta}{}^{a}h_{{\cal Q},\epsilon)}{}^{b} - 12(\gamma_{ab})_{(\alpha\kappa}\sum_{{\cal P}\cup{\cal Q}\cup{\cal S}={\cal X}}h_{{\cal P},\beta}{}^{\kappa}h_{{\cal Q},\delta}{}^{a}h_{{\cal R},\epsilon)}{}^{b}, & \cr
& &\eomwithh 
}
$$
which in pure spinor superspace takes the simple form
\eqnn \eompurespinors
$$
\eqalignno{
Q U_{{\cal X}}^{(3)} &= 6(\lambda\gamma^{ab}\lambda)\sum_{{\cal P}\cup{\cal Q}={\cal X}}{\it \Phi}_{{\cal P}}^{a}{\it \Phi}_{{\cal Q}}^{b} + 12\sum_{{\cal P}\cup{\cal Q}\cup{\cal R}={\cal X}}(\lambda\gamma_{ab}{\it \Phi}_{{\cal P}}){\it \Phi}_{{\cal Q}}^{a}{\it \Phi}_{{\cal R}}^{b}
, & \eompurespinors
}
$$
where ${\it \Phi}_{{\cal P}}^{A} = \lambda^{\alpha}{\it h}_{{\cal P},\alpha}{}^{A}$. Note that ${\it \Phi}^{A}_{i} = \phi_{i}^{A}$ for any single-particle label $i$.

\medskip
Therefore, the perturbiner approach predicts the following 4-particle amplitude
\eqnn \mfouramplitude
$$
\eqalignno{
{\cal M}_{4} &= 6\bigg[\left\langle(\lambda\gamma_{ab}\lambda){\it \Phi}_{12}^{a}{\it \Phi}_{3}^{b}U_{4}^{(3)}\right\rangle + \left\langle(\lambda\gamma_{ab}\lambda){\it \Phi}_{13}^{a}{\it \Phi}_{2}^{b}U_{4}^{(3)}\right\rangle + \left\langle(\lambda\gamma_{ab}\lambda){\it \Phi}_{23}^{a}{\it \Phi}_{1}^{b}U_{4}^{(3)}\right\rangle & \cr
& + 2\left\langle(\lambda\gamma_{ab}{\it \Phi}_{1}){\it \Phi}^{a}_{2}{\it \Phi}_{3}^{b}U_{4}^{(3)}\right\rangle + 2\left\langle(\lambda\gamma_{ab}{\it \Phi}_{2}){\it \Phi}^{a}_{1}{\it \Phi}_{3}^{b}U_{4}^{(3)}\right\rangle
+ 2\left\langle(\lambda\gamma_{ab}{\it \Phi}_{3}){\it \Phi}^{a}_{1}{\it \Phi}_{2}^{b}U_{4}^{(3)}\right\rangle\bigg]. & \cr
& & \mfouramplitude
}
$$
In order to reproduce the four-particle amplitude \allchannels\ from the perturbiner formula \perturbinerfourpoint\ restricted to the case of four external labels, one needs to make the rescaling ${\it \Phi}_{ij}^{A} = {1 \over s_{ij}}\phi^{A}_{ij}$, which is consistent with our discussion in Appendix B. After doing so, one concludes that ${\cal M}_{4} = {\cal A}_{4}$, as promised.

\newsec Tree-level N-point scattering amplitudes in 11D

\seclab\secfour

This section generalizes the preceding construction and introduces a framework for computing tree-level N-point correlators in 11D supergravity using the pure spinor worldline formalism.

\medskip
One begins by selecting a skeleton diagram consisting of a worldline that connects two particles positioned at $\pm \infty$, designated as particles $1$ and $N$. The remaining external particles are inserted via ghost number zero vertex operators, together with a single ghost number one insertion. Exploiting worldline translation invariance, one fixes the position of the ghost number one vertex at $\tau = 0$. To compute the amplitude, we integrate over the insertion positions and subsequently perform the path integral over the worldline fields, incorporating the standard pure spinor measure $\langle\lambda^{7}\theta^{9}\rangle = 1$. Explicitly,

\eqnn\elevenNpointpath
$$\eqalignno{
{\cal A}_N = \sum_{k,P_i}&\Bigg\langle U^{(3)}_1(\infty) \prod_{i = 1}^{k-1} \left( \int d\tau_{P_i} U^{(0)}_{P_i} \right) V_{P_{k}}^{(1)}(0) U^{(3)}_{N}(-\infty) \Bigg\rangle,&\elevenNpointpath
}$$
where the sum runs over integers $k \in \{1,\dots, N-2\}$ and corresponding multi-particle labels $P_i$, such that $P_1 \cup P_2 \cup \dots \cup P_k = \{2,\dots ,N-1\}$, and $P_{k} \cap \{N-1\} \neq \emptyset$. 

\medskip
The evaluation of the correlator \elevenNpointpath\ proceeds in two stages: First, one contracts all momenta as dictated by eqn. \wick. This step is equivalent to replacing the momenta with their respective differential operator counterparts
\eqnn \momentumoperatormap
$$ \eqalignno{
& P_a \to \partial_a, \quad d_{\alpha} \to D_{\alpha}, \quad \omega_{\alpha} \to \partial_{\lambda^{\alpha}}, & \momentumoperatormap
}$$
up to factors of $\sigma_{ij} = \half sign(\tau_i - \tau_j)$. Finally, one performs the integration over the zero modes of the worldline fields. The spacetime coordinates in the exponential factors $e^{kX}$ give rise to the kinematical poles of the amplitude, while the zero modes of $\theta$ and $\lambda$ are saturated by the projector $\langle \lambda^{7}\theta^{9}\rangle = 1$.



\subsec BRST and gauge invariance

As in the 4-point case, we now show that the prescription \elevenNpointpath\ is BRST-closed and invariant under gauge transformations. For clarity and simplicity, we restrict our analysis to the case of $N=5$ external particles, noting that the generalization to arbitrary $N$ is expected to follow straightforwardly by analogous reasoning.


\medskip
\noindent The 5-point amplitude ${\cal A}_5$ is given by
\eqnn\fivepoint
$$\eqalignno{
{\cal A}_5 & = \int_{-\infty}^\infty d\tau_2\int_{-\infty}^{\infty} d\tau_3\left\langle U_1^{(3)}(\infty)U_2^{(0)}(\tau_2)U_3^{(0)}(\tau_3)U_4^{(1)}(0)U_5^{(3)}(-\infty) \right\rangle& \cr
& + \int_{-\infty}^\infty d\tau\left\langle U_1^{(3)}(\infty)U_2^{(0)}(\tau)V_{34}^{(1)}(0)U_5^{(3)}(-\infty) \right\rangle +  \int_{-\infty}^\infty d\tau\left\langle U_1^{(3)}(\infty)U_3^{(0)}(\tau)V_{24}^{(1)}(0)U_5^{(3)}(-\infty) \right\rangle\cr
& + \int_{-\infty}^\infty d\tau\left\langle U_1^{(3)}(\infty)U_{23}^{(0)}(\tau)U_{4}^{(1)}(0)U_5^{(3)}(-\infty)\right\rangle + \left\langle U_1^{(3)}(\infty)V_{234}^{(1)}(0)U_5^{(3)}(-\infty)\right\rangle. & \fivepoint
}$$
In this case, we observe the presence of a new object, namely the ghost number zero multi-particle operator $U^{(0)}_{ij}$, which unlike the ghost number one operator, it does not possess a definite symmetry under particle exchange.


\medskip
To show BRST invariance, one can compute the BRST variation of each line on the right-hand side of eqn. \fivepoint. 
The first line transforms as
\eqnn\QAone
$$\eqalignno{
-&\int_{-\infty}^\infty d\tau_2\int_{-\infty}^{\infty} d\tau_3\left\langle U_1^{(3)}(\infty)\left(\partial_{\tau_2} U_2^{(1)}(\tau_2)U_3^{(0)}(\tau_3) + U_2^{(0)}(\tau_2)\partial_{\tau_3}U_3^{(1)}(\tau_3)\right)U_4^{(1)}(0)U_5^{(3)}(-\infty) \right\rangle\cr
=&\int_{-\infty}^\infty d\tau\left\langle U_1^{(3)}(\infty)\left( U_2^{(1)}(\tau+\epsilon)U_3^{(0)}(\tau) - U_2^{(1)}(\tau -\epsilon)U_3^{(0)}(\tau) \right) U_4^{(1)}(0)U_5^{(3)}(-\infty)\right\rangle\cr
+& \int_{-\infty}^\infty d\tau\left\langle U_1^{(3)}(\infty)\left( U_3^{(1)}(\tau+\epsilon)U_2^{(0)}(\tau) - U_3^{(1)}(\tau -\epsilon)U_2^{(0)}(\tau) \right) U_4^{(1)}(0)U_5^{(3)}(-\infty)\right\rangle\cr
+&\int_{-\infty}^{\infty} d\tau_3\left\langle U_1^{(3)}(\infty)\left(U_2^{(1)}(\epsilon)U_4^{(1)}(0) - U_2^{(1)}(-\epsilon)U_4^{(1)}(0)\right)U_3^{(0)}(\tau_3)U_5^{(3)}(-\infty) \right\rangle&\cr
+&\int_{-\infty}^\infty d\tau_2 \left\langle U_1^{(3)}(\infty)U_2^{(0)}(\tau_2)\left(U_3^{(1)}(\epsilon)U_4^{(1)}(0) - U_3^{(1)}(-\epsilon)U_4^{(1)}(0)\right)U_5^{(3)}(-\infty) \right\rangle, & \QAone
}$$
where $\epsilon$ represents an arbitrary small parameter, and we dropped the terms where $U^{(1)}$ approaches $\pm\infty$. After appropriate normal ordering, the parentheses in the second and third lines of eqn. \QAone\ can be written as operator insertions
\eqnn\normalordone
\eqnn\normalordonepointtwo
$$\eqalignno{
&  U_2^{(1)}(\tau+\epsilon)U_3^{(0)}(\tau) - U_2^{(1)}(\tau-\epsilon)U_3^{(0)}(\tau) = [U^{(1)}_2,U^{(0)}_3](\tau), &\normalordone\cr
&  U_3^{(1)}(\tau+\epsilon)U_2^{(0)}(\tau) - U_3^{(1)}(\tau-\epsilon)U_2^{(0)}(\tau) = [U^{(1)}_3,U^{(0)}_2](\tau). &\normalordonepointtwo
}$$

The same procedure can be applied to the last two lines of eqn. \QAone, so that the subtractions within the parentheses give rise to operator insertions. In this manner, eqn. \QAone\ reduces to
\eqnn\QAoneagain
$$\eqalignno{
&\int_{-\infty}^\infty d\tau\left\langle U_1^{(3)}(\infty)\left( [U^{(1)}_2,U^{(0)}_3] + [U^{(1)}_3,U^{(0)}_2] \right)(\tau) U_4^{(1)}(0)U_5^{(3)}(-\infty)\right\rangle\cr
+&\int_{-\infty}^{\infty} d\tau_3\left\langle U_1^{(3)}(\infty)\left\{U_2^{(1)},U_4^{(1)}\right\}(0)U_3^{(0)}(\tau_3)U_5^{(3)}(-\infty) \right\rangle\cr
+&\int_{-\infty}^\infty d\tau_2 \left\langle U_1^{(3)}(\infty)U_2^{(0)}(\tau_2)\left\{U_3^{(1)},U_4^{(1)}\right\}(0)U_5^{(3)}(-\infty) \right\rangle.&\QAoneagain
}$$

Analogously, the BRST variation of the second line in \fivepoint\ can be cast as
\eqnn\QAtwo
$$\eqalignno{
& - \int_{-\infty}^\infty d\tau\left\langle U_1^{(3)}(\infty)U_2^{(0)}(\tau)\left\{U_{3}^{(1)},U_4^{(1)}\right\}(0)U_5^{(3)}(-\infty) \right\rangle\cr
&-\left\langle U_1^{(3)}(\infty)\left(U_2^{(1)}(\epsilon)V_{34}^{(1)}(0) - U_2^{(1)}(-\epsilon)V_{34}^{(1)}(0)\right)U_5^{(3)}(-\infty) \right\rangle \cr
& - \int_{-\infty}^\infty d\tau\left\langle U_1^{(3)}(\infty)U_3^{(0)}(\tau)\left\{U_{2}^{(1)},U_{4}^{(1)}\right\}(0)U_5^{(3)}(-\infty) \right\rangle&\cr
& -\left\langle U_1^{(3)}(\infty)\left( U_3^{(1)}(\epsilon)V_{24}^{(1)}(0) - U_3^{(1)}(-\epsilon)V_{24}^{(1)}(0)\right)U_5^{(3)}(-\infty) \right\rangle, & \QAtwo
}$$
where we used eqn. \Qranktwo\ for the BRST variation of the pinching operator. We observe that the first and third lines of eqn. \QAtwo\ cancel against the last two lines of eqn. \QAoneagain. Once more, one can bring eqn. \QAtwo\ into normal-ordered form, so that the subtractions in the parentheses of the second and fourth lines generate operator insertions. Taken together, the BRST variation of the first and second lines of \fivepoint\ can be expressed as
\eqnn\QAonetwo
$$\eqalignno{
&\int_{-\infty}^\infty d\tau\left\langle U_1^{(3)}(\infty)\left( [U^{(1)}_2,U^{(0)}_3] + [U^{(1)}_3,U^{(0)}_2] \right)(\tau) U_4^{(1)}(0)U_5^{(3)}(-\infty)\right\rangle\cr
- & \left\langle U_1^{(3)}(\infty)\left\{U_2^{(1)},V_{34}^{(1)}\right\}(0)U_5^{(3)}(-\infty) \right\rangle -\left\langle U_1^{(3)}(\infty)\left\{U_3^{(1)},V_{24}^{(1)}\right\}(0)U_5^{(3)}(-\infty) \right\rangle.\ \ \ &\QAonetwo
}$$

To compute the variation of the last line in \fivepoint, one needs to make use of the relations
\eqnn\QUijzero
\eqnn\QUijkone
$$
\eqalignno{
& [Q, U^{(0)}_{23}] = \partial_\tau U^{(1)}_{23} + [U^{(0)}_{2},U^{(1)}_{3}] + [U^{(0)}_{3},U^{(1)}_{2}], &\QUijzero\cr
& \{Q,V^{(1)}_{234}\} = \{U_{2}^{(1)},V_{34}^{(1)}\} + \{U_{3}^{(1)},V_{24}^{(1)}\} +  \{U_{4}^{(1)},V_{23}^{(1)}\}.&\QUijkone
}
$$
which can be directly obtained from eqn. \expansionsuperfieldseleven\ and the consistency conditions \consistenceBRSTone\ and \consistenceBRSTzero. For such a variation, one then finds
\eqnn\QAthree
$$\eqalignno{
-&\left\langle U_1^{(3)}(\infty)\left(V_{23}^{(1)}(\epsilon)U_{4}^{(1)}(0) - V_{23}^{(1)}(-\epsilon)U_{4}^{(1)}(0)\right)U_5^{(3)}(-\infty)\right\rangle\cr
-&\int_{-\infty}^\infty d\tau\left\langle U_1^{(3)}(\infty)\left( [U^{(0)}_{2},U^{(1)}_{3}] + [U^{(0)}_{3},U^{(1)}_{2}] \right)(\tau)U_{4}^{(1)}(0)U_5^{(3)}(-\infty)\right\rangle&\cr
+&\left\langle U_1^{(3)}(\infty)\left(\{U_{2}^{(1)},V_{34}^{(1)}\} + \{U_{3}^{(1)},V_{24}^{(1)}\} +  \{U_{4}^{(1)},V_{23}^{(1)}\} \right)(0)U_5^{(3)}(-\infty)  \right\rangle. & \QAthree
}$$
Upon normal ordering the first line of eqn. \QAthree, this expression simplifies to
\eqnn\QAthreeagain
$$\eqalignno{
-&\left\langle U_1^{(3)}(\infty)\{U_{4}^{(1)},V_{23}^{(1)}\}(0)U_5^{(3)}(-\infty)\right\rangle\cr
-&\int_{-\infty}^\infty d\tau\left\langle U_1^{(3)}(\infty)\left( [U^{(0)}_{2},U^{(1)}_{3}] + [U^{(0)}_{3},U^{(1)}_{2}] \right)(\tau) U_{4}^{(1)}(0)U_5^{(3)}(-\infty)\right\rangle\cr
+&\left\langle U_1^{(3)}(\infty)\left(\{U_{2}^{(1)},V_{34}^{(1)}\} + \{U_{3}^{(1)},V_{24}^{(1)}\} +  \{U_{4}^{(1)},V_{23}^{(1)}\} \right)(0)U_5^{(3)}(-\infty)  \right\rangle.&\QAthreeagain
}$$
Therefore, one concludes that the BRST variation of the first two lines of the 5-point function \fivepoint, given by \QAonetwo, exactly cancels the BRST variation of the last line of \fivepoint, given by \QAthreeagain. This proves the BRST closure of the 5-point function.

\medskip
Let us now investigate the variation of the 5-point function \fivepoint\ under gauge transformations. These read 
\eqnn \gaugefive
$$
\eqalignno{
&\delta U^{(3)} = \{Q,\Sigma\} \ , \ \delta U_4^{(1)}=[Q,\Lambda_4] \ , \ \delta U_i^{(0)}=\partial_{\tau}\Lambda_i+ [Q,\Omega_i] \ , \ i=2,3\cr
& \delta U_{23}^{(0)} = \partial_\tau\Omega_{23}  + 2[U^{(0)}_{(2},\Lambda_{3)}] + 2\{\Omega_{(2},U^{(1)}_{3)}\}+ [Q,\Sigma_{23}] ,&\cr
& \delta V_{i4}^{(1)} = [\Lambda_4,U_{i}^{{(1)}}] + [\Lambda_i, U_4^{(1)}] + [Q,\Omega_{i4}]\ , \ i=2,3\cr
& \delta V_{234}^{(1)} = [V_{23}^{{(1)}},\Lambda_{4}] + [\Lambda_{3},V_{24}^{{(1)}}] + [\Lambda_{2},V_{34}^{{(1)}}] + [U_2^{(1)},\Omega_{34}] + [U_3^{(1)},\Omega_{24}] + [U_4^{(1)},\Omega_{23}] + [Q,\Omega_{234}], & \cr
& & \gaugefive
}
$$
for arbitrary functions $\Sigma, \Sigma_{23}, \Lambda_4, \Lambda_3, \Lambda_2, \Omega_2, \Omega_{23}, \Omega_{24}, \Omega_{34}, \Omega_{234}$. The invariance of \fivepoint\ under the gauge symmetry associated to the parameter $\Sigma$ immediately follows from the same argument as that used in the proof of BRST closure below eqn. \fivepoint.

\medskip
The action of the transformation generated by $\Lambda_4$ in eqn. \gaugefive, induces the following variation in the correlator \fivepoint
\eqnn\gaugefiveone
$$\eqalignno{
\delta_{\Lambda_4}{\cal A}_5 = \ &\int_{-\infty}^\infty d\tau\left\langle U_1^{(3)}\left([U^{(0)}_{2}, U^{(1)}_{3}] + [U^{(0)}_{3}, U^{(1)}_{2}]\right)(\tau) \Lambda_4(0)U_5^{(3)}\right\rangle\cr
+ & \int_{-\infty}^\infty d\tau\left\langle U_1^{(3)}\left( U^{(0)}_{2}(\tau) [U^{(1)}_3,\Lambda_4](0) + U^{(0)}_{3}(\tau) [U^{(1)}_2,\Lambda_4](0)\right)U_5^{(3)}\right\rangle\cr
-& \int_{-\infty}^\infty d\tau\left\langle U_1^{(3)}U_2^{(0)}(\tau)[U_3^{(1)},\Lambda_4](0)U_5^{(3)} \right\rangle - \int_{-\infty}^\infty d\tau\left\langle U_1^{(3)}U_3^{(0)}(\tau)[U_2^{(1)},\Lambda_4](0)U_5^{(3)} \right\rangle\cr
-&\int_{-\infty}^\infty d\tau\left\langle U_1^{(3)}\left([U^{(0)}_{2}, U^{(1)}_{3}] + [U^{(0)}_{3}, U^{(1)}_{2}]\right)(\tau)\Lambda_{4}(0)U_5^{(3)}\right\rangle -\left\langle U_1^{(3)}[V^{(1)}_{23},\Lambda_{4}](0)U_5^{(3)}\right\rangle\cr
&+ \left\langle U_1^{(3)}[V_{23}^{(1)},\Lambda_4](0)U_5^{(3)}\right\rangle=0.&\gaugefiveone
}$$
The first two lines of eqn. \gaugefiveone\ arise from the gauge variation of the first line in eqn. \fivepoint, once it is brought into normal-ordered form. The third line corresponds to the variation of the second line in eqn. \fivepoint, while the fourth line reflects the gauge variation of the first term in the third line of eqn. \fivepoint, again after appropriate normal ordering. Finally, the last line originates from the gauge transformation of the final term in eqn. \fivepoint.

\medskip
Similarly, the gauge variations governed by the parameters $\Lambda_2$ and $\Lambda_3$ in eqn. \gaugefive, act on ${\cal A}_5$ in the following way:
\eqnn\gaugefivetwo
$$\eqalignno{
\delta_{\Lambda_{i}}{\cal A}_5 & = -2\int_{-\infty}^\infty d\tau\left\langle U_1^{(3)}[U^{(0)}_{(2},\Lambda_{3)}](\tau) U^{(1)}_4(0)U_5^{(3)}\right\rangle&\cr
&- \int_{-\infty}^\infty d\tau\left\langle U_1^{(3)}\left( U^{(0)}_{2}(\tau) [\Lambda_3,U^{(1)}_4](0) + U^{(0)}_{3}(\tau) [\Lambda_2,U^{(1)}_4](0)\right)U_5^{(3)}\right\rangle\cr
& - \left\langle U_1^{(3)}[\Lambda_2,V_{34}^{(1)}](0)U_5^{(3)} \right\rangle -  \left\langle U_1^{(3)}[\Lambda_3,V_{24}^{(1)}](0)U_5^{(3)} \right\rangle\cr
& + \int_{-\infty}^\infty d\tau\left\langle U_1^{(3)}U_2^{(0)}(\tau)[\Lambda_3,U^{(1)}_4](0)U_5^{(3)} \right\rangle +  \int_{-\infty}^\infty d\tau\left\langle U_1^{(3)}U_3^{(0)}(\tau)[\Lambda_2,U_4^{(1)}](0)U_5^{(3)} \right\rangle\cr
& + 2\int_{-\infty}^\infty d\tau\left\langle U_1^{(3)}[U^{(0)}_{(2},\Lambda_{3)}](\tau)U_{4}^{(1)}(0)U_5^{(3)}\right\rangle\cr
&+ \left\langle U_1^{(3)}\left([\Lambda_2,V_{34}^{(1)}]+[\Lambda_3,V_{24}^{(1)}]\right)(0)U_5^{(3)}\right\rangle=0. &\gaugefivetwo
}$$
The first two lines of eqn. \gaugefivetwo\ originate from the gauge variation of the first line in eqn. \fivepoint, after applying appropriate normal ordering. The third and fourth lines arise from the variation of the second line in eqn. \fivepoint, with normal ordering applied in the third line. Finally, the last two lines result from the variation of the last line in eqn. \fivepoint.

\medskip
Furthermore, under the transformations generated by $\Omega_2$ and $\Omega_3$ in \gaugefive, the correlator \fivepoint\ transforms as follows
\eqnn\gaugefivethree
$$\eqalignno{
\delta_{\Omega_i}{\cal A}_5 = \ &2\int_{-\infty}^\infty d\tau\left\langle U_1^{(3)}\{\Omega_{(2},U^{(1)}_{3)}\}(\tau) U^{(1)}_4(0)U_5^{(3)}\right\rangle& \cr
+ & \int_{-\infty}^\infty d\tau\left\langle U_1^{(3)}\left( \Omega_{2}(\tau) \{U^{(1)}_3,U^{(1)}_4\}(0) + \Omega_{3}(\tau) \{U^{(1)}_2,U^{(1)}_4\}(0)\right)U_5^{(3)}\right\rangle\cr
-& \int_{-\infty}^\infty d\tau\left\langle U_1^{(3)}\Omega_2(\tau)\{U_3^{(1)},U^{(1)}_4\}(0)U_5^{(3)} \right\rangle - \int_{-\infty}^\infty d\tau\left\langle U_1^{(3)}\Omega_3^{(0)}(\tau)\{U_2^{(1)},U^{(1)}_4\}(0)U_5^{(3)} \right\rangle\cr
-&2\int_{-\infty}^\infty d\tau\left\langle U_1^{(3)}\{\Omega_{(2},U^{(1)}_{3)}\}(\tau)U^{(4)}_{1}(0)U_5^{(3)}\right\rangle =0. & \gaugefivethree
}$$
The first two lines of eqn. \gaugefivethree\ arise from the variation of the first line in eqn. \fivepoint, after applying appropriate normal ordering. The third line results from the variation of the second line in eqn. \fivepoint, after IBP the BRST operator. Finally, the fourth line reflects the gauge transformation of the first term in the third line of eqn. \fivepoint, again after suitable normal ordering.

\medskip
Likewise, the gauge transformations associated to the parameters $\Omega_{24}$ and $\Omega_{34}$ in \gaugefive, lead to the following change in the 5-point function
\eqnn\gaugefivefour
$$\eqalignno{
\delta_{\Omega_{i4}}{\cal A}_5 =& -\left\langle U_1^{(3)}[U_2^{(1)},\Omega_{34}](0)U_5^{(3)} \right\rangle +  \left\langle U_1^{(3)}[U_3^{(1)},\Omega_{24}](0)U_5^{(3)} \right\rangle\cr
& + \left\langle U_1^{(3)}\left( [U_2^{(1)},\Omega_{34}] + [U_3^{(1)},\Omega_{24}] \right)(0)U_5^{(3)}\right\rangle=0.&\gaugefivefour
}$$
The first line in eqn. \gaugefivefour\ corresponds to the gauge transformation of the second line in eqn. \fivepoint, after IBP the BRST operator and using normal ordering, and the last line results from the gauge transformation of the last term in eqn. \fivepoint.

\medskip
Finally, the gauge transformations generated by the parameters $\Omega_{23}$ and $\Omega_{234}$ in \gaugefive\ can be shown to leave the 5-point amplitude \fivepoint\ invariant in a trivial way. This concludes our proof of gauge invariance of the 5-point amplitude \fivepoint.



\newsec Discussions

\seclab \secfive

We have developed a pure spinor prescription for computing N-particle scattering amplitudes in 11D supergravity, which is consistent with both BRST symmetry and gauge invariance. In particular, the 4-point function derived within this framework exhibits a remarkably compact and simple form in pure spinor superspace. This expression not only manifests full invariance under the permutation group of external particles, but also aligns precisely with the amplitude formula obtained via the perturbiner method.

\medskip
To evaluate higher-point correlation functions, one must construct multi-particle vertex operators of various ghost numbers that satisfy the consistency relations \QUijzero\ and \QUijkone. For example, the vertex operators $U_{23}^{(0)}$ and $V_{234}^{(1)}$ introduced in section 4 can be derived from the (anti)commutation relations obeyed by the one- and two-particle vertex operators appearing on the right-hand side of eqns.\ \QUijzero\ and \QUijkone. Their explicit forms follow directly from eqns.\ \ghostnumberonevo, \vertexoperatorgzero, and \pinchinexpression. Since the relevant expressions are BRST-exact by construction, a straightforward comparison yields the results for $U_{23}^{(0)}$ and $V_{234}^{(1)}$. Scattering amplitudes may then be extracted from the contraction relations satisfied by these operators and their lower-rank counterparts. As a non-trivial check, the resulting amplitudes can be compared with those obtained from the perturbiner method for an arbitrary number of external particles.

\medskip
\noindent In principle, this procedure extends to the computation of any tree-level pure spinor correlator in 11D. However, its implementation might be obstructed by subtleties associated with the pole $\eta$ in the integrated vertex operator. These subtleties originate from the structure of the pure spinor measure $[dZ]$ in the projector $\langle \cdot \rangle$, which schematically scales as
\eqnn \disscussionsone
$$
\eqalignno{
[dZ] &\sim \lambda^{-7}\, d^{23}\lambda\, d^{23}\bar{\lambda}\, d^{23}r & \disscussionsone
}
$$
(see Appendix~A for details). Consequently, the integrand must decay more slowly than $\lambda^{-16}\bar{\lambda}^{-23}$. This restriction limits the number of integrated vertex operators that may appear in a correlation function, since these operators explicitly involve negative powers of $\eta$. To illustrate this point, define $\xi = \frac{\bar{\lambda}\theta}{\lambda\bar{\lambda} + r\theta}$
which satisfies $Q\xi = 1$. A naive argument would then suggest $\langle M \rangle = \langle Q(\xi M) \rangle = 0$, where $M = \lambda^{7}\theta^{9}$ represents the top cohomology of the 11D pure spinor BRST charge. This conclusion, however, is only valid if $\xi M$ diverges more slowly than $\lambda^{-16}\bar{\lambda}^{-22}$. In fact, $\xi M$ contains terms of the form $\lambda^{-16}\bar{\lambda}^{-22}\theta^{32}r^{22}$, invalidating the naive reasoning. As in 10D \berkovitsnekrasov, these difficulties can be circumvented by introducing a regularized version of the integrated vertex operator in which the problematic poles are absent. While a full treatment of this regularization lies beyond the scope of the present manuscript, we intend to return to this question in future work.

\medskip
Once the pure spinor correlator has been evaluated for a given set of supergravity external states, the corresponding physical scattering amplitudes are obtained by projecting the result to ordinary spacetime using the measure $\langle \lambda^{7}\theta^{9}\rangle=1$. This measure was derived in \maximalloopcederwall\ within the pure spinor superfield framework, however, to the best of our knowledge, it has not yet been used in explicit computations. In forthcoming work, we plan to analyze 11D pure spinor expressions compatible with the $\lambda^{7}\theta^{9}$ structure which, together with the $\theta$-expansions of \BenShahar, will allow us to compute the 4-point function presented in eqn. \allchannels\ in ordinary spacetime.

\medskip
It would also be interesting to explore how our worldline approach can provide deeper insights into the worldvolume dynamics of the supermembrane. In particular, the vertex operator recently introduced in \ghostnumberzero\ and succesfully used in this work, radically differs from that proposed in \pssupermembrane\ in the particle limit. Likewise, unlike the supermembrane scattering prescription in \pssupermembrane, which requires the presence of the ghost number four pure spinor superfield containing 11D supergravity antifields, our framework only makes use of physical pure spinor superfields. We leave these problems for future investigation.

\medskip
Another possible direction for future work is the construction of a pure spinor supertwistor description of 11D supergravity using our results as a basis, in analogy with the 10D case \refs{\nmmax,\maxdiegoone,\maxdiegotwo,\guillenchiral}. This may in turn lead to an ambitwistorial framework that exploits the power of CFT for computing supergravity scattering amplitudes. We intend to study this in the near future.

\medskip
Finally, a further natural direction is the extension of our approach to loop-level interactions. Similar concepts have recently been employed in field theory contexts, and given the antifield structure of the pure spinor formalism, these methods might be adaptable within our framework.


\medskip

\medskip


\bigskip \noindent{\bf Acknowledgements:} MG would like to thank Maor Ben-Shahar, Nathan Berkovits, Henrik Johansson, Renann Jusinskas and Oliver Schlotterer for inspiring discussions on related topics. MS wants to thanks Nathan Berkovits for helpful discussions. EV wants to thank Kostas Rigatos, Thiago Fleury, Rennan Jusinskas and Humberto Gomez for useful discussions on this topic. We are also grateful to ICTP-SAIFR for their hospitality during the workshop on Modern Amplitude Methods for Gauge and Gravity Theories, and to Nordita for organizing Eurostrings 2025, during which this work was finalized. The work of MG was partially funded by the European Research Council under ERC-STG-804286 UNISCAMP, and by the Knut and Alice Wallenberg Foundation under the grant KAW 2018.0162 (Exploring a Web of Gravitational Theories through Gauge-Theory Methods) and the Wallenberg AI, Autonomous Systems and Software Program (WASP). The work of MS was partially supported by  CAPES (88887.608907/2021-00) and CNPQ (164239/2022-7). The work of EV was supported in part by ICTP-SAIFR FAPESP grant 2019/21281-4 and by FAPESP grant 2022/00940-2.



\medskip

\appendix{A}{Integration by parts}

\seclab\appendixa

In the computation of the 4-point function presented in Section 3, we used the fact that the actions of the ghost number zero and one operators can be integrated by parts, as illustrated in \intpartsone. This greatly simplifies the calculations involved in the correlator. While, in principle, one should consider all possible contractions between the vertex operators, the integration by parts argument tells us that only a specific subset of contractions needs to be evaluated, as is shown in eqn. \postfourpoint. The reasoning underlying this argument is explained in detail in this Appendix. As a preliminary step, we begin by examining the structure and key properties of the correlator used in Section 3. This foundational analysis will then enable a precise formulation of the integration by parts argument.

One defines the quantum mechanical correlator $\langle {\cal O} \rangle$ as the path integral over the fields with insertion ${\cal O}$,
\eqnn\correlator
$$\eqalignno{
    \langle{\cal O}\rangle & = \int {\cal D} \Phi \, {\cal N} e^{- S} {\cal O} , & \correlator
}$$
where ${\cal D} \Phi$ represents the measure for integration over all fields and ${\cal N}$ is a regularization factor that ensures that the integral over the bosonic fields is finite. After integrating out $x^m$ and the various non-zero modes using the contractions, one is left with an integral over the zero modes of $\lambda^{\alpha}$, $\bar{\lambda}_{\alpha}$, $r_{\alpha}$ and $\theta^{\alpha}$, 
\eqnn\nonminimalzeromode
$$\eqalignno{
    \langle{\cal O}\rangle  & = \int d^{32}\theta[d\lambda][d\bar{\lambda}][dr] \, {\cal N} f(\lambda, \bar{\lambda}, r, \theta), & \nonminimalzeromode
}$$
for some function $f$. The holomorphic top-form on the respective spaces gives the zero-mode integration measure,
\eqnn\measurelambda
\eqnn\measurelambdabar
\eqnn\measurer
$$\eqalignno{
&[d\lambda] \lambda^{\alpha_1} \cdots \lambda^{\alpha_7}  = (\epsilon T^{-1})^{\alpha_1 \cdots \alpha_7}_{\ \ \ \ \ \ \ \beta_1 \dots \beta_{23}} d \lambda^{\beta_1} \dots d \lambda^{\beta_{23}}, & \measurelambda \cr
&[d\bar{\lambda}] \bar{\lambda}_{\alpha_1} \cdots \bar{\lambda}_{\alpha_7}  = (\epsilon T)_{\alpha_1 \cdots\alpha_7}^{\ \ \ \ \ \ \ \beta_1 \dots \beta_{23}} d \bar{\lambda}_{\beta_1} \dots d \bar{\lambda}_{\beta_{23}}, & \measurelambdabar\cr
&[dr]  = (\epsilon T^{-1})^{\alpha_1 \cdots \alpha_7}_{\ \ \ \ \ \ \ \beta_1 \dots \beta_{23}} \bar{\lambda}_{\alpha_1} \cdots \bar{\lambda}_{\alpha_7} \left( {\partial \over \partial r_{\beta_1}} \right) \dots \left( {\partial \over \partial r_{\beta_{11}}} \right), & \measurer
}$$
where the tensors $(\epsilon T^{-1})^{\alpha_1 \cdots \alpha_7}_{\ \ \ \ \ \ \ \beta_1 \dots \beta_{23}}$ and $(\epsilon T)_{\alpha_1 \cdots\alpha_7}^{\ \ \ \ \ \ \ \beta_1 \dots \beta_{23}}$ are symmetric in $(\alpha_1, \cdots, \alpha_{7})$ and antisymmetric in $[\beta_1, \dots, \beta_{23}]$. Their explicit form can be found in \measureeleven, but this will not be needed here. The presence of the regulator ${\cal N}$ is necessary because the integral over the unbounded bosonic ghosts diverges in its absence. This regulator ${\cal N}$ is BRST trivial and thus cannot change the value of the amplitude. It is given by
\eqnn\Nregulator
$$\eqalignno{
    {\cal N} & = \exp\{ - \{Q, \bar{\lambda}_{\alpha} \theta^{\alpha} \} \} = \exp \{ - \bar{\lambda}_{\alpha} \lambda^{\alpha} - r_{\alpha} \theta^{\alpha}\}, & \Nregulator
}$$
where only the zero-modes of the fields appearing in the last equation contribute. Due to BRST invariance, the amplitude does not depend on the specific choice of regulator. Plugging eqn. \Nregulator\ into \nonminimalzeromode, the integration furnishes exactly the rule of the equation $\langle\lambda^7\theta^9\rangle=1$ \refs{\NMPS,\MafraNMoneloop}.

Having reviewed the path integral definition of the correlator used in section 3, we now turn to the integration by parts argument, which is based on the following key idea. Consider the correlator
\eqnn\intpartsone
$$\eqalignno{
& \langle{\cal O}_1U{\cal O}_2\rangle =\int {\cal D}\Phi{\cal N}e^{-S}\left({\cal O}_1U{\cal O}_2\right), &\intpartsone
}$$
where ${\cal O}_i$ are arbitrary operators inserted at times $\tau_i$ and $U$ is a vertex operator inserted at time $\tau$, with $\tau_1>\tau>\tau_2$. For notational simplicity, we omit explicit time dependence. This structure characterizes the type of correlator encountered in the 4-point amplitude, as given in \fourpointeleven. The goal is to demonstrate that the contraction of $U$ with only one of the operators ${\cal O}_i$ suffices to evaluate the correlator. For example
\eqnn\apA
$$\eqalignno{
&\langle{\cal O}_1U{\cal O}_2\rangle  = \langle[{\cal O}_1,U\}{\cal O}_2\rangle,&\apA
}$$
where the brackets $[A,B\}$ means the contraction between $A$ and $B$, as defined in \mapcommutatorspathintegral. In our context, $U$ may be a single-particle operator with ghost number zero or one $-$ denoted $U^{(0)}$ and $U^{(1)}$, respectively $-$ or a two-particle operator with ghost number one $V_{ij}^{(1)}$. We begin by analyzing the case where $U$ is a single-particle operator; the extension to the two-particle case then follows in a straightforward manner.

\medskip
The first step in the proof consists of performing gauge transformations on the operators $U^{(0)}$ and $U^{(1)}$ such that the resulting operators commute with the regulator measure \Nregulator. As will be shown below, this condition ensures that the action of the vertex operators can be integrated by parts. The required gauge transformations are given by \transformU
\eqnn\gt
$$\eqalignno{
&U^{(0)} \to \tilde{U}^{(0)} = \{b, U^{(1)}\}, \quad \ and\ \quad U^{(1)} \to \tilde{U}^{(1)} = [\hat{\Sigma}, U^{(3)}],&\gt
}$$
where the operators $b$ and $\hat{\Sigma}$ are defined in eqns. \bghost\ and \Sigmahat, respectively. To verify that the transformed operators $\tilde{U}^{(0)}$ and $\tilde{U}^{(1)}$ indeed commute with ${\cal N}$, we begin by observing that both $b$ and $\hat{\Sigma}$ themselves commute with it. This follows from the identities
\eqnn\bNregulatorb
\eqnn\bNregulatorsigma
$$\eqalignno{
&[b, \{Q, \bar{\lambda}_{\alpha} \theta^{\alpha}\}]  = - [Q, \{b, \bar{\lambda}_{\alpha} \theta^{\alpha}\}]=0, & \bNregulatorb\cr
&[\hat\Sigma, \{Q, \bar{\lambda}_{\alpha} \theta^{\alpha}\}]  = - [Q, \{\hat\Sigma, \bar{\lambda}_{\alpha} \theta^{\alpha}\}] = 0, & \bNregulatorsigma
}$$
which can be readily verified by employing the non-minimal pure spinor constraints. As a direct consequence, it follows that
\eqnn\bScommute
$$\eqalignno{
&[b, {\cal N}] = [\hat\Sigma, {\cal N}] = 0.&\bScommute
}$$

We now use this result to show that the gauge-transformed operators $\tilde U^{(0)}$ and  $\tilde U^{(1)}$ also commute with ${\cal N}$. We begin with the ghost number one case. Using Jacobi identity, we write the commutator between $\tilde U^{(1)}$ and ${\cal N}$ as
\eqnn\bcommutes
$$\eqalignno{
&\left[{\cal N},\tilde U^{(1)}\right] = \left[{\cal N},[\hat\Sigma,U^{(3)}]\right] = \left[U^{(3)},[\hat\Sigma,{\cal N}]\right] + \left[\hat\Sigma,[U^{(3)},{\cal N}]\right].&\bcommutes
}$$
The first term on the right-hand side vanishes due to eqn. \bScommute. The second term also vanishes because $U^{(3)}$ commutes with ${\cal N}$, which follows directly from 
\eqnn\bcommutescont
$$\eqalignno{
& \left[U^{(3)},\bar\lambda_\alpha\lambda^\alpha + r_\alpha\theta^\alpha\right] =  0,&\bcommutescont
}$$
where we have used that $U^{(3)}$ does not contain momentum variables. This establishes the first part of our claim, that the vertex operator $\tilde U^{(1)}$ commutes with the pure spinor measure
\eqnn\UoneN
$$\eqalignno{
&[\tilde U^{(1)},{\cal N}]=0.&\UoneN
}$$

The proof that $[\tilde U^{(0)},{\cal N}]=0$ proceeds in a similar manner. In this case, we express $\tilde U^{(0)}$ as $[b,\tilde U^{(1)}]$, which is justified because this corresponds to a gauge transformation of $\tilde U^{(0)}$. To see this, we use the fact that there exists an operator $\Omega$ such that $\tilde U^{(1)} = U^{(1)} + Q\Omega$. Consequently, we can write
\eqnn\bUgauge
$$\eqalignno{
&[b,\tilde U^{(1)}] = [b,U^{(1)} + Q\Omega] =  [b,U^{(1)}] + \partial\Omega = \tilde U^{(0)} + \partial\Omega, &\bUgauge
}$$
which shows that $[b,\tilde U^{(1)}]$ is a gauge transformation of $\tilde U^{(0)}$. Writing $\tilde{U}^{(0)}$ in this form allows us to straightforwardly prove that it commutes with the pure spinor measure, as follows
\eqnn\bcommutes
$$\eqalignno{
&\left[{\cal N},\tilde U^{(0)}\right] = \left[{\cal N},[b,\tilde U^{(1)}]\right] = \left[\tilde U^{(1)},[b,{\cal N}]\right] + \left[b,[\tilde U^{(1)},{\cal N}]\right] = 0.&\bcommutes
}$$
The first term on the right-hand side vanishes due to eqn. \bScommute, while the second term is zero because of eqn. \UoneN. Hence, we conclude that
\eqnn\UzeroN
$$\eqalignno{
&[\tilde{U}^{(0)}, {\cal N}] = 0.&\UzeroN
}$$

We can now proceed to the final step of the proof, which consists of performing integration by parts with respect to $\tilde{U}$ in the correlator, where $\tilde U$ is either $\tilde U^{(0)}$ or $\tilde U^{(1)}$. We begin by writing all contractions of the operator $\tilde{U}$ within the correlator \intpartsone,
\eqnn\contractionsap
$$\eqalignno{
& \langle{\cal O}_1\tilde U{\cal O}_2\rangle  = \half \langle[{\cal O}_1,\tilde U\}{\cal O}_2\rangle - \half\langle{\cal O}_1[\tilde U,{\cal O}_2\}\rangle. & \contractionsap
}$$
The integration by parts is then carried out on the second term on the right-hand side of \contractionsap,
\eqnn\ibp
$$\eqalignno{
\langle{\cal O}_1[\tilde U,{\cal O}_2\}\rangle & = \int [d\Phi]{\cal N}\left({\cal O}_1[\tilde U,{\cal O}_2\}\right) &\ibp\cr
& =-\int [d\Phi]\left([{\cal N}{\cal O}_1,\tilde U\}{\cal O}_2\right)\cr
& =-\int [d\Phi]{\cal N}\left([{\cal O}_1,\tilde U\}{\cal O}_2\right) - \int [d\Phi]\left([{\cal N},\tilde U\}{\cal O}_1{\cal O}_2\right) = -\langle[{\cal O}_1,\tilde U\}{\cal O}_2\rangle,
}$$
where we have used the path integral formula in \nonminimalzeromode, with the integration measure defined as $[d\Phi] = d^{32}\theta\, [d\lambda]\, [d\bar{\lambda}]\, [dr]$. In the final equality above, we have used the fact that $\tilde{U}$ commutes with ${\cal N}$, as established in eqns. \UoneN\ and \UzeroN.
\medskip
An important final observation is that the multi-particle vertex operator $V_{ij}^{(1)}$ admits an analogous integration by parts procedure. Specifically, gauge transformations acting on the single-particle vertex operators induce a corresponding transformation in the pinching operator. Concretely, the transformed pinching operator assumes the form
\eqnn\pinchtransf
$$\eqalignno{
&\tilde V^{(1)}_{ij} = - {2\over s_{ij}}\left[\tilde U^{(1)}_{(i}, \tilde U^{(0)}_{j)}\right].&\pinchtransf
}
$$

Since both $\tilde U^{(1)}_{i}$ and $\tilde U^{(0)}_{j}$ commute with the pure spinor measure, the transformed pinching operator $\tilde V^{(1)}_{ij}$ given by the expression above also commutes with the measure ${\cal N}$. We can then apply eqn. \ibp\ for the diagram that include the pinching operator. This completes the proof: the correlator \intpartsone\ can indeed be evaluated by contracting $U$ with only one of the operators ${\cal O}_i$. In summary, we have,
\eqnn\ibpresult
$$\eqalignno{
\langle{\cal O}_1U{\cal O}_2\rangle = \langle{\cal O}_1\tilde U{\cal O}_2\rangle &= \half \langle[{\cal O}_1,\tilde U]{\cal O}_2\rangle - \half\langle{\cal O}_1[\tilde U,{\cal O}_2]\rangle\cr
&= \langle[{\cal O}_1,\tilde U]{\cal O}_2\rangle = \langle[{\cal O}_1,U]{\cal O}_2\rangle,  &\ibpresult
}$$
where the first and last equalities follow from the BRST invariance of the correlator, while the first equality on the second line uses the integration-by-parts argument. This result is what was used in Section 3, leading to eqn. \postfourpoint.

\appendix{B}{11D supergravity}

\seclab\appendixb

In this Appendix, we will review the necessary equations of 11D supergravity. We will start with the linearized theory and briefly address the non-linear regime.

\subsec Linearized supergravity

\subseclab\appendixbone

To formulate supergravity in superspace, we introduce the vielbein 1-form, $E^A =  dZ^M E_{M}^{\; \; A}$, and also the connection $\Omega_{A}^{\;\;B} = dZ^M\Omega_{M A}^{\quad B}$, which is a 1-form that takes values in the Lorentz Lie-algebra. The connection is used to define the super-covariant exterior derivative, which acts on Lorentz tensors as
\eqnn \covariantderivative
$$
\eqalignno{
{\cal D}F_{A_1\dots A_m}{}^{B_1 \dots B_n} & = dF_{A_1\dots A_m}{}^{B_1 \dots B_n} - \Omega_{A_1}^{\;\;C} F_{C\dots A_m}{}^{B_1 \dots B_n} \cr
& + F_{A_1\dots A_m}{}^{C \dots B_n} \Omega_{C}^{\;\;B_1} + \dots,& \covariantderivative\
}
$$
\noindent
where $d$ is the usual exterior derivative. The two non-trivial tensors that can be constructed from these objects are the torsion 2-form, $T^A = {\cal D} E^A$, and the curvature 2-form, given by $R_A{}^B = d \Omega_A{}^B + \Omega_A{}^C \Omega_C{}^B$. These supergeometric objects are constrained by the super-Bianchi identities,
\eqnn \elevendgeometry
$$
\eqalignno{
{\cal D}T^{A} = E^{B}R_{B}{}^{A}&, \quad {\cal D}R_{A}{}^{B} = 0. & \elevendgeometry\cr
}
$$
As usual, the equations of motion satisfied by the fields are obtained by solving the Bianchi identities under the appropriate set of constraints. 

We also define $\nabla_A = E_A{}^M \partial_M$. Using eqn. \covariantderivative\ and the definition of the torsion tensor, the identity $d^2 = 0$ implies that 
\eqnn \torsiondef
$$
\eqalignno{
[\nabla_{A}, \nabla_{B}\} &= - T_{AB}{}^{C}\nabla_{C} - 2\Omega_{[AB\}}{}^{C}\nabla_{C}. & \torsiondef
}
$$
Here, $[\, ,\,\}$ denotes the graded (anti)commutator. Finally, the field content of 11D supergravity contains a 3-form superfield $F = E^C E^B E^A F_{ABC}$, which is defined modulo gauge transformations $\delta F = dL$, with $L$ an arbitrary super-2-form. The corresponding field strength is given by $G = dF$, and satisfies the identity $dG = 0$. 

In linearized supergravity, we expand the operator $\nabla_A$ as the flat-space covariant derivative plus first-order corrections, namely
\eqnn \introducingh
$$
\eqalignno{
\nabla_{A} &= D_{A} - h_{A}{}^{B}D_{B},& \introducingh 
}
$$
with $D_{A} = \hat{E}_{A}{}^{M}\partial_{M}$ and $h_{A}{}^{B} = \hat{E}_{A}{}^{M}E_{M}^{(1)B} = -E^{(1)M}_{A}\hat{E}_{M}{}^{B}$, where ($\hat{E}_{A}{}^{M}$, $\hat{E}_{M}{}^{B}$) are the flat-space values of the vielbeins, and ($E_{A}^{(1)M}$, $E_{M}^{(1)A}$) are their corresponding first-order perturbations. The fields of the theory satisfy the dynamical constraints
\eqnn\dinamicalconst
$$\eqalignno{
&T_{\alpha\beta}{}^a = (\gamma^a)_{\alpha\beta}, \quad G_{\alpha\beta ab} = (\gamma_{ab})_{\alpha\beta},&\dinamicalconst
}$$
and conventional constraints
\eqnn\convconst
$$\eqalignno{
&T_{\alpha\beta}{}^{\delta} = T_{a\alpha}{}^{c} = T_{ab}{}^{c} = G_{\alpha\beta\delta\epsilon} = G_{a\alpha\beta\delta} = G_{abc\alpha} = 0.&\convconst
}$$
After substituting \introducingh\ into the eqn. \torsiondef\ and using the constraints, one obtains the following set of linearized equations of motion \refs{\maxmasoncasaliberkovits, \maxthesis}
\eqnn \eomone
\eqnn \eomtwo
\eqnn \eomthree
\eqnn \eomfour
\eqnn \eomfive
\eqnn \eomsix
$$
\eqalignno{
2D_{(\alpha}h_{\beta)}{}^{a} - 2h_{(\alpha}{}^{\delta}(\gamma^{a})_{\beta)\delta} + h_{b}{}^{a}(\gamma^{b})_{\alpha\beta} &= 0, & \eomone \cr
2D_{(\alpha}h_{\beta)}{}^{\delta} - 2\Omega_{(\alpha\beta)}{}^{\delta} + (\gamma^{a})_{\alpha\beta}h_{a}{}^{\delta} &= 0, & \eomtwo\cr
\partial_{a}h_{\alpha}{}^{\beta} - D_{\alpha}h_{a}{}^{\beta} - T_{a\alpha}{}^{\beta} - \Omega_{a\alpha}{}^{\beta} &= 0, & \eomthree\cr
\partial_{a}h_{\alpha}{}^{b} - D_{\alpha}h_{a}{}^{b} - h_{a}{}^{\beta}(\gamma^{b})_{\beta\alpha} + \Omega_{\alpha a}{}^{b} &= 0, & \eomfour\cr
\partial_{a}h_{b}{}^{\alpha} - \partial_{b}h_{a}{}^{\alpha} - T_{ab}{}^{\alpha} &= 0, & \eomfive\cr
\partial_{a}h_{b}{}^{c} - \partial_{b}h_{a}{}^{c} - 2\Omega_{{ab}}{}^{c} &= 0. & \eomsix 
}
$$

It is straightforward to check that these equations are invariant under the linearized gauge transformations,
\eqnn \gtfull
$$
\eqalignno{
\delta h_{\alpha}{}^{a} = D_{\alpha}\Lambda^{a} + (\gamma^{a})_{\alpha\beta}\Lambda^{\beta} \ \ , \ \
\delta h_{\alpha}{}^{\beta} &= D_{\alpha}\Lambda^{\beta} + \Lambda_{\alpha}{}^{\beta} \ \ , \ \
\delta \Omega_{\alpha\beta}{}^{\epsilon} = D_{\alpha}\Lambda_{\beta}{}^{\epsilon} \ , &  \cr
\delta h_{a}{}^{b} = \partial_{a}\Lambda^{b} + \Lambda_{a}{}^{b} \ \ , \ \
\delta h_{a}{}^{\beta} &= \partial_{a}\Lambda^{\beta}\ \ , \ \ \delta \Omega_{a\alpha}{}^{\beta} = \partial_{a}\Lambda_{\alpha}{}^{\beta}  \ , & \gtfull
}
$$
where $\Lambda^{a}$, $\Lambda^{\alpha}$, $\Lambda_{\alpha}{}^{\beta} = {1\over 4}(\gamma^{ab})_{\alpha}{}^{\beta}\Lambda_{ab}$ are arbitrary gauge parameters. 

We also derive linearized equations of motion of the 3-form superfield. They are deduced from the 4-form superfield $H$ defined from the field strength $G$ as 
\eqnn \tensorcapitalh
$$
\eqalignno{
H_{ABCD} &= \hat{E}_{[D}{}^{Q}\hat{E}_{C}{}^{P}\hat{E}_{B}{}^{N}\hat{E}_{A\}}{}^{M}G_{MNPQ}. & \tensorcapitalh 
}
$$
The 4-form can also be written in terms of the flat-space value of the torsion $\hat{T}_{AB}{}^C$ as
\eqnn\Hrewrite
$$
\eqalignno{
& H_{ABCD} = 4 D_{[A} C_{BCD\}} + 6 \hat{T}_{[AB}{}^{E} C_{ECD\}}, &\Hrewrite
}
$$
where $C_{ABC} = \hat{E}_{[C}{}^{P}\hat{E}_{B}{}^{N}\hat{E}_{A\}}{}^M F_{MNP}$. Using the constraints \convconst\ and \dinamicalconst, eqn. \Hrewrite\ can be written in components as
\eqnn \eomseven
\eqnn \eomeight
\eqnn \eomnine
\eqnn \eomten
$$
\eqalignno{
4D_{(\alpha}C_{\beta\delta\epsilon)} + 6(\gamma^{a})_{(\alpha\beta}C_{a\delta\epsilon)} &= 0, & \eomseven\cr
\partial_{a}C_{\alpha\beta\delta} - 3D_{(\alpha}C_{a\beta\delta)} + 3(\gamma^{b})_{(\alpha\beta}C_{ba\delta)}  &= 3(\gamma_{ab})_{(\alpha\beta}h_{\delta)}{}^{b}, & \eomeight \cr
2\partial_{[a}C_{b]\alpha\beta} + 2D_{(\alpha}C_{\beta) ab} + (\gamma^{c})_{\alpha\beta}C_{cab} &= 2(\gamma_{[b}{}^{c})_{\alpha\beta}h_{a]c} + 2(\gamma_{ab})_{(\alpha\delta}h_{\beta)}{}^{\delta}, & \eomnine\cr
3\partial_{[a}C_{bc]\alpha} - D_{\alpha}C_{abc} & =  3(\gamma_{[ab})_{\alpha\beta}h_{c]}{}^{\beta}. & \eomten
}
$$

Again, it is easy to show that these equations are invariant under the linearized gauge transformations
\eqnn \gtone
\eqnn \gttwo
\eqnn \gtthree
\eqnn \gtfour
$$
\eqalignno{
\delta C_{\alpha\beta\epsilon} &= D_{(\alpha}\Lambda_{\beta\epsilon)} + (\gamma^{a})_{(\alpha\beta}\Lambda_{a\epsilon)}  \ ,& \gtone\cr
\delta C_{a\alpha\epsilon} &= {1\over 3}\partial_{a}\Lambda_{\alpha\epsilon} + {2\over 3}D_{(\alpha}\Lambda_{\epsilon)a} + 
{1\over 3}(\gamma^{b})_{\alpha\epsilon}\Lambda_{ba} + (\gamma_{ab})_{\alpha\epsilon}\Lambda^{b}\ , & \gttwo\cr
\delta C_{ab\alpha} &= {2
\over 3}\partial_{[a}\Lambda_{b]\alpha} + {1\over 3}D_{\alpha}\Lambda_{ab} -(\gamma_{ab})_{\alpha\beta}\Lambda^{\beta} \ ,& \gtthree\cr
\delta C_{abc} &= \partial_{[a}\Lambda_{bc]}  \ ,& \gtfour
}
$$
where $\Lambda_{\beta\alpha}, \Lambda_{a\alpha}, \Lambda_{ab}, \Lambda^{a}, \Lambda^{\alpha}$ are arbitrary gauge parameters.


\subsec Non-linear supergravity: pinching operator

\subseclab\appendixbtwo

The definition of the 4-point scattering amplitude in Section 3 involves a two-particle vertex operator, referred to as the {\it pinching operator}, defined in eqn. \secondorderop. To evaluate the 4-point amplitude, we expressed this operator in terms of the 11D supergravity fields, as written in \pinchinexpression. In this section, we derive the expression \pinchinexpression\ by studying the equations of motion of 11 supergravity up to second-order terms in the expansion of the vielbein around flat space.

We start by considering the superfields $h_\alpha{}^A= (h_{\alpha}{}^{a}, h_{\alpha}{}^{\beta})$ and $\Omega_{\alpha\,\beta}{}^{\delta}$ up to second order in the nonlinear expansion. To construct these, we begin by writing the component expression of the torsion $T_{MN}{}^{A}$,
\eqnn \equationtorsion
$$
\eqalignno{
T_{MN}{}^{A} &= \partial_{M}E_{N}{}^{A} + (-)^{MN + 1}\partial_{N}E_{M}{}^{A} - (-)^{M(N+B)}E_{N}{}^{B}\Omega_{M\,B}{}^{A} - (-)^{BN+1}E_{M}{}^{B}\Omega_{N\,B}{}^{A}, & \cr
& &\equationtorsion
}
$$
where the letters in the exponents denote the degree of the respective indices. Using the fact that $T_{MN}{}^{A} = (-)^{M(N+B)}E_{N}{}^{B}E_{M}{}^{C}T_{CB}{}^{A}$, one can contract \equationtorsion\ with $(-)^{N+1}\hat{E}_{\beta}{}^{N}\hat{E}_{\alpha}{}^{M}$, to obtain the equation of motion for the field $h_\alpha{}^A$ at second order in the non-linear expansion,
\eqnn \equationhalphabeta
$$
\eqalignno{
&2D_{(\alpha}h_{\beta)}{}^{A} + (\gamma^{c})_{\alpha\beta}h_{c}{}^{A} - 2\Omega_{(\alpha\,\beta)}{}^{A} - (-)^{B+1}2h_{(\alpha}{}^{B}\Omega_{\beta)\,B}{}^{A}&\cr
&\ \ \ \ \ \ \ \ \ \ \ \ \ = (-)^{B+1}h_{\beta}{}^{B}T_{\alpha\,B}{}^{A} + h_{\alpha}{}^{C}T_{C\,\beta}{}^{A},\ \ & \equationhalphabeta
}
$$
where $\Omega_{A\,B}{}^{C} = \hat{E}_{A}{}^{M}\Omega_{M\,B}{}^{C}$. One can now contract \equationhalphabeta\ with $\lambda^\alpha\lambda^\beta$, to obtain the following second-order equations,
\eqnn \qphia
\eqnn \qphialpha
$$
\eqalignno{
\{Q,{\it \Phi}^{a}\} &= (\lambda\gamma^{a}{\it \Phi}) - {\it \Phi}^{b}{\it \Omega}_{b}{}^{a} +{1\over 2}({\it \Phi}\gamma^{a}{\it \Phi}), & \qphia\cr
\{Q,{\it \Phi}^{\alpha}\} &= {\it \Omega}^\alpha + {\it \Phi}^{\delta}{\it \Omega}_{\delta}{}^{\alpha} + {\it \Phi}^{c}T_{c\delta}{}^{\alpha}\lambda^{\delta} , & \qphialpha
}
$$
where ${\it \Omega}^{\alpha} = \lambda^{\beta}\lambda^{\delta}\Omega_{\beta\,\delta}{}^{\alpha}$, ${\it \Omega}_{a}{}^{b} = \lambda^{\alpha}\Omega_{\alpha\,a}{}^{b}$, and ${\it \Phi}^A = \lambda^\alpha h_\alpha{}^A$. Now, we use eqns. \qphia\ and \qphialpha\ to write the BRST transformations of the two-particle superfields ${\it \Phi}^a_{12}$ and ${\it \Phi}^\alpha_{12}$ as
\eqnn\aqphia
\eqnn\aqphialpha
$$
\eqalignno{
\{Q,{\it \Phi}^{a}_{12}\} &= (\lambda\gamma^{a}{\it \Phi}_{12}) - \phi^{b}_{1}\Omega_{2,b}{}^{a} - \phi^{b}_{2}\Omega_{1,b}{}^{a} + (\phi_{1}\gamma^{a}\phi_{2}), & \aqphia\cr
\{Q,{\it \Phi}^{\alpha}_{12}\} &= {\it \Omega}_{12}^\alpha + 2\phi_{(1}^{\delta}\Omega_{2),\delta}{}^{\alpha} + 2\phi_{(1}^{c}T_{2),c\,\delta}{}^{\alpha}\lambda^{\delta}. & \aqphialpha
}
$$
In eqns. \aqphia, \aqphialpha\ we used that ${\it \Phi}_{i}^{A}=\phi_{i}^{A}$, ${\it \Omega}_{i}^{A}=\Omega_{i}^{A}$ for single-particle superfields. It is possible to find an expression for ${\it \Phi}_{12}^a$ in terms of the linearized 11D supergravity fields, satisfying the BRST-invariance \aqphia. This solution can be written as ${\it \Phi}_{12}^{a} = {1\over s_{12}}\phi_{12}^{a} $, with $\phi_{12}^{a}$ defined by 
\eqnn \honetwoa
$$
\eqalignno{
\phi_{12}^{a} &= \bigg[-k_{2}^{c}h_{1,c}{}^{d}\Omega_{2,d}{}^{a} - k_{2}^{c}\Omega_{1,c}{}^{d}h_{2,d}{}^{a} + k_{2}^{c}(\phi_{2}\gamma^{a})_{\delta}h_{1,c}{}^{\delta} + h_{2,c}{}^{\beta}(\gamma^{a})_{\delta\beta}T_{1,c\alpha}{}^{\delta}\lambda^{\alpha} & \cr
&\ \ \ \  + k_{2}^{c}\phi_{2}^{d}\Omega_{1,cd}{}^{a} + h_{2,c}{}^{d}R_{1,\alpha cd}{}^{a}\lambda^{\alpha} + (1 \leftrightarrow 2)\bigg]. & \honetwoa
}
$$
As a check, one can apply the BRST operator on the rhs of eqn. \honetwoa, the use of the equation of motion \eomseven-\eomten\ and \eomone-\eomsix, allows one to conclude that ${\it \Phi}_{12}^{a}$ indeed satisfies \aqphia. Similarly, the 2-particle superfields ${\it \Phi}_{12}^{\beta}$, ${\it \Omega}_{12,c}{}^{d}$ can be expressed in the form ${\it \Phi}_{12}^{\beta} = {1\over s_{12}}\phi_{12}^{\beta}$, ${\it \Omega}_{12,c}{}^{d} = {1\over s_{12}}\Omega_{12,c}{}^{d}$, where
\eqnn \itphitwoalpha
\eqnn \itomegatwoab
$$
\eqalignno{
\phi^{\beta}_{12} &= \bigg[k_{1}^{a}\phi_{1}^{\delta}\Omega_{2,a\,\delta}{}^{\beta} + h_{1,a}{}^{\delta}k_{2}^{a}\Omega_{2,\delta}{}^{\beta} + k_{12}^{a}T_{12,a\alpha}{}^{\beta}\lambda^{\alpha} + k_{1}^{a}\phi_{1}^{b}T_{2,ab}{}^{\beta} + \phi_{1}^{b}k_{2}^{a}T_{2,ab}{}^{\alpha} & \cr
& + k_{1}^{a}\phi_{1}^{\delta}T_{2,a\delta}{}^{\beta} + \phi_{1}^{\delta}k_{2}^{a}T_{2,a\delta}{}^{\beta} + h_{1,a}{}^{c}k_{2}^{a}T_{2,c\alpha}{}^{\beta}\lambda^{\alpha} + (1 \leftrightarrow 2)\bigg], &\itphitwoalpha \cr
\Omega_{12, c}{}^{d} &= \bigg[k_{1}^{a}\Omega_{1,c}{}^{b}\Omega_{2,a\,b}{}^{d} - \Omega_{1,a\,c}{}^{b}k_{2}^{a}\Omega_{2,b}{}^{d} + k_{12}^{a}\lambda^{\alpha}R_{12,a\alpha\,c}{}^{d} &\cr 
& + h_{1,a}{}^{\beta}k_{2}^{a}R_{2,\beta\alpha\,c}{}^{d}\lambda^{\alpha} + h_{1,a}{}^{b}k_{2}^{a}R_{2,b\alpha\,c}{}^{d}\lambda^{\alpha} + k_{1}^{a}\phi_{1}^{\delta}R_{2,a\delta\,c}{}^{d} + k_{1}^{a}\phi_{1}^{b}R_{2,ab\,c}{}^{d} &\cr
& + \phi_{1}^{b}k_{2}^{a}R_{2,ab\,c}{}^{d} + + \phi_{1}^{\delta}k_{2}^{a}R_{2,a\delta\,c}{}^{d} + (1\leftrightarrow 2) \bigg]. &\itomegatwoab
}
$$
The superfields ${\phi}^a_{12},\phi^\alpha_{12},\Omega^\alpha_{12}$ appear explicitly in the expression for the pinching operator \pinchinexpression, with the field $\phi^a_{12}$ appearing in the result of the 4-point amplitude \allchannels.

\medskip
Now, we turn to the explicit formula of the pinching operator in terms of 11D supergravity fields. Let us first write it generically as
\eqnn\pinchingC
$$\eqalignno{
&V^{(1)}_{23} = {2\over s_{23}}\left(P_a\tilde\phi^a_{23} + d_\alpha\tilde\phi_{23}^\alpha - \half\tilde\phi_{23,ab}N^{ab}\right).&\pinchingC
}$$
We aim to determine explicit expressions for the fields $\tilde{\phi}^a_{23}$, $\tilde{\phi}^\alpha_{23}$, and $\tilde{\phi}_{23,ab}$ in terms of the fundamental 11D supergravity fields. To this end, we utilize the defining property of the pinching operator encoded by the relation
\eqnn\QVijC
$$\eqalignno{
&\{Q,V^{(1)}_{23}\} = \{U_2^{(1)},U_3^{(1)}\},&\QVijC
}$$
alongside the equations of motion satisfied by the supergravity fields, \aqphia\ and \aqphialpha. Our strategy proceeds by explicitly computing both sides of the eqn. \QVijC. The comparison of these two results will yield concrete formulae for the component fields $\tilde{\phi}^a_{23}$, $\tilde{\phi}^\alpha_{23}$, and $\tilde{\phi}_{23,ab}$. To begin, we evaluate the right-hand side of \QVijC\ by computing the anticommutator of the two ghost number one vertex operators $U_2^{(1)}$ and $U_3^{(1)}$. We obtain 
\eqnn\OPEghostnumberoneC
$$\eqalignno{
{s_{23}\over2}\{U_{2}^{(1)},U_{3}^{(1)}\} &= P_{a}\tilde{\chi}_{23}^{a} + d_{\alpha}\tilde{\chi}_{23}^{\alpha} - w_{\alpha}\tilde{\omega}_{23}^{\alpha},\ \ \ \ &\OPEghostnumberoneC
}$$
where the component fields $\tilde{\chi}_{23}^{a}, \tilde{\chi}_{23}^{\alpha}, \tilde{\omega}_{23}^{\alpha}$ are explicitly given by
\eqnn\chitilde
$$\eqalignno{
&\tilde{\chi}_{23}^{a} = \{Q,\hat\phi^a_{23}\} - (\lambda\gamma^a\hat\phi_{23}),\ \ \tilde{\chi}_{23}^{\alpha} = [Q,\hat\phi^\alpha_{23}] - \hat\Omega^\alpha_{23},\ \ \tilde{\omega}_{23}{}^{\alpha} = \{Q,\hat\Omega_{23}^\alpha\},\ \ \ &\chitilde
}$$
where 
\eqnn\chiaCc
\eqnn\chialphaCc
\eqnn\omegaCc
$$\eqalignno{
& \hat{\phi}_{23}^{a} =\phi_{23}^{a} - \frac{s_{23}}{2}\left(\phi_{2}^{c}h_{3,c}{}^{a} + \phi_{3}^{c}h_{2,c}{}^{a} + \phi_{2}^{\delta}h_{3,\delta}{}^{a} + \phi_{3}^{\delta}h_{2,\delta}{}^{a}\right),&\chiaCc\cr
& \hat{\phi}_{23}^{\alpha} = \phi_{23}^{\alpha} - \frac{s_{23}}{2}\left(\phi_{2}^{c}h_{3,c}{}^{\alpha} + \phi_{3}^{c}h_{2,c}{}^{\alpha} + \phi_{2}^{\delta}h_{3,\delta}{}^{\alpha} + \phi_{3}^{\delta}h_{2,\delta}{}^{\alpha}\right), &\chialphaCc\cr
& \hat{\Omega}_{23}{}^{\alpha} = \Omega_{23}{}^{\alpha} - s_{23}\left(\phi^{a}_{(1}\Omega_{2),a}{}^{\alpha} + \phi^{c}_{(1}\lambda^{\beta}T_{2),\beta c}{}^{\alpha}\right).&\omegaCc
}$$
Next, we compute the left-hand side of eqn. \QVijC 
\eqnn\lhs
$$\eqalignno{
& {s_{23}\over2}\{Q,V^{(1)}_{23}\} = P_a\left(\{Q,\tilde\phi^a_{23}\} - (\lambda\gamma^a\tilde\phi_{23})\right) + d_\alpha\left([Q,\tilde\phi^\alpha_{23}] - \tilde\Omega^\alpha_{23}\right) - w_\alpha \{Q,\tilde\Omega_{23}^\alpha\}.\ \ &\lhs
}$$
Then, comparing the left hand side with the right hand side of eqn. \QVijC\ by the use of eqns. \chitilde-\omegaCc\ and \lhs, we find 
\eqnn\tildaphia
\eqnn\tildaphialpha
\eqnn\tildaomega
$$\eqalignno{
&\tilde{\phi}_{23}^{a} = \phi_{23}^{a} - \frac{s_{23}}{2}\Big(\phi_{2}^{c}h_{3,c}{}^{a} + \phi_{3}^{c}h_{2,c}{}^{a} + \phi_{2}^{\delta}h_{3,\delta}{}^{a} + \phi_{3}^{\delta}h_{2,\delta}{}^{a}\Big),&\tildaphia\cr
&\tilde{\phi}_{23}^{\alpha} = \phi_{23}^{\alpha} - \frac{s_{23}}{2}\Big(\phi_{2}^{c}h_{3,c}{}^{\alpha} + \phi_{3}^{c}h_{2,c}{}^{\alpha} + \phi_{2}^{\delta}h_{3,\delta}{}^{\alpha} + \phi_{3}^{\delta}h_{2,\delta}{}^{\alpha}\Big),&\tildaphialpha\cr
&\tilde{\Omega}_{23}{}^{\alpha} = (\lambda\gamma^{ab})^\alpha\tilde\phi_{23,ab} = \Omega_{23}{}^{\alpha} - s_{23}\Big(\phi^{a}_{(1}\Omega_{2),a}{}^{\alpha} - \phi^{c}_{(1}\lambda^{\beta}T_{2),\beta c}{}^{\alpha}\Big).&\tildaomega
}$$

\noindent To conclude, it is simple to see that the BRST variations of these fields are 
\eqnn\Qtildaphia
\eqnn\Qtildaphialpha
$$\eqalignno{
&\{Q,\tilde\phi^a_{23}\} = \lambda\gamma^a\tilde\phi_{23} - s_{23}\Big(\phi_{(2}^c\partial_c\phi_{3)}^a + \phi_{(2}^\delta D_\delta\phi_{3)}^a - \Omega_{(2}^\alpha h^a_{3),\alpha}\Big),&\Qtildaphia\cr
&[Q,\tilde\phi^\alpha_{23}] = (\lambda\gamma^{ab})^\alpha\tilde\phi_{23,ab}- s_{23}\Big(\phi_{(2}^c\partial_c\phi_{3)}^\alpha + \phi_{(2}^\delta D_\delta\phi_{3)}^\alpha - \Omega_{(2}^\beta h^\alpha_{3),\beta}\Big).&\Qtildaphialpha
}$$

\appendix{C}{Explicit computations of the 4-point function}

\seclab\appendixc

The computation of the 4-point function correlator in section 3 involves non-trivial algebraic manipulations that were omitted in the main text. The expressions \schannel, \tchannel, \stchannel\ and \uchannel\ were calculated by performing several contractions between superfields. In this appendix, we will show how to derive these expressions in detail. We will start with the $u$-channel, given by \uchannel, which is obtained from the following contractions 
\eqnn\Duchannel
$$\eqalignno{
{\cal A}_4^u &= \left\langle \{U_1^{(3)},V_{23}^{(1)}\}U_4^{(3)}\right\rangle\cr
&= {2\over s_{23}}\left(\left\langle \partial_aU_1^{(3)}\tilde\phi^a_{23}U_4^{(3)}\right\rangle + \left\langle D_\alpha U_1^{(3)}\tilde\phi^\alpha_{23}U_4^{(3)}\right\rangle - \half\left\langle [N^{ab},U_1^{(3)}]\tilde\phi_{23,ab}U_4^{(3)}\right\rangle\right),\ \ \ \ &\Duchannel
}$$
where the overall factor $2/s_{23}$ comes from the integration of the Koba-Nielsen measure. The calculation of this correlator will make use of the equations of motion for $U_1^{(3)}$, given by \partialUthree\ and \DUthree. The first term in the correlator gives 
\eqnn\Duchannelone
$$\eqalignno{
\left\langle \partial_aU_{1}^{(3)}\tilde\phi_{23}^a U_4^{(3)}\right\rangle\
& = 3\left\langle (\lambda\gamma_{ab}\lambda)\phi^a_{1}\tilde\phi_{23}^b U_4^{(3)}\right\rangle + 3\left\langle [Q,C_{1,a}]\tilde\phi_{23}^a U_4^{(3)}\right\rangle\cr
& = 3\left\langle (\lambda\gamma_{ab}\lambda)\phi^a_{1}\tilde\phi_{23}^b U_4^{(3)}\right\rangle +3 \left\langle C_{1,a}\{Q,\tilde\phi_{23}^a\} U_4^{(3)}\right\rangle.&\Duchannelone
}$$
where we also integrated $Q$ by parts. The second term becomes 
\eqnn\Duchanneltwo
$$\eqalignno{
\left\langle D_\alpha U_{1}^{(3)}\tilde\phi_{23}^\alpha U_4^{(3)}\right\rangle & = -3\left\langle \{Q,C_{1,\alpha}\}\tilde\phi_{23}^\alpha U_4^{(3)}\right\rangle -3\left\langle (\lambda\gamma^a)_\alpha C_{1,a} \tilde\phi_{23}^\alpha U_4^{(3)}\right\rangle\cr
& =-3\left\langle C_{1,\alpha}[Q,\tilde\phi_{23}^\alpha] U_4^{(3)}\right\rangle -3\left\langle  C_{1,a} (\lambda\gamma^a\tilde\phi_{23}) U_4^{(3)}\right\rangle.&\Duchanneltwo
}$$
Finally, the third term is 
\eqnn\Duchannelthree
$$\eqalignno{
-\half\left\langle [N^{ab},U_{1}^{(3)}]\tilde\phi_{23,ab} U_4^{(3)}\right\rangle
& = 3\left\langle (\lambda\gamma^{ab})^\alpha C_{1,\alpha}\tilde\phi_{23, ab} U_4^{(3)}\right\rangle = 3\left\langle C_{1,\alpha}\tilde\Omega^\alpha_{3} U_4^{(3)}\right\rangle.&\Duchannelthree
}$$
The final expression for the $u$-channel is given by
\eqnn\Duchannelfinal
$$\eqalignno{
{\cal A}_4^u& = {2\over s_{23}}\Bigg( 3\left\langle (\lambda\gamma_{ab}\lambda)\phi^a_{1}\tilde\phi_{23}^b U_4^{(3)}\right\rangle +3\left\langle C_{1,a}(\{Q,\tilde\phi_{23}^a\} - (\lambda\gamma^a\tilde\phi_{23}) )U_4^{(3)}\right\rangle \cr
& \quad \quad \quad - 3\left\langle C_{1,\alpha}([Q,\tilde\phi_{23}^\alpha] - \tilde\Omega^\alpha_{23} )U_4^{(3)}\right\rangle\Bigg). &\Duchannelfinal
}$$
Using eqns. \tildaphia, \Qtildaphia\ and \Qtildaphialpha, one obtains the final expression \uchannel. 

Let us now perform the same procedure for the $s$-channel, given in eqn. \schannel. Note that the evaluation of the $t$-channel follows a very similar procedure, so we do not reproduce it here. The $s$-channel corresponds to the following contractions 
\eqnn\apschannel
$$\eqalignno{
{\cal A}_4^s &= \left\langle \Big\{\Big[U_1^{(3)}(\tau_1),U_2^{(0)}\Big],U_3^{(1)}(\tau_3)\Big\} U_4^{(3)}(\tau_2)\right\rangle\cr
&= {2\over s_{12}}\Bigg(\left\langle \partial_aU_{12}^{(3)}\phi^a_{3}U_4^{(3)}\right\rangle + \left\langle D_\alpha U_{12}^{(3)}\phi^\alpha_{3}U_4^{(3)}\right\rangle \cr
& - \half\left\langle [N^{ab},U_{12}^{(3)}]\phi_{3,ab}U_4^{(3)}\right\rangle\Bigg) +3\left\langle \{C_{12}, U_3^{(1)}\} U_4^{(3)}\right\rangle, &\apschannel
}$$
where we contract $U^{(3)}$ with $U^{(0)}$ using eqn. \contrationsUUone, with $C_{ij} = C_{i,a}\phi_{j}^{a} + C_{i,\alpha}\phi_{j}^{\alpha}$. Now, let us apply the equations of motion for $U_{12}^{(3)}$ \eqmUijone\ and \eqmUijtwo. The first term of eqn. \apschannel\ becomes 
\eqnn\apschannelone
$$\eqalignno{
\left\langle \partial_aU_{12}^{(3)}\phi_{3}^a U_4^{(3)}\right\rangle& = 3\left\langle (\lambda\gamma_{ba}\lambda)\phi^b_{12}\phi_{3}^a U_4^{(3)}\right\rangle + 3\left\langle [Q,C_{12,a}]\phi_{3}^a U_4^{(3)}\right\rangle + \frac{3s_{12}}{2}\left\langle M_{12,a}\phi_{3}^a U_4^{(3)}\right\rangle\cr
& = 3\left\langle (\lambda\gamma_{ba}\lambda)\phi^b_{12}\phi_{3}^a U_4^{(3)}\right\rangle + 3\left\langle C_{12,a}\{Q,\phi_{3}^a\} U_4^{(3)}\right\rangle + \frac{3s_{12}}{2}\left\langle M_{12,a}\phi_{3}^a U_4^{(3)}\right\rangle, \cr
& \quad  &\apschannelone
}$$
where $M_{12,a} = 2(\lambda\gamma_{ab}\phi_{(1})\phi_{2)}^{b}+ 2(\lambda\gamma_{bc}\lambda)\phi_{(1}^c h^b_{2),a}$. The second term reads 
\eqnn\apschanneltwo
$$\eqalignno{
\left\langle D_\alpha U_{12}^{(3)}\phi_{3}^\alpha U_4^{(3)}\right\rangle& = -3\left\langle \{Q,C_{12,\alpha}\}\phi_{3}^\alpha U_4^{(3)}\right\rangle - 3\left\langle (\lambda\gamma^a)_\alpha C_{12,a} \phi_{3}^\alpha U_4^{(3)}\right\rangle + \frac{3s_{12}}{2}\left\langle M_{12,\alpha}\phi_{3}^\alpha U_4^{(3)}\right\rangle\cr
& = -3\left\langle C_{12,\alpha}[Q,\phi_{3}^\alpha] U_4^{(3)}\right\rangle -  3\left\langle  C_{12,a} (\lambda\gamma^a\phi_{3}) U_4^{(3)}\right\rangle+ \frac{3s_{12}}{2}\left\langle M_{12,\alpha}\phi_{3}^\alpha U_4^{(3)}\right\rangle, \cr
& \quad &\apschanneltwo
}$$
where $M_{12,\alpha} = 2(\lambda\gamma_{ab})_{\alpha}\phi^{a}_{1}\phi_{2}^{b} + 2(\lambda\gamma_{ab}\lambda)h_{(1,\alpha}^{a}\phi_{2)}^{b}$. Finally, the third term gives 
\eqnn\apschannelthree
$$\eqalignno{
-\half\left\langle [N^{ab},U_{12}^{(3)}]\phi_{3,ab} U_4^{(3)}\right\rangle = &\ 3\left\langle (\lambda\gamma^{ab})^\alpha C_{12,\alpha}\phi_{3, ab} U_4^{(3)}\right\rangle = 3\left\langle C_{12,\alpha}\tilde\Omega^\alpha_{3} U_4^{(3)}\right\rangle.&\apschannelthree
}$$
Using the equations of motion for the supergravity fields \eomone-\eomsix, one gets the expression for the $s$-channel \schannel, given by
\eqnn\AsC
$$\eqalignno{
{\cal A}_4^s & = {6 \over s_{12}} \left\langle (\lambda\gamma_{ab}\lambda)\phi^a_{12}\phi_3^b U_4^{(3)}\right\rangle + 3\left\langle \{C_{12}, U_3^{(1)}\} U_4^{(3)}\right\rangle \cr
& + 3\left\langle (\lambda\gamma_{ba}\phi_1)\phi^b_{2}\phi_3^a U_4^{(3)}\right\rangle + 3\left\langle (\lambda\gamma_{ba}\phi_2)\phi^b_{1}\phi_3^a U_4^{(3)}\right\rangle + 6\left\langle (\lambda\gamma_{ba}\phi_3)\phi^b_{1}\phi_2^a U_4^{(3)}\right\rangle \cr
& + 3\left\langle (\lambda\gamma_{cb}\lambda)\phi_1^ch^b_{2,a}\phi_3^a U_4^{(3)}\right\rangle + 3\left\langle (\lambda\gamma_{cb}\lambda)\phi_2^ch^b_{1,a}\phi_3^a U_4^{(3)}\right\rangle\cr
& + 3\left\langle (\lambda\gamma_{ab}\lambda)\phi_2^a h_{1,\alpha}^{b}\phi_{3}^{\alpha}  U_4^{(3)}\right\rangle + 3\left\langle (\lambda\gamma_{ab}\lambda)\phi_1^ah_{2,\alpha}^b\phi_3^\alpha U_4^{(3)}\right\rangle.&\AsC
}$$
The $t$-channel is given by eqn. \tchannel. The sum of the $s$- and $t$-channels contains the following term
\eqnn\Cijcorrelator
$$\eqalignno{
&3\left\langle \{C_{12}, U_3^{(1)}\} U_4^{(3)}\right\rangle + (2\leftrightarrow3)\cr
={}&3\left\langle \{(C_{1a}\phi^a_2),U_3^{(1)}\} U_4^{(3)}\right\rangle + 3\left\langle \{(C_{1\alpha}\phi^\alpha_2), U_3^{(1)}\} U_4^{(3)}\right\rangle + (2\leftrightarrow3).&\Cijcorrelator
}$$
We will now perform these remaining contractions. The contraction of $U_3^{(1)}$ with $\phi$ gives 
\eqnn\Cijphi
$$\eqalignno{
&3\left\langle C_{1a}\{\phi^a_2,U_3^{(1)}\} U_4^{(3)}\right\rangle + 3\left\langle C_{1\alpha}[\phi^\alpha_2,U_3^{(1)}] U_4^{(3)}\right\rangle + (2\leftrightarrow3)\cr
={} & 6\left\langle C_{1a}\Big(\phi_{(2}^c\partial_c\phi^a_{3)}+\phi_{(2}^\delta D_\delta\phi^a_{3)}-\Omega_{(2}^\alpha h^a_{3),\alpha}\Big)U_4^{(3)}\right\rangle\cr
& - 6\left\langle C_{1\alpha}\Big(\phi_{(2}^c\partial_c\phi^\alpha_{3)}+\phi_{(2}^\delta D_\delta\phi^\alpha_{3)}-\Omega_{(2}^\beta h^\alpha_{3),\beta}\Big)U_4^{(3)}\right\rangle.&\Cijphi
}$$
The contraction of $U_3^{(1)}$ with $C$ gives 
\eqnn\CijD
$$\eqalignno{
&3\left\langle \phi^a_2[C_{1a},U_3^{(1)}] U_4^{(3)}\right\rangle- 3\left\langle \phi^\alpha_2\{C_{1\alpha},U_3^{(1)}\} U_4^{(3)}\right\rangle + (2\leftrightarrow3)\cr
=\ & 6\left\langle (\phi^a_{(2}\phi_{3)}^c\partial_c C_{1a} + \phi^a_{(2}\phi_{3)}^\alpha D_\alpha C_{1a} + \phi^a_{(2}\Omega_{3)}^\alpha C_{1,\alpha a} )U_4^{(3)}\right\rangle\cr
=\ & 6\left\langle (\lambda\gamma_{ab}\phi_{1})\phi_{2}^a\phi_3^b U_4^{(3)}\right\rangle + 3\left\langle (\lambda\gamma_{ab}\phi_{2})\phi_{3}^a\phi_1^b U_4^{(3)}\right\rangle + 3\left\langle (\lambda\gamma_{ab}\phi_{3})\phi_{2}^a\phi_1^b U_4^{(3)}\right\rangle\cr
& -6\left\langle (\lambda\gamma_{ab}\lambda)\phi_{(2}^a\phi_{3)}^ch_{1,c}^b U_4^{(3)}\right\rangle - 6\left\langle (\lambda\gamma_{ab}\lambda)\phi_{(2}^a\phi_{3)}^\alpha h_{1,\alpha}^b U_4^{(3)}\right\rangle\cr
&+6\left\langle \left\{Q,\Big( \phi^a_{(2}\phi^b_{3)}C_{1,ab} + \phi^a_{(2}\phi^\alpha_{3)}C_{1,a\alpha} + \phi^\alpha_{(2}\phi^\beta_{3)}C_{1,\alpha\beta} \Big) U_4^{(3)}\right\}\right\rangle.&\CijD
}$$
The final result is 
\eqnn\Cijcorrelatorresult
$$\eqalignno{
&3\left\langle \{C_{12}, U_3^{(1)}\} U_4^{(3)}\right\rangle + (2\leftrightarrow3)\cr
=\ & 6\left\langle (\lambda\gamma_{ab}\phi_{1})\phi_{2}^a\phi_3^b U_4^{(3)}\right\rangle + 3\left\langle (\lambda\gamma_{ab}\phi_{2})\phi_{3}^a\phi_1^b U_4^{(3)}\right\rangle + 3\left\langle (\lambda\gamma_{ab}\phi_{3})\phi_{2}^a\phi_1^b U_4^{(3)}\right\rangle\cr
& -6\left\langle (\lambda\gamma_{ab}\lambda)\phi_{(2}^a\phi_{3)}^ch_{1,c}^b U_4^{(3)}\right\rangle - 6\left\langle (\lambda\gamma_{ab}\lambda)\phi_{(2}^a\phi_{3)}^\alpha h_{1,\alpha}^b U_4^{(3)}\right\rangle
\cr
& +6\left\langle C_{1a}\Big(\phi_{(2}^c\partial_c\phi^a_{3)}+\phi_{(2}^\delta D_\delta\phi^a_{3)}-\Omega_{(2}^\alpha h^a_{3),\alpha}\Big)U_4^{(3)}\right\rangle\cr
& -6\left\langle C_{1\alpha}\Big(\phi_{(2}^c\partial_c\phi^\alpha_{3)}+\phi_{(2}^\delta D_\delta\phi^\alpha_{3)}-\Omega_{(2}^\beta h^\alpha_{3),\beta}\Big)U_4^{(3)}\right\rangle\cr
& +6\left\langle \left\{Q,\Big( \phi^a_{(2}\phi^b_{3)}C_{1,ab} + \phi^a_{(2}\phi^\alpha_{3)}C_{1,a\alpha} + \phi^\alpha_{(2}\phi^\beta_{3)}C_{1,\alpha\beta} \Big) U_4^{(3)}\right\}\right\rangle.&\Cijcorrelatorresult
}$$
Combining this with the expressions \schannel\ and \tchannel, one obtains the result given in eqn. \stchannel.

\subsec Compact form for the 4-point function

\subseclab\appendixcone

In this subsection, we will prove the alternative compact formula for the 4-point amplitude in 11D, given in Section 3 by eqn. \ciclicity. For this purpose, we start from eqn. \allchannels\ and then use the equations of motion for $U^{(3)}$ given by \partialUthree, \DUthree, \eqmUijone, \eqmUijtwo. For a better reading, let us repeat the equations of motion we will use 
\eqnn\Eone
\eqnn\Etwo
\eqnn\Ethree
\eqnn\Efour
$$\eqalignno{
&\partial_a U^{(3)} - 3[Q,C_a] = 3(\lambda\gamma_{ab}\lambda)\phi^b,&\Eone\cr
&D_\alpha U^{(3)} + 3\{Q,C_\alpha\} = -3(\lambda\gamma^a)_\alpha C_a,&\Etwo\cr
&\partial_{a}U^{(3)}_{12} - 3[Q,C_{a,12}] =3(\lambda\gamma_{ab}\lambda)\phi^{b}_{12}+\frac{3s_{12}}{2}M_{a,12},&\Ethree\cr
&D_\alpha U_{12}^{(3)} + 3\{Q,C_{\alpha,12}\} = -3(\lambda\gamma^a)_\alpha C_{a,12} + \frac{3s_{12}}{2}M_{\alpha,12},&\Efour
}$$
where 
\eqnn\ME
$$\eqalignno{
&M_{a,12} = (\lambda\gamma_{ab}\phi_{1})\phi_{2}^{b} + (\lambda\gamma_{ab}\phi_{2})\phi_{1}^{b} + (\lambda\gamma_{bc}\lambda)\phi_1^ch^b_{2,a} + (\lambda\gamma_{bc}\lambda)\phi_2^ch^b_{1,a},\cr
&M_{\alpha,12}=2(\lambda\gamma_{ab})_{\alpha}\phi^{a}_{1}\phi_{2}^{b} + (\lambda\gamma_{ab}\lambda)h_{1,\alpha}^{a}\phi_{2}^{b} + (\lambda\gamma_{ab}\lambda)h_{2,\alpha}^{a}\phi_{1}^{b}.&\ME
}$$
Let us start with the first term in \ciclicity,
\eqnn\startE
$$\eqalignno{
&\left\langle(\lambda\gamma_{ab}\lambda)\phi_{12}^a\phi_3^aU_4^{(3)}\right\rangle = -{1\over3}\left\langle\partial_aU^{(3)}_{12}\phi_3^aU_4^{(3)}\right\rangle +\left\langle [Q,C_{a,12}]\phi_3^aU_4^{(3)}\right\rangle + \frac{s_{12}}{2}\left\langle M_{a,12}\phi_3^aU_4^{(3)}\right\rangle\cr
={} & {1\over3}\left\langle U^{(3)}_{12}\phi_3^a\partial_aU_4^{(3)}\right\rangle - \left\langle C_{a,12}\{Q,\phi_3^a\}U_4^{(3)}\right\rangle + \frac{s_{12}}{2}\left\langle M_{a,12}\phi^a_3U_4^{(3)}\right\rangle\cr
={} & \left\langle U^{(3)}_{12}(\lambda\gamma_{ab}\lambda)\phi_3^a\phi_4^b\right\rangle +\left\langle U_{12}^{(3)}\phi_3^a[Q,C_{a,4}]\right\rangle - \left\langle C_{a,12}(\lambda\gamma^a\phi_3))U_4^{(3)}\right\rangle + \frac{s_{12}}{2}\left\langle M_{a,12}\phi^a_3U_4^{(3)}\right\rangle,\cr
&&\startE
}$$
where we used eqn. \Ethree\ in the first line, we integrated by parts $\partial_a$ and $Q$ in the second line, discarding the boundary terms, and finally used $Q\phi^a=(\lambda\gamma^a\phi)$ and eqn. \Eone\ in the last line. Now, we notice that the first term in the last line of \startE\ is the kind of expression we want. Indeed, by using \QUij\ we find that 
\eqnn\termE
$$\eqalignno{
&\left\langle U^{(3)}_{12}(\lambda\gamma_{ab}\lambda)\phi_3^a\phi_4^b\right\rangle = {2\over3}\left\langle U_{12}^{(3)}\left\{Q,\left(U_{34}^{(3)}/s_{34}\right)\right\}\right\rangle = {2\over3}\left\langle U_{12}^{(3)}\{Q,V_{34}^{(3)}\}\right\rangle.&\termE
}$$
Having identified this important piece of the computation, next we integrate $Q$ by parts in the second term of the last line in \startE. We get 
\eqnn\secondE
$$\eqalignno{
\left\langle(\lambda\gamma_{ab}\lambda)\phi_{12}^a\phi_3^aU_4^{(3)}\right\rangle = & {2\over 3}\left\langle U_{12}^{(3)}\{Q,V_{34}^{(3)}\}\right\rangle+ \left\langle U_{12}^{(3)}(\lambda\gamma^a\phi_3)C_{a,4}\right\rangle - \left\langle C_{a,12}(\lambda\gamma^a\phi_3)U_4^{(3)} \right\rangle\cr
&{}- \frac{3s_{12}}{2}\left\langle (\lambda\gamma_{ab}\lambda)\phi^a_1\phi^b_2\phi_3^eC_{e,4} \right\rangle + \frac{s_{12}}{2}\left\langle M_{a,12}\phi_3^aU_4^{(3)}\right\rangle.&\secondE
}$$
Now, using \Etwo\ and \Efour\ in the second line of the above expression, we get 
\eqnn\thirdE
$$\eqalignno{
\left\langle(\lambda\gamma_{ab}\lambda)\phi_{12}^a\phi_3^aU_4^{(3)}\right\rangle  =&  {2\over 3}\left\langle U_{12}^{(3)}\{Q,V_{34}^{(3)}\}\right\rangle + \frac{s_{12}}{2}\left\langle (M_{a,12}\phi_3^a-M_{\alpha,12}\phi_3^\alpha)U_4^{(3)}\right\rangle\ \ \ \ \cr
& -\frac{3s_{12}}{2}\left\langle (\lambda\gamma_{ab}\lambda)\phi^a_1\phi^b_2(\phi_3^eC_{e,4}+\phi_3^\alpha C_{\alpha,4}) \right\rangle.&\thirdE
}$$
We will now write $C_a$ and $C_\alpha$ in terms of the physical operators ${\bf C}_a$ and ${\bf C}_\alpha$ defined in eqns. \chatalphaexp-\chataexp
\eqnn\nmE
$$\eqalignno{
&C_{a,4} = \{{\bf C}_a,U_4^{(3)}\} + \cdots\ ,\ \ \ \ \ \  C_{\alpha,4} = [{\bf C}_\alpha, U_4^{(3)}] + \cdots, &\nmE
}$$
where $\cdots$ are expressions involving non-minimal variables, which are guaranteed to cancel at the end of the computation. We then use this to manipulate the last line of \thirdE,
\eqnn\NMmanip
$$\eqalignno{
&\ \ \ \ - \frac{3s_{12}}{2}\left\langle (\lambda\gamma_{ab}\lambda)\phi^a_1\phi^b_2(\phi_3^eC_{e,4}+\phi_3^\alpha C_{\alpha,4}) \right\rangle = -\frac{3s_{12}}{2}\left\langle (\lambda\gamma_{ab}\lambda)\phi^a_1\phi^b_2\left(\phi_3^c{\bf C}_c + \phi_3^\alpha{\bf C}_\alpha\right)U_4^{(3)}\right\rangle\cr
& = \frac{3s_{12}}{2}\left\langle\Big[\left( \phi_3^c{\bf C}_c-\phi_3^\alpha{\bf C}_\alpha\right),(\lambda\gamma_{ab}\lambda)\phi^a_1\phi^b_2\Big]U_4^{(3)}\right\rangle\cr
& = -\frac{s_{12}}{2}\left\langle (M_{a,12}\phi_3^a-M_{\alpha,12}\phi_3^\alpha)U_4^{(3)}\right\rangle\cr
&\ \ \ -s_{12}\left\langle(\lambda\gamma_{ab}\phi_3)\phi^a_1\phi^b_2U_4^{(3)}\right\rangle - \frac{s_{12}}{2}\left\langle(\lambda\gamma_{ab}\phi_1)\phi^a_2\phi^b_3U_4^{(3)}\right\rangle - \frac{s_{12}}{2}\left\langle(\lambda\gamma_{ab}\phi_2)\phi^a_1\phi^b_3U_4^{(3)}\right\rangle.\cr
&&\NMmanip
}$$

\noindent Now, we use \thirdE\ together with \NMmanip\ to rewrite the first three terms of \allchannels\ as
\eqnn\threechannelsE
$$\eqalignno{
&{1\over s_{12}} \left\langle (\lambda\gamma_{ab}\lambda)\phi^b_{12}\phi_3^a U_4^{(3)}\right\rangle + {1\over s_{13}} \left\langle (\lambda\gamma_{ab}\lambda)\phi^b_{13}\phi_2^a U_4^{(3)}\right\rangle + {1\over s_{23}} \left\langle (\lambda\gamma_{ab}\lambda)\phi^b_{23}\phi_1^a U_4^{(3)}\right\rangle \cr
= \ &  {2\over3}\left\langle V_{12}^{(3)}\left\{Q,V_{34}^{(3)}\right\}\right\rangle + {2\over3}\left\langle V_{13}^{(3)}\left\{Q,V_{24}^{(3)}\right\}\right\rangle + {2\over3}\left\langle V_{23}^{(3)}\left\{Q,V_{14}^{(3)}\right\}\right\rangle\cr
&\ \ -  2\left\langle (\lambda\gamma_{ab}\phi_1)\phi^b_{2}\phi_3^a U_4^{(3)}\right\rangle -  2\left\langle (\lambda\gamma_{ab}\phi_3)\phi^b_{1}\phi_2^a U_4^{(3)}\right\rangle -   2\left\langle (\lambda\gamma_{ab}\phi_2)\phi^b_{3}\phi_1^a U_4^{(3)}\right\rangle.\cr
&&\threechannelsE
}$$
Therefore, it follows that
\eqnn\ciclE
$$\eqalignno{
{\cal A}_4 &\ = 4\left\langle V_{12}^{(3)}\left\{Q,V_{34}^{(3)}\right\}\right\rangle + 4\left\langle V_{13}^{(3)}\left\{Q,V_{24}^{(3)}\right\}\right\rangle + 4\left\langle V_{23}^{(3)}\left\{Q,V_{14}^{(3)}\right\}\right\rangle,  &\ciclE
}$$
as desired.

\seclab\appendixb

\listrefs

\bye

%% file: harvmacMv3.tex


\input amssym.tex 

\def\unredoffs{}
\tolerance=1000\hfuzz=2pt
\catcode`\@=11 
\ifx\hyperdef\UNd@FiNeD\def\hyperdef#1#2#3#4{#4}\def\hyperref#1#2#3#4{#4}\def\href#1#2{#2}\fi
\magnification=1200\unredoffs\baselineskip=16pt plus 2pt minus 1pt
\def\Date#1{\vfill\leftline{#1}\tenpoint\supereject%
\footline={\hss\tenrm\hyperdef\hypernoname{page}\folio\folio\hss}}%

{\count255=\time\divide\count255 by 60 \xdef\hourmin{\number\count255}
 \multiply\count255 by-60\advance\count255 by\time
 \xdef\hourmin{\hourmin:\ifnum\count255<10 0\fi\the\count255}
}
\def\date{\number\day.\number\month.\number\year\ at \hourmin}


\def\nolabels{\def\wrlabeL##1{}\def\eqlabeL##1{}\def\reflabeL##1{}}
\def\writelabels{\def\wrlabeL##1{\leavevmode\vadjust{\rlap{\smash%
{\line{{\escapechar=` \hfill\rlap{\sevenrm\hskip.03in\string##1}}}}}}}%
\def\eqlabeL##1{{\escapechar-1\rlap{\sevenrm\hskip.05in\string##1}}}%
\def\reflabeL##1{\noexpand\llap{\noexpand\sevenrm\string\string\string##1}}}
\nolabels

\global\newcount\secno \global\secno=0
\global\newcount\meqno \global\meqno=1
\def\s@csym{}

\def\newsec#1\par{\global\advance\secno by1%
{\toks0{#1}\message{(\the\secno. \the\toks0)}}%
\global\subsecno=0\eqnres@t\let\s@csym\secsym\xdef\secn@m{\the\secno}\noindent
{\bf\hyperdef\hypernoname{section}{\the\secno}{\the\secno.} #1}%
\writetoca{{\string\hyperref{}{section}{\the\secno}{\bf \the\secno\quad}} {\bf #1}}\par%
\nobreak\medskip\nobreak\noindent\ignorespaces}
\def\eqnres@t{\xdef\secsym{\the\secno.}\global\meqno=1\bigbreak\bigskip}
\def\sequentialequations{\def\eqnres@t{\bigbreak}}\xdef\secsym{}

\global\newcount\subsecno \global\subsecno=0
\def\subsec#1\par{\global\advance\subsecno by1%
{\toks0{#1}\message{(\s@csym\the\subsecno. \the\toks0)}}%
\global\subsubsecno=0%
\ifnum\lastpenalty>9000\else\bigbreak\fi
\noindent{\it\hyperdef\hypernoname{subsection}{\secn@m.\the\subsecno}%
{\secn@m.\the\subsecno.} #1}\writetoca{\string\hskip1.45cm
{\string\hyperref{}{subsection}{\secn@m.\the\subsecno}{\secn@m.\the\subsecno.}}
{#1}}\par\nobreak\medskip\nobreak\noindent\ignorespaces}

\global\newcount\subsubsecno \global\subsubsecno=0
\def\subsubsec#1\par{\global\advance\subsubsecno by1%
{\toks0{#1}\message{(\secn@m.\the\subsecno.\the\subsubsecno. \the\toks0)}}%
\global\subsubsubsecno=0%
\ifnum\lastpenalty>9000\else\bigbreak\fi
\noindent{\it\hyperdef\hypernoname{subsubsection}{\secn@m.\the\subsecno\the\subsubsecno}%
{\secn@m.\the\subsecno.\the\subsubsecno.} #1}
\par\nobreak\medskip\nobreak\noindent\ignorespaces}

\global\newcount\subsubsubsecno \global\subsubsubsecno=0
\def\subsubsubsec#1\par{\global\advance\subsubsubsecno by1%
{\toks0{#1}\message{(\secn@m.\the\subsecno.\the\subsubsecno.\the\subsubsubsecno \the\toks0)}}%
\ifnum\lastpenalty>9000\else\bigbreak\fi
\noindent{\it\hyperdef\hypernoname{subsubsection}{\secn@m.\the\subsecno\the\subsubsecno\the\subsubsubsecno}%
{\secn@m.\the\subsecno.\the\subsubsecno.\the\subsubsubsecno.} #1}%
\par\nobreak\medskip\nobreak\noindent\ignorespaces}


\def\newnewsec#1#2\par{\global\advance\secno by1%
{\toks0{#2}\message{(\secn@m. \the\toks0)}}%
\global\subsecno=0\global\subsubsecno=0\eqnres@t\let\s@csym\secsym\xdef\secn@m{\the\secno}\noindent
\ifnum\lastpenalty>9000\else\bigbreak\fi
\noindent{\bf\hyperdef\hypernoname{section}{\secn@m}{\secn@m.} #2}%
\writetoca{{\string\hyperref{}{section}{\the\secno}{\bf \the\secno\quad}} {\bf #2}}
\DefWarn#1%
\xdef#1{\noexpand\hyperref{}{section}{\the\secno}%
{\the\secno}}\writedef{#1\leftbracket#1}\wrlabeL{#1=#1}%
\par\nobreak\medskip\nobreak\noindent\ignorespaces}

\def\newsubsec#1#2\par{\global\advance\subsecno by1%
{\toks0{#2}\message{(\secn@m.\the\subsecno. \the\toks0)}}%
\global\subsubsecno=0%
\ifnum\lastpenalty>9000\else\bigbreak\fi
\noindent{\it\hyperdef\hypernoname{subsection}{\secn@m.\the\subsecno}%
{\secn@m.\the\subsecno.} #2}
\DefWarn#1%
\xdef#1{\noexpand\hyperref{}{subsection}{\secn@m.\the\subsecno}%
{\secn@m.\the\subsecno}}\writedef{#1\leftbracket#1}\wrlabeL{#1=#1}%
\writetoca{\string\hskip1.45cm
{\string\hyperref{}{subsection}{\secn@m.\the\subsecno}{\secn@m.\the\subsecno.}}
{#2}}%
\par\nobreak\medskip\nobreak\noindent\ignorespaces}

\def\newsubsubsec#1#2\par{\global\advance\subsubsecno by1%
{\toks0{#2}\message{(\secn@m.\the\subsecno.\the\subsubsecno. \the\toks0)}}%
\global\subsubsubsecno=0%
\ifnum\lastpenalty>9000\else\bigbreak\fi
\noindent{\it\hyperdef\hypernoname{subsubsection}{\secn@m.\the\subsecno\the\subsubsecno}%
{\secn@m.\the\subsecno.\the\subsubsecno.} #2}
\DefWarn#1%
\xdef#1{\noexpand\hyperref{}{subsubsection}{\secn@m.\the\subsecno.\the\subsubsecno}%
{\secn@m.\the\subsecno.\the\subsubsecno}}\writedef{#1\leftbracket#1}\wrlabeL{#1=#1}%
\par\nobreak\medskip\nobreak\noindent\ignorespaces}

\def\newsubsubsubsec#1#2\par{\global\advance\subsubsubsecno by1%
{\toks0{#2}\message{(\secn@m.\the\subsecno.\the\subsubsecno.\the\subsubsubsecno \the\toks0)}}%
\ifnum\lastpenalty>9000\else\bigbreak\fi
\noindent{\it\hyperdef\hypernoname{subsubsection}{\secn@m.\the\subsecno\the\subsubsecno\the\subsubsubsecno}%
{\secn@m.\the\subsecno.\the\subsubsecno.\the\subsubsubsecno.} #2}
\DefWarn#1%
\xdef#1{\noexpand\hyperref{}{subsubsubsection}{\secn@m.\the\subsecno.\the\subsubsecno.\the\subsubsubsecno}%
{\secn@m.\the\subsecno.\the\subsubsecno.\the\subsubsubsecno}}\writedef{#1\leftbracket#1}\wrlabeL{#1=#1}%
\par\nobreak\medskip\nobreak\noindent\ignorespaces}

\def\appendix#1#2{\global\meqno=1\global\subsecno=0\global\subsubsecno=0\xdef\secsym{\hbox{#1.}}%
\bigbreak\bigskip\noindent{\bf Appendix \hyperdef\hypernoname{appendix}{#1}%
{#1.} #2}{\toks0{(#1. #2)}\message{\the\toks0}}%
\xdef\s@csym{#1.}\xdef\secn@m{#1}%
\writetoca{{\string\hyperref{}{appendix}{#1}{\bf {#1}\quad}} {\bf #2}}%
\par\nobreak\medskip\nobreak}

%
\def\checkm@de#1#2{\ifmmode{\def\f@rst##1{##1}\hyperdef\hypernoname{equation}%
{#1}{#2}}\else\hyperref{}{equation}{#1}{#2}\fi}
\def\eqnn#1{\DefWarn#1\xdef #1{(\noexpand\relax\noexpand\checkm@de%
{\s@csym\the\meqno}{\secsym\the\meqno})}%
\wrlabeL#1\writedef{#1\leftbracket#1}\global\advance\meqno by1}
\def\f@rst#1{\c@t#1a\em@ark}\def\c@t#1#2\em@ark{#1}
\def\eqna#1{\DefWarn#1\wrlabeL{#1$\{\}$}%
\xdef #1##1{(\noexpand\relax\noexpand\checkm@de%
{\s@csym\the\meqno\noexpand\f@rst{##1}1}{\hbox{$\secsym\the\meqno##1$}})}
\writedef{#1\numbersign1\leftbracket#1{\numbersign1}}\global\advance\meqno by1}
\def\eqn#1#2{\DefWarn#1%
\xdef #1{(\noexpand\hyperref{}{equation}{\s@csym\the\meqno}%
{\secsym\the\meqno})}$$#2\eqno(\hyperdef\hypernoname{equation}%
{\s@csym\the\meqno}{\secsym\the\meqno})\eqlabeL#1$$%
\writedef{#1\leftbracket#1}\global\advance\meqno by1}
\def\xeqn{\expandafter\xe@n}\def\xe@n(#1){#1}
\def\xeqna#1{\expandafter\xe@n#1}
\def\eqns#1{(\e@ns #1{\hbox{}})}
\def\e@ns#1{\ifx\UNd@FiNeD#1\message{eqnlabel \string#1 is undefined.}%
\xdef#1{(?.?)}\fi{\let\hyperref=\relax\xdef\next{#1}}%
\ifx\next\em@rk\def\next{}\else%
\ifx\next#1\xeqn#1\else\def\n@xt{#1}\ifx\n@xt\next#1\else\xeqna#1\fi
\fi\let\next=\e@ns\fi\next}
\def\DefWarn#1{}
%
\newskip\footskip\footskip14pt plus 1pt minus 1pt 
\def\footnotefont{\ninepoint}\def\f@t#1{\footnotefont #1\@foot}
\def\f@@t{\baselineskip\footskip\bgroup\footnotefont\aftergroup\@foot\let\next}
\setbox\strutbox=\hbox{\vrule height9.5pt depth4.5pt width0pt}
\global\newcount\ftno \global\ftno=0
\def\foot{\global\advance\ftno by1\def\foot@rg{\hyperref{}{footnote}%
{\the\ftno}{\the\ftno}\xdef\foot@rg{\noexpand\hyperdef\noexpand\hypernoname%
{footnote}{\the\ftno}{\the\ftno}}}\footnote{$^{\foot@rg}$}}
%
%
%
\global\newcount\refno \global\refno=1
\newwrite\rfile
\def\ref{[\hyperref{}{reference}{\the\refno}{\the\refno}]\nref}
\def\nref#1{\DefWarn#1%
\xdef#1{[\noexpand\hyperref{}{reference}{\the\refno}{\the\refno}]}%
\writedef{#1\leftbracket#1}%
\ifnum\refno=1\immediate\openout\rfile=\jobname.refs\fi
\chardef\wfile=\rfile\immediate\write\rfile{\noexpand\item{[\noexpand\hyperdef%
\noexpand\hypernoname{reference}{\the\refno}{\the\refno}]\ }%
\reflabeL{#1\hskip.31in}\pctsign}\global\advance\refno by1\findarg}
\def\findarg#1#{\begingroup\obeylines\newlinechar=`\^^M\pass@rg}
{\obeylines\gdef\pass@rg#1{\writ@line\relax #1^^M\hbox{}^^M}%
\gdef\writ@line#1^^M{\expandafter\toks0\expandafter{\striprel@x #1}%
\edef\next{\the\toks0}\ifx\next\em@rk\let\next=\endgroup\else\ifx\next\empty%
\else\immediate\write\wfile{\the\toks0}\fi\let\next=\writ@line\fi\next\relax}}
\def\striprel@x#1{} \def\em@rk{\hbox{}}
\def\lref{\begingroup\obeylines\lr@f}
\def\lr@f#1#2{\DefWarn#1\gdef#1{\let#1=\UNd@FiNeD\ref#1{#2}}\endgroup\unskip}

\def\addref#1{\immediate\write\rfile{\noexpand\item{}#1}} 
\def\listrefs{\vfill\supereject\immediate\closeout\rfile\writestoppt
\baselineskip=\footskip\centerline{{\bf References}}\bigskip{\parindent=20pt%
\frenchspacing\escapechar=` \input \jobname.refs\vfill\eject}\nonfrenchspacing}
\def\startrefs#1{\immediate\openout\rfile=\jobname.refs\refno=#1}
\def\xref{\expandafter\xr@f}\def\xr@f[#1]{#1}
\def\refs#1{\count255=1[\r@fs #1{\hbox{}}]}
\def\r@fs#1{\ifx\UNd@FiNeD#1\message{reflabel \string#1 is undefined.}%
\nref#1{need to supply reference \string#1.}\fi%
\vphantom{\hphantom{#1}}{\let\hyperref=\relax\xdef\next{#1}}%
\ifx\next\em@rk\def\next{}%
\else\ifx\next#1\ifodd\count255\relax\xref#1\count255=0\fi%
\else#1\count255=1\fi\let\next=\r@fs\fi\next}
%

%
\newwrite\ffile\global\newcount\figno \global\figno=1
\def\fig{fig.~\hyperref{}{figure}{\the\figno}{\the\figno}\nfig}
\def\nfig#1{\DefWarn#1%
\xdef#1{fig.~\noexpand\hyperref{}{figure}{\the\figno}{\the\figno}}%
\writedef{#1\leftbracket fig.\noexpand~\xfig#1}%
\ifnum\figno=1\immediate\openout\ffile=\jobname.figs\fi\chardef\wfile=\ffile%
{\let\hyperref=\relax
\immediate\write\ffile{\noexpand\medskip\noexpand\item{Fig.\ %
\noexpand\hyperdef\noexpand\hypernoname{figure}{\the\figno}{\the\figno}. }
\reflabeL{#1\hskip.55in}\pctsign}}\global\advance\figno by1\findarg}
\def\xfig{\expandafter\xf@g}\def\xf@g fig.\penalty\@M\ {}
\def\figs#1{figs.~\f@gs #1{\hbox{}}}
\def\f@gs#1{{\let\hyperref=\relax\xdef\next{#1}}\ifx\next\em@rk\def\next{}\else
\ifx\next#1\xfig #1\else#1\fi\let\next=\f@gs\fi\next}
%
\def\figin{\epsfcheck\figin}\def\figins{\epsfcheck\figins}
\def\epsfcheck{\ifx\epsfbox\UnDeFiNeD
\message{(NO epsf.tex, FIGURES WILL BE IGNORED)}
\gdef\figin##1{\vskip2in}\gdef\figins##1{\hskip.5in}
\else\message{(FIGURES WILL BE INCLUDED)}%
\gdef\figin##1{##1}\gdef\figins##1{##1}\fi}
\def\figinsert{\goodbreak\topinsert}
\def\ifig#1#2#3{\DefWarn#1\xdef#1{fig.~\the\figno}
\writedef{#1\leftbracket fig.\noexpand~\the\figno}%
\figinsert\figin{\centerline{#3}}
\smallskip
\leftskip=0pt \rightskip=0pt
\baselineskip12pt\noindent
{{\bf Fig.~\the\figno}\ \ninepoint #2}
\medskip
\global\advance\figno by1\par\endinsert}
\newwrite\lfile
{\escapechar-1\xdef\pctsign{\string\%}\xdef\leftbracket{\string\{}
\xdef\rightbracket{\string\}}\xdef\numbersign{\string\#}}
\def\writedefs{\immediate\openout\lfile=label.defs \def\writedef##1{%
{\let\hyperref=\relax\let\hyperdef=\relax\let\hypernoname=\relax
 \immediate\write\lfile{\string\checkdef\string##1\rightbracket}}}}%
\def\writestop{\def\writestoppt{\immediate\write\lfile{\string\pageno
 \the\pageno\string\startrefs\leftbracket\the\refno\rightbracket
 \string\def\string\secsym\leftbracket\secsym\rightbracket
 \string\secno\the\secno\string\meqno\the\meqno}\immediate\closeout\lfile}}
\def\writestoppt{}\def\writedef#1{}

\def\seclab#1\par{\DefWarn#1%
\xdef #1{\noexpand\hyperref{}{section}{\the\secno}{\the\secno}}%
\writedef{#1\leftbracket#1}\wrlabeL{#1=#1}\par%
\nobreak\medskip\nobreak\noindent\ignorespaces}
\def\subseclab#1\par{\DefWarn#1%
\xdef #1{\noexpand\hyperref{}{subsection}{\the\secno.\the\subsecno}%
{\the\secno.\the\subsecno}}\writedef{#1\leftbracket#1}\wrlabeL{#1=#1}\par%
\nobreak\medskip\nobreak\noindent\ignorespaces}
\def\subsubseclab#1\par{\DefWarn#1%
\xdef#1{\noexpand\hyperref{}{subsubsection}{\the\secno.\the\subsecno.\the\subsubsecno}%
{\the\secno.\the\subsecno.\the\subsubsecno}}\writedef{#1\leftbracket#1}\wrlabeL{#1=#1}\par%
\nobreak\medskip\nobreak\noindent\ignorespaces}
\def\applab#1\par{\DefWarn#1%
\xdef#1{\noexpand\hyperref{}{appendix}{\secn@m}{\secn@m}}%
\writedef{#1\leftbracket#1}\wrlabeL{#1=#1}%
\par\nobreak\medskip\nobreak\noindent\ignorespaces}
\def\appsublab#1{\DefWarn#1%
\xdef #1{\noexpand\hyperref{}{appendix}{\secn@m.\the\subsecno}{\secn@m.\the\subsecno}}%
\writedef{#1\leftbracket#1}\wrlabeL{#1=#1}}
\newwrite\tfile \def\writetoca#1{}
\def\leaderfill{\leaders\hbox to 1em{\hss.\hss}\hfill}
\def\writetoc{\immediate\openout\tfile=\jobname.toc
   \def\writetoca##1{{\edef\next{\write\tfile{\noindent ##1
   \string\leaderfill{
   \string\hyperref{}{page}{\noexpand\number\pageno}%
   {\noexpand\number\pageno}} \par}}\next}}
}
\newread\ch@ckfile
\def\listtoc{\immediate\closeout\tfile\immediate\openin\ch@ckfile=\jobname.toc
\ifeof\ch@ckfile\message{no file \jobname.toc, no table of contents this pass}%
\else\closein\ch@ckfile\centerline{\bf Contents}\nobreak\medskip%
{\baselineskip=15.5pt\footnotefont\parskip=0pt\catcode`\@=11\input\jobname.toc
\catcode`\@=12\bigbreak\bigskip}\fi}
\catcode`\@=12 
\def\tenpoint{\def\rm{\fam0\tenrm}
\textfont0=\tenrm \scriptfont0=\sevenrm \scriptscriptfont0=\fiverm
\textfont1=\teni  \scriptfont1=\seveni  \scriptscriptfont1=\fivei
\textfont2=\tensy \scriptfont2=\sevensy \scriptscriptfont2=\fivesy
\textfont\itfam=\tenit \def\it{\fam\itfam\tenit}\def\footnotefont{\ninepoint}%
\textfont\bffam=\tenbf \def\bf{\fam\bffam\tenbf}\def\sl{\fam\slfam\tensl}\rm}
\font\ninerm=cmr9 \font\sixrm=cmr6 \font\ninei=cmmi9 \font\sixi=cmmi6
\font\ninesy=cmsy9 \font\sixsy=cmsy6 \font\ninebf=cmbx9
\font\nineit=cmti9 \font\ninesl=cmsl9 \skewchar\ninei='177
\skewchar\sixi='177 \skewchar\ninesy='60 \skewchar\sixsy='60
\def\ninepoint{\def\rm{\fam0\ninerm}
\textfont0=\ninerm \scriptfont0=\sixrm \scriptscriptfont0=\fiverm
\textfont1=\ninei \scriptfont1=\sixi \scriptscriptfont1=\fivei
\textfont2=\ninesy \scriptfont2=\sixsy \scriptscriptfont2=\fivesy
\textfont\itfam=\ninei \def\it{\fam\itfam\nineit}\def\sl{\fam\slfam\ninesl}%
\textfont\bffam=\ninebf \def\bf{\fam\bffam\ninebf}\rm}
%
\hyphenation{anom-aly anom-alies coun-ter-term coun-ter-terms}

\def\tikzcaption#1#2{\DefWarn#1\xdef#1{Fig.~\the\figno}
\writedef{#1\leftbracket Fig.\noexpand~\the\figno}%
{
\smallskip
\leftskip=20pt \rightskip=20pt \baselineskip12pt\noindent
{{\bf Fig.~\the\figno}\ \ninepoint #2}
\bigskip
\global\advance\figno by1 \par}}

\def\ntoalpha#1{%
\ifcase#1%
@%
\or A\or B\or C\or D\or E\or F\or G\or H\or I\or J\or K\or L\or M%
\fi
}

\global\newcount\appno \global\appno=1
\def\applab#1{\xdef #1{\ntoalpha{\appno}}\writedef{#1\leftbracket#1}\wrlabeL{#1=#1}
\global\advance\appno by1}

\def\preprint#1 #2\par{\rightline{\vbox{\baselineskip12pt\hbox{#1}\hbox{#2}}}\vskip2cm}
%
\def\title#1\par{\centerline{\bf #1}\nopagenumbers\pageno=0}
\def\author#1\par{\bigskip\bigskip\centerline{#1}}

\newcount\addressno

\def\email#1#2{
\footnote{\null}{\kern-\parindent \llap{$^#1$\hskip1pt}email: #2}}

\def\startcenter{%
  \par
  \begingroup
  \leftskip=0pt plus 1fil
  \rightskip=\leftskip
  \parindent=0pt
  \parfillskip=0pt
}
\def\stopcenter{\endgroup}

\def\address{\bigskip%
  \ifnum\the\addressno=0\else\stopcenter\endgroup\fi
  \advance\addressno by 1%
  \begingroup
  \startcenter
  \it
  \obeylines
  \addressAux
}
\def\addressAux#1{#1}

\def\abstract{\stopcenter\endgroup\bigskip\bigskip\noindent}

\def\Dsl{\,\raise.15ex\hbox{/}\mkern-13.5mu D} 
\def\dsl{\raise.15ex\hbox{/}\kern-.57em\partial}
 
\def\boxeqn#1{\vcenter{\vbox{\hrule\hbox{\vrule\kern3pt\vbox{\kern3pt
	\hbox{${\displaystyle #1}$}\kern3pt}\kern3pt\vrule}\hrule}}}


\def\half{{1\over 2}}

\def\bar{\overline}
\def\({\left(}
\def\){\right)}




\def\sfrac#1/#2{\kern.1em\raise.5ex\hbox{\the\scriptfont0 #1}%
\kern-.1em/\kern-.15em\lower.25ex\hbox{\the\scriptfont0 #2}}



\def\qed{\hbox{\hskip 3pt
\vbox{\hrule\hbox to 7pt{\vrule height 7pt\hfill\vrule}
\hrule}}\hskip3pt}

\overfullrule=0pt\relax

\frenchspacing

\def\checkdef#1#2{
\ifx\UndeFined#1%
	\def#1{#2}
\else
	\immediate\write16{*** BUG ***: the label \string#1 is already defined ***}
\fi
}
\newread\instream

\openin\instream= label.defs
\ifeof\instream\message{No labels in advance yet. Wait till next pass.}
\else\closein\instream \input label.defs
\fi
\writedefs

\def\arXiv:#1].{\hepthStrip#1 \nil}
\def\hepthStrip#1 #2\nil{\href{http://arxiv.org/abs/#1}{arXiv:#1 #2\unskip}].}


\def\tikzcaption#1#2{\DefWarn#1\xdef#1{Fig.~\the\figno}
\writedef{#1\leftbracket Fig.\noexpand~\the\figno}%
{
\smallskip
\leftskip=20pt \rightskip=20pt \baselineskip12pt\noindent
{{\bf Fig.~\the\figno}\ \ninepoint #2}
\bigskip
\global\advance\figno by1 \par}}

\def\ntoalpha#1{%
\ifcase#1%
@%
\or A\or B\or C\or D\or E\or F\or G\or H\or I\or J\or K\or L\or M%
\fi
}

\global\newcount\appno \global\appno=1
\def\applab#1{\xdef #1{\ntoalpha{\appno}}\writedef{#1\leftbracket#1}\wrlabeL{#1=#1}
\global\advance\appno by1}

\def\preprint#1 #2\par{\rightline{\vbox{\baselineskip12pt\hbox{#1}\hbox{#2}}}\vskip2cm}
%
\def\title#1\par{\centerline{\bf #1}\nopagenumbers\pageno=0}
\def\author#1\par{\bigskip\bigskip\centerline{#1}}

\newcount\addressno

\def\email#1#2{
\footnote{\null}{\kern-\parindent \llap{$^#1$\hskip1pt}email: #2}}

\def\startcenter{%
  \par
  \begingroup
  \leftskip=0pt plus 1fil
  \rightskip=\leftskip
  \parindent=0pt
  \parfillskip=0pt
}
\def\stopcenter{\endgroup}

\def\address{\bigskip%
  \ifnum\the\addressno=0\else\stopcenter\endgroup\fi
  \advance\addressno by 1%
  \begingroup
  \startcenter
  \it
  \obeylines
  \addressAux
}
\def\addressAux#1{#1}

\def\abstract{\stopcenter\endgroup\bigskip\bigskip\noindent}

\def\Dsl{\,\raise.15ex\hbox{/}\mkern-13.5mu D} 
\def\dsl{\raise.15ex\hbox{/}\kern-.57em\partial}
 
\def\boxeqn#1{\vcenter{\vbox{\hrule\hbox{\vrule\kern3pt\vbox{\kern3pt
	\hbox{${\displaystyle #1}$}\kern3pt}\kern3pt\vrule}\hrule}}}


\def\half{{1\over 2}}

\def\bar{\overline}
\def\({\left(}
\def\){\right)}


\def\bU{{\Bbb U}}



\def\sfrac#1/#2{\kern.1em\raise.5ex\hbox{\the\scriptfont0 #1}%
\kern-.1em/\kern-.15em\lower.25ex\hbox{\the\scriptfont0 #2}}



\def\qed{\hbox{\hskip 3pt
\vbox{\hrule\hbox to 7pt{\vrule height 7pt\hfill\vrule}
\hrule}}\hskip3pt}

\overfullrule=0pt\relax

\frenchspacing

\def\checkdef#1#2{
\ifx\UndeFined#1%
	\def#1{#2}
\else
	\immediate\write16{*** BUG ***: the label \string#1 is already defined ***}
\fi
}